\documentclass[a4paper,fleqn,usenatbib]{mnras}

\usepackage{newtxtext,newtxmath}

\usepackage[T1]{fontenc}
\usepackage{ae,aecompl}

\usepackage{graphicx}	
\usepackage{amsmath}	
\usepackage{amssymb}	
\usepackage{footnote}
\usepackage{pdflscape}
\usepackage{tablefootnote}

\newcommand\kms{km\,s$^{-1}$}

\title[Environmental Effects on Galaxy Kinematics]{Searching for Environmental Effects on Galaxy Kinematics in Groups and Clusters at $z\sim1$ from the ORELSE Survey}

\author[D. Pelliccia et al.]{Debora Pelliccia$^{1}$\thanks{E-mail: dpelliccia@ucdavis.edu},
Brian C. Lemaux$^{1}$, Adam R. Tomczak$^{1}$, Lori M. Lubin$^{1}$, \newauthor Lu Shen$^{1}$, Beno\^{i}t Epinat$^{2}$, Po-Feng Wu$^{3}$, Roy R. Gal$^{4}$, Nicholas Rumbaugh$^{5}$, \newauthor Dale D. Kocevski$^{6}$,  Laurence Tresse$^{7}$, Gordon Squires$^{8}$
\\
$^{1}$Department of Physics, University of California, Davis, 1 Shields Avenue, Davis CA 95616, USA\\
$^{2}$Aix Marseille Univ, CNRS, LAM, Laboratoire d'Astrophysique de Marseille,  F-13013 Marseille, France \\
$^{3}$Max-Planck Institut f\"{u}r Astronomie, K\"{o}nigstuhl 17, D-69117, Heidelberg, Germany\\
$^{4}$ University of Hawai'i, Institute for Astronomy, 2680 Woodlawn Drive, Honolulu, HI 96822, USA\\
$^{5}$ National Center for Supercomputing Applications, University of Illinois, 1205 West Clark St., Urbana, IL 61801, USA \\
$^{6}$ Department of Physics and Astronomy, Colby College, Waterville, ME 04961, USA\\
$^{7}$ Univ Lyon, Univ Lyon1, Ens de Lyon, CNRS, Centre de Recherche Astrophysique de Lyon UMR5574, F-69230 Saint-Genis-Laval, France \\
$^{8}$ Spitzer Science Centre, California Institute of Technology, M/S 220-6, 1200 E. California Blvd., Pasadena, CA 91125, USA
}

\date{Accepted 2018 October 21. Received 2018 October 12; in original form 2018 July 12}

\pubyear{2018}

\begin{document}
\label{firstpage}
\pagerange{\pageref{firstpage}--\pageref{lastpage}}
\maketitle
\begin{abstract}
We present an investigation of the dependence of galaxy kinematics on the environment for a sample of 94 star-forming galaxies at $z\sim0.9$ from the ORELSE survey. ORELSE is a large photometric and spectroscopic campaign dedicated to mapping out and characterizing galaxy properties across a full range of environments in 15 fields containing large-scale structures (LSSs) in a redshift range of $0.6 < z < 1.3$. We constrained the rotation velocity for our kinematic sample in an ORELSE field, containing the SC1604 supercluster, by fitting high-resolution semi-analytical models to the data. We constructed the stellar-mass/B-band Tully-Fisher relation and found no dependence of the intrinsic scatter on both local and global environment. Moreover, we compared the stellar-to-dynamical mass ratio ($M_\ast/M_{dyn}$) of SC1604 galaxies to those residing in less dense local environment by leveraging data from the HR-COSMOS sample.
We found that, at fixed stellar mass,  SC1604 galaxies have  $\sim30\%$ smaller dynamical masses on average. By comparing the distributions of the galaxy parameters that define  $M_{dyn}$ (i.e., circular velocity and the characteristic radius $r_{2.2}$) between SC1604 and HR-COSMOS, we found that  smaller dynamical masses are mostly caused by smaller $r_{2.2}$ for SC1604 galaxies. We also observed that SC1604 galaxies in general show $\sim20\%$ lower stellar specific angular momentum ($j_\ast$) with respect to the HR-COSMOS sample. Adopting literature estimates for (1) the excess rate of galaxy-galaxy mergers in intermediate/high-density environments and (2) the average amount of $j_\ast$ loss per merger event, we investigated the possibility that galaxy mergers are mainly responsible for the loss of angular momentum in higher density environments.

\end{abstract}




\section{Introduction}
Galaxy evolution is one of the most debated topic of modern astrophysics. Although many discoveries have improved its investigation, much is left to be understood about the processes that drive the total mass assembly of galaxies.
It is well known that galaxies go through various transformations during their lifetime, as suggested by the observed time evolution of their cosmic star formation rate density \citep[e.g.,][]{MadauDickinson2014}, stellar and molecular gas mass density \citep[e.g.,][]{Dickinson2003, Tomczak2014, Scoville2017}, size \citep[e.g.,][]{vanderWel2014}, and morphology \citep[e.g.,][]{vandenBergh2002, Mortlock2013}.
However it is still not well understood how galaxies assemble their total mass, i.e., how galaxies grow according to their total mass components (stars, gas, and dark matter), mainly because of the impossibility to directly observe the dark matter.

Galaxy internal kinematics can help in this respect, since galaxy rotation velocity traces the radial distribution of the galaxy's dynamical (i.e., total) mass, allowing for indirect estimation of dark matter mass, once stellar and gas mass are known. Such an estimation contributes to the understanding of the interplay between galaxy total mass components during its assembly. The Tully-Fisher relation \citep[TFR,][]{Tully1977} is a well-established scaling relation between galaxy luminosity and rotation velocity and is one of the best tool to investigate galaxy evolution. The same relation can be expressed  substituting galaxy luminosity with its stellar mass, called the stellar-mass Tully-Fisher relation (smTFR), which is observed to be tighter than the luminosity form of this relation and provides a direct connection between the stellar and total  (probed by the velocity) mass of the galaxy. Constraining the relation at different cosmic epochs would, therefore, allow us to trace the evolution of this connection and to investigate  the galaxy total mass growth. Moreover, the TFR (or smTFR) is largely used as a test to verify the theoretical models of galaxy formation and evolution  \citep[e.g.,][]{Dalcanton1997, Sommer-Larsen2003}. 
To date, quite a few studies have investigated the evolution of the smTFR with time and, although a large fraction of these studies \citep[e.g.,][]{Conselice2005, Miller2011, Pelliccia2017, Harrison2017} found a non-evolution of the smTFR up to redshift $z\sim 1$, others have claimed to observe an evolution at  $z\sim 1$ \citep[e.g.,][]{Tiley2016}, or at higher redshifts \citep[e.g.,][]{Cresci2009, Gnerucci2011, Straatman2017}.

The evolution of galaxies is also determined by the environment in which they reside. The idea is that once galaxies from a field-like environment enter a denser region, they would be affected by the interaction with other galaxies and/or by the gravitational potential of the overarching halo, resulting in a different evolutionary path with respect to the galaxies that remain in lower density environments. It has been observed that in the local Universe denser environments (i.e., cluster-like environments) are dominated by red, passive, early-type galaxies whereas less dense regions are preferentially populated by blue, star-forming, late-type systems \citep[e.g.,][]{Dressler1980, Peng2010}. These trends still hold at higher redshifts \citep[e.g.,][]{Cooper2007, Scoville2013}, although they become less clear at these epochs. 
Several processes taking place in higher density environments are able to affect galaxy evolution. Amongst others, galaxy-galaxy interactions (i.e., mergers or tidal interaction during high-speed encounters), ram pressure stripping due to the pressure exerted by the intracluster medium \citep{GunnGott1972}, and ``starvation'' \citep{Larson1980, Balogh2000}, which is an exhaustion of the galaxy's gas due to the cut off of the gas reservoir, are some of the processes acting on galaxies in dense regions. Each of these processes are effective in overlapping regions in cluster environments \citep[e.g.,][]{Treu2003, Moran2007}, with ram pressure stripping being more effective closer to the cluster core, starvation and tidal interactions being important at larger distances, and mergers being dominant in the outskirt of the cluster or prior to entering the cluster environment completely, where ``pre-processing'' can occur in galaxy groups which will be infalling into the cluster.

It is still an open question whether environment is as effective at influencing galaxy kinematics as for other galaxy properties. 
Several effects might impact the position in the smTFR plane of a galaxy in a dense environment. A temporary increase of the star formation activity or feedback can shift a galaxy along the stellar-mass axis or kinematic asymmetry induced by interactions between galaxies in dense environments would shift galaxies along the velocity axis.
\mbox{N-body/hydrodynamical} simulations \citep{Kronberger2008} showed that ram pressure stripping can introduce distortions in the gas rotation velocity at large radii ($\sim$12\,kpc), although this effect is considered low compared to the distortions caused by tidal interactions and is difficult to observe at intermediate redshift in seeing limited observations. However, ground-based IFU observations at low redshift have been able to observe the stripped ionized gas and to provide information on the kinematical and physical properties of this ionized gas as well as of stars \citep{Poggianti2017}.
A few works have studied the dependence of the TFR and smTFR on the environment. At $z\geq0.1$ some authors investigated the effects of the environment, defined as ``cluster'' or ``field'' \citep{Bosch2013, PerezMartinez2017} or by local density \citep{Pelliccia2017}, on the smTFR and found no variations in the relation. This may be a consequence of small sample of galaxies, especially in the higher density regions (i.e., cluster core or highest local overdensity). However, changes in velocity and stellar mass may happen on the same timescale along the relation, while galaxy luminosity, e.g., rest-frame B band, can change in shorter timescale, since it probes recent episodes of star formation.  For this reason, past works have explored the B-band TFR as a function of the environment, however finding discordant results. Some authors \citep[e.g.,][]{Ziegler2003, Nakamura2006, Jaffe2011} have found no effect of the environment either on the fitted relation, nor on the scatter around the relation, while other authors \citep{Bamford2005, Bosch2013, PerezMartinez2017} have identified the effect of environment in a shift of the relation towards higher or lower luminosity for galaxies in dense environment. This discordance may come as a consequence of not measuring kinematics and/or environment in a consistent manner, and of small galaxy samples combined with highly stochastic processes. 

The sizes of galaxies are also affected by environment. It has been observed at intermediate redshift that spiral galaxies in dense environments are, in general, more compact \citep{Maltby2010, CebrianTrujillo2014, Kuchner2017} than their counterpart in less dense environments. Ram pressure stripping is a process that can explain these observations in clusters, by stripping the gas in the outskirt of the galaxy (where the gas is less bound), and effectively reducing the galaxy size. It has been also proposed \citep[e.g.,][]{Bekki1998, Querejeta2015} that major mergers can produce S0 galaxy remnants, which are characterized by more concentrated bulge and a faded disk. This process would in practice reduce the measured half-light radius (radius containing half of the total galaxy light) of galaxies. 

Galaxy specific angular momentum $j$ (angular momentum per stellar mass) is one of the most fundamental quantity to describe a galaxy \citep[e.g.,][]{Fall1983}, and it can be estimated from the measurements of galaxy rotation curves and sizes. In $\Lambda$CDM cosmology, dark matter haloes acquire rotation from tidal torques \citep{Hoyle1951}, which is then transferred to the baryonic matter before the process of galaxy formation starts. 
Galaxies are supposed to maintain their specific angular momentum if they do not undergo transformations that are able to reduce/increase the galaxy angular momentum. Reduction in $j$ has been observed and seems to correlate with galaxy morphology. \citet{Fall1983} discovered the existence of a fundamental relation between the galaxy stellar specific angular momentum $j_\ast$ and stellar mass $M_\ast$, valid for both spiral and elliptical galaxies ($j_\ast\propto M_\ast^{\sim2/3}$), but at a given $M_\ast$ the specific angular momentum of ellipticals is approximately five times smaller than that of spirals. By comparing $j_\ast$ predicted by theory in the case where galaxies do not lose their initial angular momentum with measured $j_\ast$ values, it is possible to estimate how much of the original angular momentum galaxies have lost. \citet{RomanowskyFall2012}, for example, found that spiral and elliptical galaxies in the local Universe have retained 80\% and 10\%, respectively, of their estimated initial specific angular momentum. One of the main processes thought to be responsible for the lost of angular momentum are mergers. \citet{Lagos2018}, using the Evolution and Assembly of GaLaxies and their Environments (EAGLE) simulations \citep{Schaye2015}, have shown that major mergers are able to reduce $j_\ast$ by $\sim 20\%$, with gas-poor mergers being more efficient than gas-rich mergers. 

In this study, we investigate the dependence of galaxy kinematics on the environment for a sample of star-forming galaxies at $z\sim0.9$, which are part of the Observations of Redshift Evolution in Large-Scale Environments survey \citep[ORELSE;][]{Lubin2009}. Due to the thousands of high-quality spectra and extensive photometry available, ORELSE allows for a 3-D mapping of the density field around 15 known large-scale structures \citep[e.g.,][]{Lemaux2017}, providing measurements of local and global environment. Combining these accurate measurements of environment with galaxy kinematics measurements, obtained using high spatial and spectral resolution semi-analytical models \citep[as done in ][]{Pelliccia2017}, allows us to constrain and then analyze the smTFR for our sample against two metrics of environments. We investigate, as well, possible environmental dependences of the galaxy stellar-to-dynamical mass ratio ($M_\ast/M_{dyn}$) and stellar specific angular momentum ($j_\ast$) to unveil different evolutionary path between galaxies in different environments. We also make use of the HR-COSMOS sample from \citet{Pelliccia2017} throughout the paper as a comparing sample residing at lower local environment, though, on one occasione we specifically sub-sample it to include only galaxies in the field environment.
The paper is organized as follows. In Section~\ref{sec:data} we introduce the data and our sample selection. In Section~\ref{sec:analysis} we describe the methods used to derive our stellar mass, kinematics and environment measurements. The results are presented in Section~\ref{sec:results}, showing our finding on the environmental dependence of smTFR, $M_\ast/M_{dyn}$, and $j_\ast$. We, then, summarize our results in Section~\ref{sec:results}.

Throughout this paper, we adopt a \citet{Chabrier2003} initial mass function and a standard $\Lambda$CDM cosmology with H$_0$ = 70 km s$^{-1}$, $\Omega_\Lambda$ = 0.7, and $\mathrm{\Omega_M}$ = 0.3. Magnitudes are given in the AB system. Distances are in proper units.

\begin{figure*}
	\includegraphics[width=\textwidth]{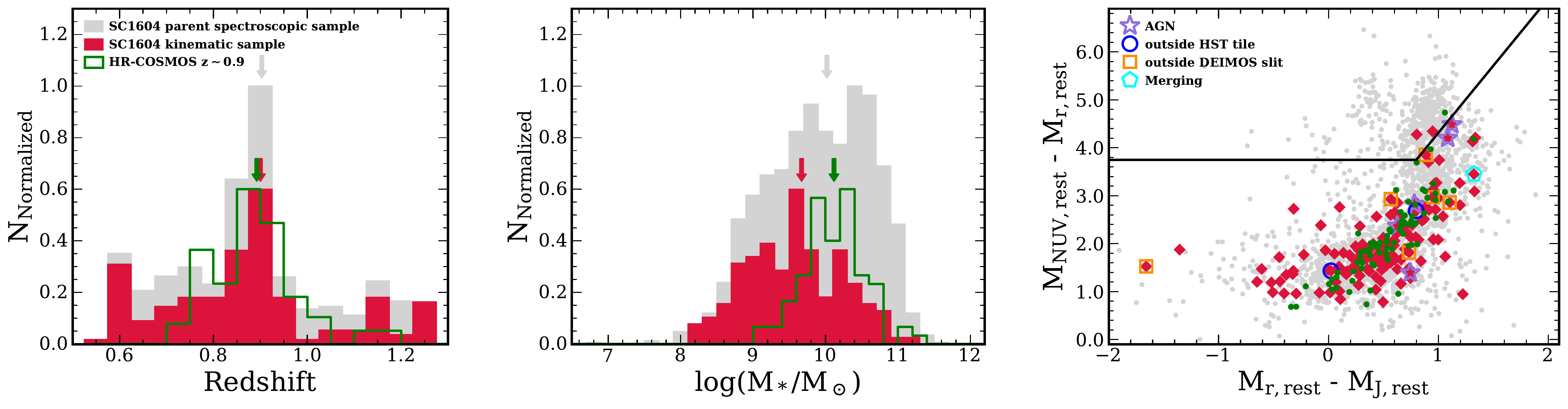}
    \caption{Redshift (left panel) and stellar mass (middle panel) distribution of our kinematic sample (red) and the parent SC1604 spectroscopic sample (gray). The distributions for the HR-COSMOS sample (green) from \citet{Pelliccia2017} are also shown for comparison. The arrows point to the median value of each distribution. The histograms are re-normalized for a better visual comparison of their spreads and peaks, therefore, their normalization does not reflect the actual scale. \textit{Right panel:} Rest-frame $\mathrm{M_{NUV}- M_r}$ versus $\mathrm{M_r - M_J}$ color-color diagram for our kinematic sample (red), the parent SC1604 spectroscopic sample (gray), and  the HR-COSMOS sample (green). Open markers indicate peculiar cases described in Sec.~\ref{subsec:kinesample}. The black line show the delineation between star-forming (below the line) and quiescent (above the line) adopted from \citet{Lemaux2014}. } 
    \label{fig:properties}
\end{figure*}
\section{Data} \label{sec:data}
\subsection{ORELSE}
This study is performed on data taken from the Observations of Redshift Evolution in Large-Scale Environments survey \citep[ORELSE;][]{Lubin2009}. ORELSE is a large multi-wavelength photometric and spectroscopic campaign covering a total of 5~deg$^2$ dedicated to map out and characterize galaxy properties in 15 fields which contain large-scale structures (LSSs) over the redshift range of $0.6 < z <1.3$. The survey aims to characterize the properties of galaxies in a wide range of environments from sparse fields to dense cluster cores. The kinematics analysis presented in this paper focused on one of the ORELSE fields, namely the SC1604. This field is dominated by SC1604 supercluster at $z\sim0.9$, which subtends roughly 13~$h_{70}^{-1}$ Mpc on the sky and 100 $h_{70}^{-1}$ Mpc in depth \citep{GalLubin2004, Gal2008}. It is one of the largest structures observed at high redshift, and it has been extensively studied by other works \citep[e.g.,][]{Lubin1998a, Postman2001, GalLubin2004, Kocevski2009, Lemaux2012, Wu2014}. It consists of three clusters and five groups that range in line of sight galaxy velocity dispersion of $\sim$300-800 \kms~corresponding to a dynamical mass range of $M_{vir}\sim 10^{13.4} - 10^{14.7} M_\odot$ \citep[see][for details on how these quantities are calculated]{Lemaux2012, Ascaso2014}. In addition, through visual inspection of the Monte Carlo Voronoi maps (see Sec.~\ref{subsubsec:localdens} for details)  two serendipitous clusters were discovered along the line of sight at $z=0.60$ and $z=1.18$ with dynamical masses of $M_{vir}= 10^{14.7} M_\odot$ and $10^{14.4} M_\odot$, respectively, calculated from $\sim$20 spectroscopic members per cluster (Hung et al \textit{in prep.}).

\subsection{Photometric and Spectroscopic Data} \label{subsec:data}
Photometric observations are available, as part of the ORELSE observing campaigns and of archival data,  across a wide range of bands, from optical to mid-infrared (\mbox{mid-IR}). This includes data from the Large Format Camera \citep[LFC;][]{Simcoe2000} on the Palomar 200-inch Hale Telescope, Suprime-Cam on the Subaru Telescope \citep{Miyazaki2002}, the Wide-field InfraRed Camera \citep[WIRCam;][]{Puget2004} on the Canada France Hawaii Telescope (CHFT), the Wide Field Camera \citep[WFCAM;][]{Casali2007} on the United Kingdom InfraRed Telescope (UKIRT), and the InfraRed Array Camera \citep[IRAC;][]{Fazio2004} on the \textit{Spitzer Space Telescope}. In \citet{Tomczak2017} the reader can find more detailed information on the optical/mid-IR photometry and its reduction. Additionally, nine out of the 15 ORELSE fields (including SC1604) have archival imaging from the Advanced Camera for Surveys (ACS) and the Wide Field Planetary Camera 2 (WFPC2) on-board the \textit{Hubble Space Telescope}(\textit{HST}). The vast majority of those fields have single or small pointing observations centered on the coordinates of the known cluster, with the exception of the SC1604 supercluster, for which almost the entire footprint is covered by \textit{HST}. 
In this study we used the 17-pointing F814W ACS mosaic imaging of SC1604 to derive morphological parameters for our kinematic sample (see Sec. \ref{subsec:kinesample}). ACS observations of SC1404 were taken with an average integration time of 1998\,s and cover a field of view of $\sim 3\arcmin\times 3\arcmin$, including 75.2\% of the entire spectroscopic sample. Moreover, F814W images are, also, entirely covered by the other photometric observations. More details on this observations and their reduction are presented in \citet{Kocevski2009}.

The spectroscopic data used in this study were obtained as part of a massive 300~hour observing campaign conducted with DEep Imaging
Multi-Object Spectrograph \citep[DEIMOS;][]{Faber2003} on Keck II 10m telescope targeting all the ORELSE fields. More details on the selection of the ORELSE survey targets, data acquisition and reduction are presented in \citet{Gal2008}, \citet{Lubin2009}, \citet{Lemaux2009, Lemaux2012} and \citet{Tomczak2017}. 
A total of 18 DEIMOS slitmasks were observed between May 2003 and June 2010 to map the SC1604 supercluster structure. All the slitmasks were observed with a typical integration time of 8329\,s per mask using the 1200-line~mm$^{-1}$ grating, blazed at 7500\,\AA, and 1\arcsec slits. This configuration allowed to obtain spectra with full width at half maximum (FWHM) spectral resolution of 1.7\,\AA\ (68\,\kms) and a typical wavelength coverage of 6385-9015\,\AA\  \citep[see][and references therein for more information on the observations of the SC1604 field]{Lemaux2012}. A total of 1397 spectra with high quality extragalactic redshift measurements \citep[Q=3-4, see][for the explanation on the quality codes]{Gal2008} were obtained, of which 448 are in the redshift range adopted for the supercluster \citep[$0.84\,\leq\,z\,\leq\,0.96$,][]{Lemaux2012}. Additional spectra from Keck I/Low-Resolution Imaging Spectrometer \citep[LRIS;][]{Oke1995} are available for SC1604 field \citep{Oke1998, GalLubin2004, Lemaux2012} providing 235 (85 in the redshift range adopted for the supercluster) additional high quality extragalactic redshift measurements, which contribute to provide accurate measurements of the local environments (see Sec.~\ref{subsubsec:localdens}). However, we decided not to include LRIS spectra in our kinematic sample due to their considerably lower spectral resolution (FWHM$\sim$8-11\AA) with respect to DEIMOS observations.  

The presence of three or more star spectra per slitmask allowed for the determination of the effect of the seeing for galaxies of interest placed on the same slitmask. Following the same technique adopted by \citet{Pelliccia2017}, we collapsed each star spectrum along the spectral direction, and determined the  FWHM by fitting a Gaussian function to its spatial profile. We adopted the mean of all the FWHMs in each slitmask as measurement of the spatial resolution for the spectra in that slitmask. The measured values  range between 0.57\arcsec and 1.45\arcsec and have been used in the construction of the kinematic models (see Sec. \ref{subsubsec:kinmodel}).

\subsection{Kinematic Sample Selection} \label{subsec:kinesample}
To investigate the effect of environment on galaxy kinematics, we selected a sample of galaxies, observed in the SC1604 field, for which it was possible to measure the kinematics. First of all, although the supercluster structure is known to be at $0.84\,\leq\,z\,\leq\,0.96$, we extended our selection to a wider redshift range, $0.6\,\leq\,z\,\leq\,1.3$,  where the vast majority of the ORELSE spectral sample lies and where our preferred emission lines for recovering kinematics are observable in the DEIMOS spectra (see later in the section for the identity of these lines). Moreover,  accurate measurements of galaxy structural parameters are crucial for the determination of  kinematics; therefore, we selected only the galaxies covered by the \textit{HST}/ACS imaging. This first-step selection left us with a sample of 703 galaxies. However, a further selection based on the galaxy morphology was necessary in order to properly measure the galaxy kinematics. 

We performed morphological measurements by using the most recent version of  SExtractor \citep[v2.19.5,][]{Bertin1996, Bertin2011}, which features the implementation of a two-dimensional model-fitting method,  in combination with PSFEx  \citep{Bertin2011}. SExtractor was run the fist time to detect the sources in the SC1604  F814W mosaic, then PSFEx identifies the detections that are likely to be point-sources and extracts precise models of the Point Spread Function (PSF). Finally,  another run of SExtractor provides structural parameters measurements, by  independently fitting each galaxy image with a Sersic+exponential disk profile model convolved with the local PSF model from PSFEx. The two-dimensional model-fitting procedure relies on the Levenberg-Marquardt minimization algorithm carried out on a modified $\chi^2$ of the residuals \citep[see Eq. 3 in][]{Bertin2011}, providing best-fit parameters, as well as estimates of uncertainties derived from Hessian matrix. The measured structural parameters of main interest for the derivation of the galaxy kinematics are: position angle (PA), defined as the angle (East of North) between the North direction in the sky and the galaxy major axis; inclination (\textit{i}), defined as the angle between the line of sight and the normal to the plane of the galaxy (i.e., $i = 0$ for face-on galaxies), and exponential scalelength ($r_s$) of the disc light profile. 

Since  ORELSE  was not originally designed for kinematic studies, the spectroscopic observations were taken with slit tilts pseudo-randomly\footnote{At some point during the target selection process, slit PAs were, to the best of our ability, aligned with the major axis of the targets as measured from the LFC imaging. However, the severely variable PSF of the LFC instrument resulted in this selection being effectively random with respect to the true PA. } chosen within the range allowed by the instrument (i.e., slit PA is typically chosen to be within 30\degr from the mask PA), and therefore, the slit PA was often not aligned with the true galaxy morphological PA.
To reduce the uncertainties introduced on the kinematic measurements by correcting for this misalignment, we require that galaxies in our sample have |$\Delta$PA|=|PA$_{galaxy}- $PA$_{slit}$|$\leq$45\degr. A selection based on the inclination was also necessary.
The inclination \textit{i} was determined from the galaxy axis ratio $b/a$ (i.e., the ratio between galaxy minor \textit{b} and major \textit{a} axis), measured on the \textit{HST} images, as:
\begin{equation}
\qquad \qquad \qquad \quad i = \arccos \sqrt{\dfrac{(b/a)^2 - q_0^2}{1-q_0^2}} \, ,
\end{equation}
where $q_0$  is the axial ratio of a galaxy viewed edge-on. In this study we assumed $q_0 = 0.19$, similar to other studies \citep[e.g.,][]{Pizagno2005, Miller2011,  Straatman2017}.  We kept in our sample only galaxies with $i\geq 20\degr$, since galaxy rotation velocity scales with $i$ (see Eq.~\ref{eq:velocity_short}) and for very small inclinations the rotation velocity is highly uncertain. This selection step reduced our sample by 54.3\% (with 53\% lost due only to the $\Delta$PA selection), leaving us with 321 galaxies.

In addition, a visual inspection of the 2D emission lines used for  deriving the kinematics (i.e., the doublet [\ion{O}{ii}]$\lambda \lambda$3726,3729\AA, H$\beta\lambda$4861\AA, and [\ion{O}{iii}]$\lambda$5007\AA) was performed for all galaxies in our sample. This allowed us to spot artifacts, non-detected emission lines or lines that were too faint. These galaxies were discarded from our sample (55\% of galaxies was lost) as it would have been impossible to measure their kinematics.

All this careful process led us to a final kinematic sample of 144 galaxies (9 with H$\beta$, 17 with [\ion{O}{iii}] and 118 with [\ion{O}{ii}] emission) for which we  measured the kinematics as described in Sec~\ref{subsec:kinemeasure}. We show in Figure~\ref{fig:properties} (left panel) the redshift distribution of this sample (red), compared to the distribution for the parent SC1604 spectroscopic sample (gray), and the HR-COSMOS sample at $z\sim0.9$ (see Sec.~\ref{subsec:hrcosmos}).
An additional visual inspection was performed on the \textit{HST} images with the DEIMOS slits superimposed. This check revealed that: seven galaxies were partially outside of the slit, making the rotation velocity measured not reliable, two galaxies were partially outside of the \textit{HST}/ACS mosaic, having, therefore, incorrect measurements of the structural parameters, and one galaxy was paired in the same slit with another galaxy (not in our kinematic sample) at the same redshift, suggesting a possible on-going merger. Since mergers are not included in our kinematic models, the rotation velocity measurement was not considered reliable for such galaxies. Moreover, we cross-matched  the kinematic sample with the ORELSE X-ray point source catalog, used for \mbox{X-ray} analysis by \citet{Rumbaugh2017}, and we found that five galaxies were present in that catalog. This suggested that those galaxies may host an active galactic nucleus (AGN), which may dominate the emission. From an inspection of the 1D spectrum of those galaxies, we confirmed that three of them indeed exhibit spectral features typical of AGN (e.g., broad [\ion{O}{ii}]/H$\beta$/ [\ion{O}{iii}] lines or presence of emission lines with high ionization energy like \ion{Ne}{V}). These 15 galaxies are referred throughout the paper as  ``peculiar'' cases and, although they are not removed from our kinematic sample (e.g., we show their position in the plots in Figure~\ref{fig:properties} and  \ref{fig:smTFR} with special markers), they are not used for the study presented in Sec.~\ref{sec:results}.

\subsection{The Comparison Sample: HR-COSMOS} \label{subsec:hrcosmos}
HR-COSMOS is a sample presented in \citet{Pelliccia2017} composed of 82 star-forming galaxies at $z\sim0.9$. Kinematic measurements have been performed using spectroscopic observations obtained with VIsible Multi-Object Spectrograph \citep[VIMOS,][]{leFevre2003} on ESO Very Large Telescope (VLT).  Stellar masses were computed using the COSMOS photometric catalog \citep{Laigle2016} derived from deep-ground and space-based imaging in 30 bands from UV to IR. Rotation velocity measurements have been obtained fitting semi-analytical models, which account for spatial and spectral resolution, to the data in a similar way as done in this study (see Sec.~\ref{subsec:kinemeasure}). After extracting the rotation velocity at a characteristic radius $r_{2.2}$ (as done in this paper, see Sec.~\ref{subsec:smTFR}), \citet{Pelliccia2017} constrained the smTFR and studied its evolution with redshift.  An environmental analysis has been also attempted by the authors, adopting local densities measurements from \cite{Scoville2013} obtained using a two-dimensional Voronoi tessellation technique. \citet{Pelliccia2017} have found no evidences for any dependence of smTFR on the environment and they argued that it may have been a consequence of HR-COSMOS galaxies not probing very high densities (see also Figure~\ref{fig:localoverdens}, right panel). In this study we use the HR-COSMOS sample as comparison sample residing at lower density environment.

Other large surveys of field galaxies obtained with IFU spectroscopy, e.g., KMOS$\mathrm{^{3D}}$ \citep{Wisnioski2015} and KROSS \citep{Stott2016}, have provided galaxy kinematic measurements \citep[e.g.,][]{Ubler2017, Harrison2017} that could be used to supplement our comparison sample in lower density 
environments.
 In this study, however, we aimed at avoiding possible biases due to the comparison of samples with measurements obtained with different techniques. We are confident that the velocity measurements for the SC1604 and HR-COSMOS samples are comparable, since the procedure adopted is almost identical (see Sec.~\ref{subsec:kinemeasure}). Moreover, we were able to make all the necessary checks (see Sec.~\ref{subsec:stellarmass} and Sec.~\ref{subsec:BbandTFR}) to verify that the different techniques and assumptions adopted in the SED fitting to measure galaxy stellar masses for the two samples should not bias the comparison. As such, and 
because the main limitation of the statistical significance of some of our results, as we will show later (Sec.~\ref{subsec:specific_angular_momentum}), comes from too few galaxies in \emph{higher} density environments, we choose to limit ourselves to just 
the HR-COSMOS sample for our lower density point of comparison.

\section{Galaxy Measurements} \label{sec:analysis}
 
 \subsection{Stellar Mass} \label{subsec:stellarmass}
Stellar mass measurements are used in this study to constrain the smTFR (Sec.~\ref{subsec:smTFR}), stellar-to-dynamical mass ratio (Sec.~\ref{subsec:stellar-to-dynamical}), and the $j_\ast - M_\ast$ diagram (Sec.~\ref{subsec:specific_angular_momentum}) for our kinematic sample.
These measurements were obtained following the same recipes described in detail in \citet{Tomczak2017}, based on the FAST \citep[Fitting and Assessment of
Synthetic Templates;][]{Kriek2009} code. In brief, FAST generates a series of model fluxes from the stellar population synthesis (SPS) package developed by \citet{Bruzual2003}. In addition, we assumed  \citet{Chabrier2003} stellar initial mass function, delayed exponentially declining star formation
histories (SFH $\propto t \times e^{-t/\tau}$), and solar metallicity. For dust extinction we adopted the \citet{Calzetti2000} attenuation curves, and we allowed A$_\mathrm{V}$ to range between $0-4$. Each galaxy is then fit by every model generated by FAST, and the model with the lowest minimum $\chi^2$ is adopted as best-fit. 
In order to account for extra uncertainties produced by the  assumptions about model parameters, we add 0.2~dex systematic error \citep[see discussion in][]{Courteau2014} in the stellar mass error budget. 
We show in Figure~\ref{fig:properties} (middle panel) the stellar mass distribution for our kinematic sample (red) in comparison to the distribution for the parental SC1604 spectroscopic sample (gray), and for the HR-COSMOS sample presented in \citet{Pelliccia2017}, which we use for comparison throughout the paper. We find that in general HR-COSMOS galaxies have higher stellar masses with respect to SC1604 ones, and show a lack of galaxies with stellar masses lower than $\sim10^9 M_\odot$. This difference in stellar mass distribution is a consequence of the different magnitude selection between the two samples. While SC1604 spectral targets were selected to have F814W magnitude as faint as $\sim25\,$mag,  HR-COSMOS was drawn from the zCOSMOS $10k-$bright sample \citep{Lilly2007}, which adopted the observing strategy to select galaxies with $I_{AB}<22.5\,$mag. The bias of the HR-COSMOS sample towards brighter galaxies explains the lack of lower stellar mass galaxies. We take into account this bias in the analysis described later (see Sec.~\ref{subsec:stellar-to-dynamical}).

Although the rotation velocity measurements for HR-COSMOS have been performed in a similar way to SC1604, some differences in the assumptions adopted in the stellar mass measurements exist.  In order to ensure that the comparison is fair we re-run FAST with parameters that matched the HR-COSMOS model assumptions. In particular, we allowed for three types of star formation histories (one exponentially declining and two with a delayed exponentially decline having a maximum star formation rate peak after 1 and 3 Gyr) and for two metallicities, i.e., solar and half-solar.
This exercise revealed that no bias was observed when comparing stellar masses measured with these two different prescriptions and a mass-independent scatter of 0.1~dex was measured, which is largely within the 0.2~dex accounted for uncertainties on the model assumptions. Therefore, we are confident that the comparison is valid.

\begin{figure*}
	\includegraphics[width=\textwidth]{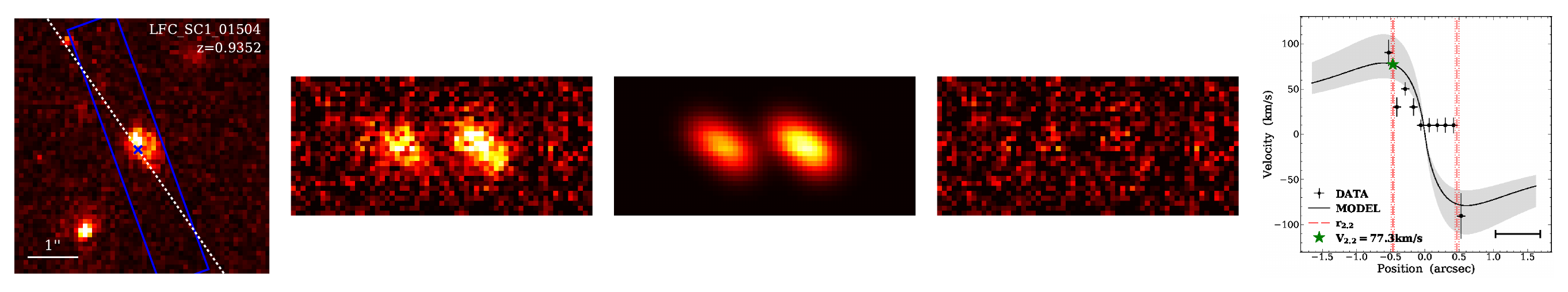}
	\includegraphics[width=\textwidth]{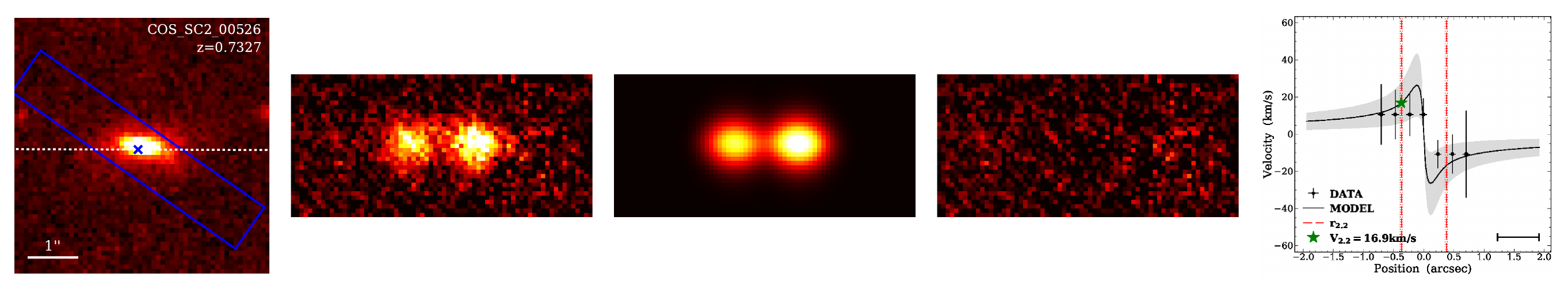}
	\includegraphics[width=\textwidth]{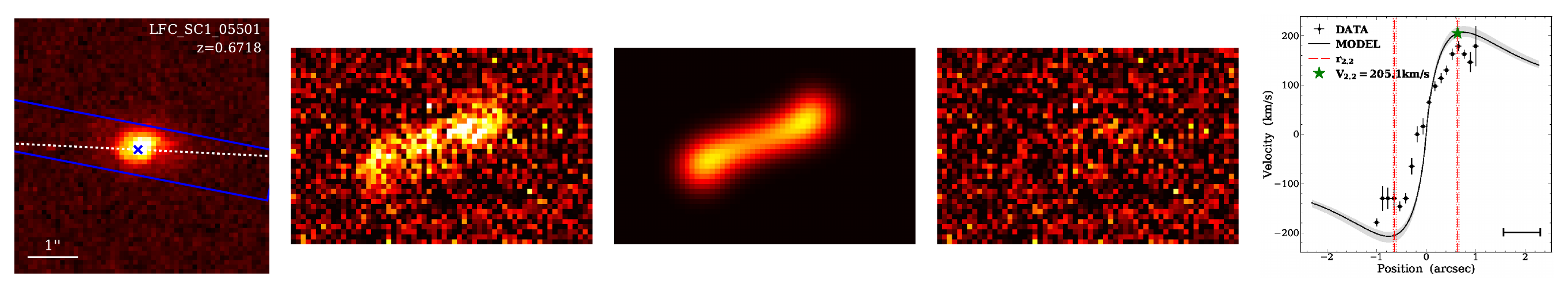}
	\includegraphics[width=\textwidth]{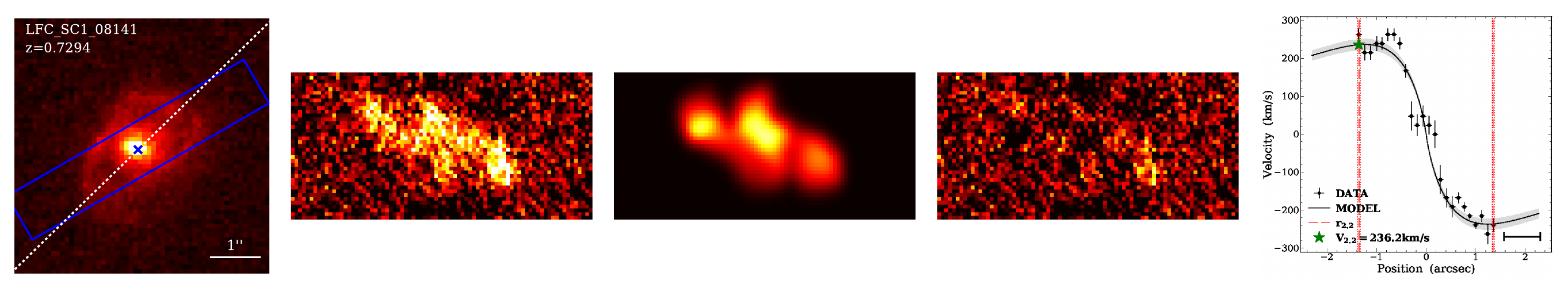}
    \caption{Examples of the kinematic modeling for four galaxies with  [\ion{O}{ii}]  and H$\beta$  emission. These galaxies have stellar masses typical of the four stellar mass bins used in the analysis in Sec.~\ref{subsec:stellar-to-dynamical}, which span the entire range from low (top) to high (bottom) $M_\ast$. \textit{Each row:} From left to right: \textit{HST}/ACS F814W postage stamp (5\arcsec$\times$5\arcsec) with superimposed in blue the DEIMOS slit and in white the orientation of the galaxy PA;  continuum-subtracted 2D spectrum centered at the emission line; best-fit kinematic model; residual image between the 2D spectrum and the best-fit model on the same intensity scale as the 2D spectrum; high-resolution rotation curve model (black line), corrected for the inclination, with 1$\sigma$ uncertainty (shaded area), compared to the observed rotation curve (black points). The red dashed and dotted lines indicate the radius $r_{2.2}$ (see Sec.~\ref{subsec:smTFR}) and its uncertainty, respectively. The horizontal black bar on the bottom right corner represents the DEIMOS spatial PSF. }
    \label{fig:kinemodel}
\end{figure*}
In the right panel of Figure~\ref{fig:properties} we compare the rest-frame $\mathrm{M_{NUV}- M_r}$ versus $\mathrm{M_r - M_J}$ color-color diagram for our kinematic sample (red), to the one of the parent SC1604 spectroscopic sample (gray), and  the HR-COSMOS sample (green). This color-color diagram is a diagnostic plot that enables us to separate star-forming from quiescent galaxies. The dividing line is adopted from \citet{Lemaux2014} for galaxies at $0.5<z<1$. The rest-frame colors for SC1604 are estimated by using the code Easy and Accurate Redshifts from Yale \citep[EAZY,][]{Brammer2008} as described in \citet{Tomczak2017}, while the rest-frame colors for HR-COSMOS were computed using the software Le Phare \citep{Arnouts2002, Ilbert2006} as described in \citet[][and references therein]{Pelliccia2017}. We verified that, although using different softwares, different templates, and different filters, the measured colors for the two samples are comparable, since the difference in $\mathrm{M_{NUV}}$, $\mathrm{M_r}$, and $\mathrm{M_J}$ is less or equal to 0.25~mag as measured for an ORELSE field, to which we were able to apply both fitting methods. The color-color diagram in Figure~\ref{fig:properties} (right panel) shows that our kinematic sample follows well the underlying parent population of star-forming galaxies and it occupies, as well, the same locus of the HR-COSMOS sample.

\subsection{Kinematics} \label{subsec:kinemeasure}
\subsubsection{Semi-analytical Model and Fitting} \label{subsubsec:kinmodel}
To derive galaxy kinematics we developed semi-analytical models. The idea is to create mock DEIMOS observations with known kinematics and directly compare them to the real data. To this end, we construct our kinematic models as described in \citet{Pelliccia2017}. Although \citet{Pelliccia2017} adopted a 2D modeling approach, we here adopt a 3D version of the same model. This approach allows us to better reproduce the misalignment between PA$_{slit}$ and PA$_{gal}$ and to properly correct for it in the measured rotation velocity. We refer to \citet{Pelliccia2017} for a detailed description of the model, which we briefly summarize here.

The mock emission for the [\ion{O}{ii}] doublet (the most frequent line) is defined as:
\begin{equation} \label{eq:model}
\begin{split}
\qquad \quad	I(r,V)=\dfrac{\Sigma(r)}{\sqrt{2\pi}\sigma(r)} \left\{\exp -\left[ \dfrac{(V_{line1} - V(r))^2}{2\sigma^2(r)}\right]  + \right. \\ 
\left. + R \exp - \left[ \dfrac{(V_{line2} - V(r))^2}{2\sigma^2(r)}\right]  \right\},
\end{split}
\end{equation}
which describes its intensity $I(r,V)$ at each radius $r$, defined at each position (\textit{x,y}) on the plane of the galaxy,  and velocity $V$. $V_{line1}$ and $V_{line2}$ are the velocities relative to each line of the doublet at longer and shorter wavelength, respectively, while $R$ is the ratio between the intensities of the two lines. $\Sigma(r)$ represents the intrinsic line-flux distribution in the plane of the disc, which we assumed to be a truncated exponential disc \citep[see Eq.~2 in][]{Pelliccia2017}, and $\sigma(r)$ describes the galaxy velocity dispersion, which we assumed to have a constant value ($\sigma_0$) as a function of the radius $r$.
For the galaxies that exhibit, instead, single emission line (i.e., H$\beta$,[\ion{O}{iii}])  the mock observation is still described by Equation \ref{eq:model}, but with $R$ equal to zero.
The velocity along the line of sight is described as:
\begin{equation} \label{eq:velocity_short}
\\ \\ \\ V(r)= V_{sys}+V_{rot}(r) \sin i \cos \theta ,
\end{equation}
where $V_{sys}$ is the systemic velocity of the entire galaxy, and $\theta$ is the azimuth angle on the plane of the galaxy. To model the intrinsic rotation velocity $V_{rot}(r)$ we chose an exponential Freeman disc \citep{Freeman1970} profile \citep[as expressed in Eq. 4-5 in][]{Pelliccia2017}, based on \citet{Pelliccia2017} finding that 60\% of their sample was better modeled by an exponential disc compared to other two functions used (flat and arctangent profiles). This high-resolution model is described by two parameters: the maximum velocity $V_t$ and the transition radius $r_t$.
We performed, then, spatial and spectral smoothing by convolving the 3D high resolution model with the seeing measured for each mask (see Sec~\ref{subsec:data}) and the spectral resolution of the instrument, and we re-binned it to match the DEIMOS sampling.
The additional step performed in this 3D modeling approach with respect to the 2D approach in \citet{Pelliccia2017} was to simulate the placement of the slit along the major axis of the galaxy and reproduce the 2D emission line by integrating the information within the slit.

The comparison with the observation was done through the $\chi^2$-minimization fitting, based on the Levenber-Marquardt least-squares technique. 
Except for the parameters $i$ and PA, which are not allowed to vary in the fitting process, in total eight (nine, in the case of doublet) are free parameters, of which the most relevant in this analysis are: $V_t$, $r_t$ and $\sigma_0$.
Least-squares fitting is known to be sensitive to the local minima; therefore, it is extremely important to choose initial guesses for the fit to be as close as possible to the global minimum. We accounted for that by carefully measuring the initial guesses. We refer the reader to Sec.~3.2 of \citet{Pelliccia2017} for a detailed description of these measurements.
Moreover, we adopted a Monte-Carlo approach in order to explore the impact of the choice of the initial guesses on the resultant best-fit parameters. To that end, we perturbed the initial guesses by sampling a Gaussian distribution and obtaining 20 combinations of initial parameters. We run iteratively the fitting process for each combination, and we computed the residuals between the best-fit model and the data; if the root mean square (rms) in the residual image was less than 20\% higher than the rms noise computed from the data background, we chose that best-model as the final best-model. 

Four examples of the comparison between the observations and the model are shown in Figure~\ref{fig:kinemodel} (second, third and fourth panels from left to right). We, also, show in the left panels the \textit{HST}/ACS F814W galaxy image indicating the orientation of the galaxy PA as measured in Sec.~\ref{subsec:kinesample} and  the superimposed DEIMOS slit, and in the right panels the best-fit high-resolution rotation curve model corrected for the inclination, compared to the rotation curve extracted from the observations. The four galaxies are representative of the population of galaxies in the four stellar mass bins adopted for the analysis in Sec.~\ref{subsec:stellar-to-dynamical}, since they have $M_\ast$ similar to the median $M_\ast$ in each bin, and span the entire range of stellar mass probed by our SC1604 kinematic sample. We show the same plots for the entire ``kinematically reliable'' sample (see Sec.~\ref{subsec:smTFR} for its definition) in Appendix~\ref{app:figures}.

\begin{figure*}
	\includegraphics[width=0.6\textwidth]{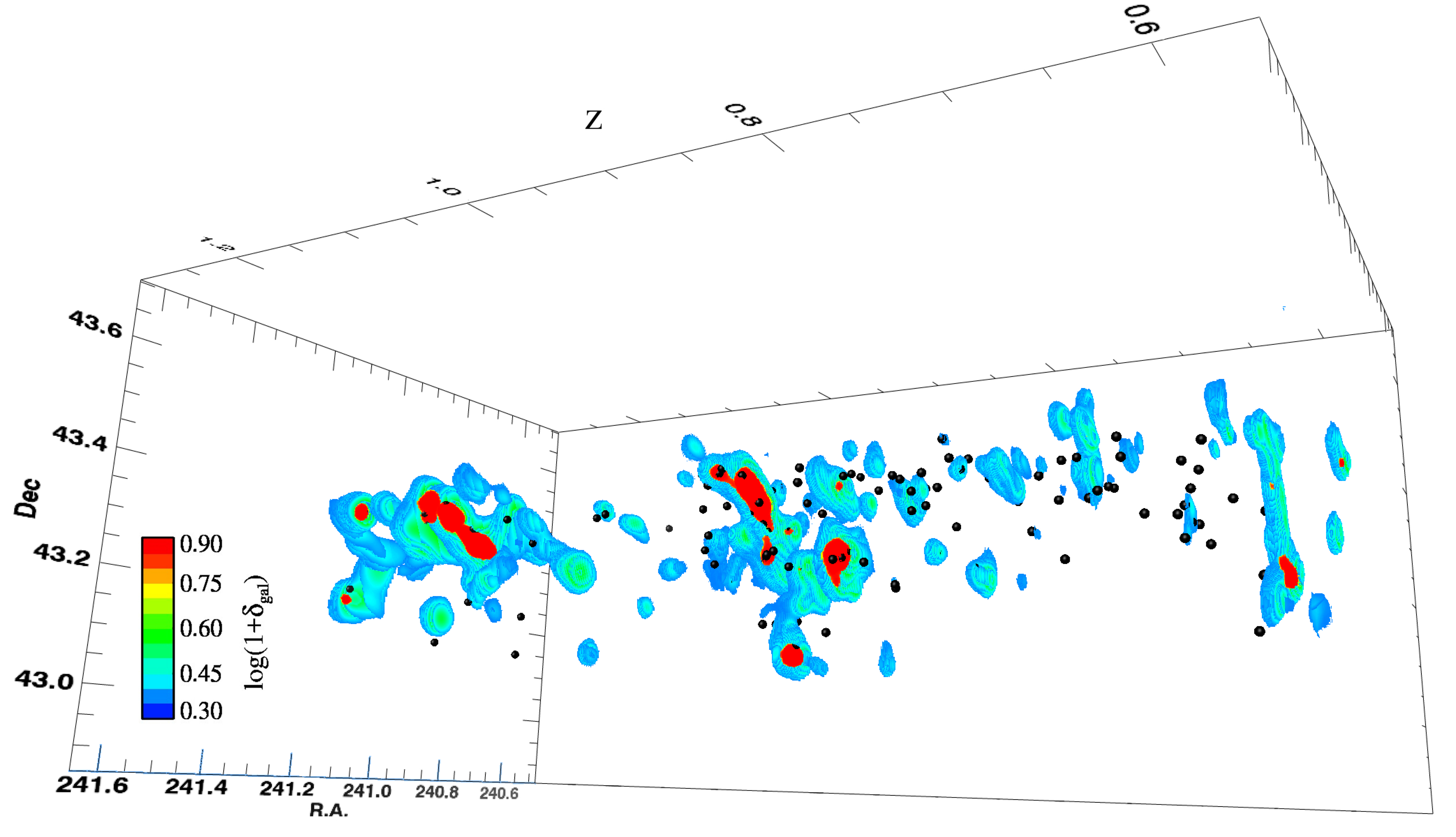}
		\includegraphics[width=0.39\textwidth]{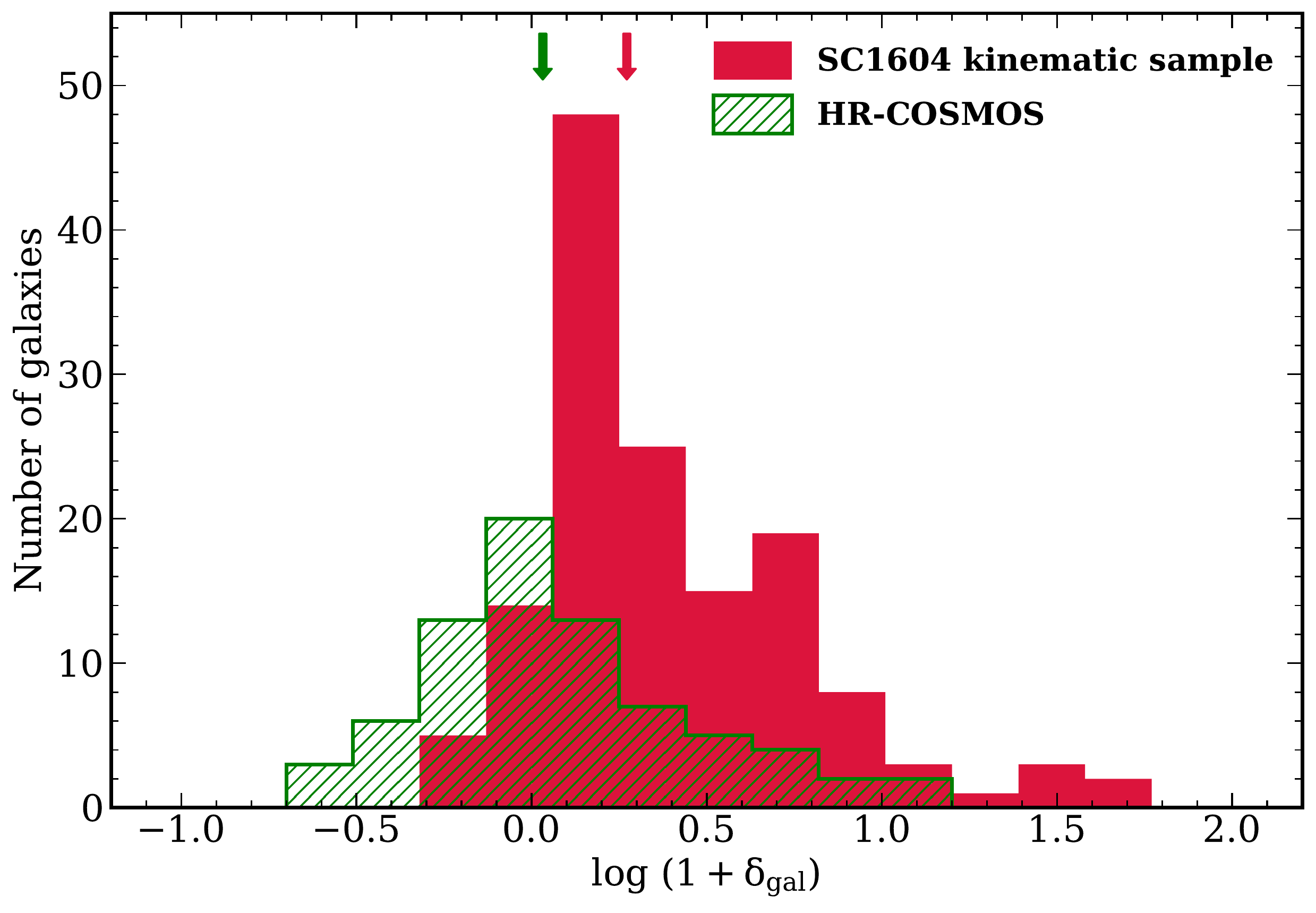}
    \caption{\textit{Left: } SC1604 LSS 3D map, color-coded according to the local overdensity measurements. This was created using the Voronoi tessellation technique described in Sec.~\ref{subsubsec:localdens}, but using a smaller (150\,\kms) step size for a smoother rendering. The black points show the location of the galaxies in our kinematic sample (see Sec.~\ref{subsec:kinesample}). Due to the way the map is constructed, some of the black points in very dense regions are embedded in those red regions and are not visible. Therefore, the number of galaxies in high density environments are underestimated in this image.  \textit{Right: } Comparison of the distributions of $log(1 + \delta_{gal})$ for our kinematic sample and the HR-COSMOS sample from \-\citet{Pelliccia2017}. The arrows point to the median value of each distribution.  }
    \label{fig:localoverdens}
\end{figure*}
\subsubsection{Parameter uncertainties} \label{subsubsec:kine_uncert}
We computed kinematic parameter uncertainties by creating 100 Monte-Carlo realizations of the data perturbed according to the 2D spectrum noise, and we re-fit the kinematic model for each realization.
We derive 1$\sigma$ uncertainties based on the 16th and 84th percentiles of the posterior distributions.
In addition, two other contributions have been considered in the error budget of the kinematic parameters.
We propagated the uncertainties on the measurement of $i$ (propagated from the uncertainties on $b/a$) and PA into the error on the parameter $V_t$. Past studies have shown that there often exists a misalignment between the morphological PA and the kinematic PA \citep{Wisnioski2015, Contini2016, Harrison2017}, which in median is equal to $\sim$13\degr. Although this misalignment is small we decided to added it in quadrature to the uncertainties on PA, which are then propagated into the error budget of $V_t$.
 Moreover, we take into account variations of the spectral resolution generally observed across the  DEIMOS detector, which are of the order of 10\%, by adding it into the error budget of $\sigma_0$.

\subsubsection{Spatially Resolved Emission} \label{subsubsec:resolved_emiss}
The reliability of the measured kinematics depends strongly on the galaxy size relative to the observations' seeing. It has been shown by \citet{Newman2013}, by comparing the kinematics of a sample  galaxies at $z = 1.0−-2.5$ observed with an IFU telescope in both seeing-limited and adaptive optics (AO) mode, that many galaxies that were considered dispersion-dominated from seeing-limited data actually showed evidence for rotation in higher resolution data. This is the result of an instrumental effect, called ``beam~smearing'', which ``smears'' the velocities coming from different part of the galaxy into a given spatial resolution element. This beam~smearing has the effect of artificially increasing the velocity dispersion in the central part of the galaxies, where the velocity gradients are in general very large. This effect is generally taken into account in our kinematic models for sizes of the emitting region observed in a given galaxy that are comparable or larger than the seeing. We compared, therefore, the intrinsic extent of the emission lines used to derive the kinematics for our sample to their spatial resolution. To this end, we collapsed the 2D spectrum in the wavelength direction over a range of about $\pm 20$\,\AA\ centered on the central $\lambda $ of the emission line, and we fit spatial profile with a Gaussian function, allowing to measure the FWHM of the galaxy emission profile (FWHM$_{em}$). Since this measurement of  FWHM$_{em}$ is convolved with the spatial resolution, in order to recover its intrinsic value we subtracted in quadrature the seeing as: $\mathrm{FWHM}_{em, intr}=\sqrt{\mathrm{FWHM}_{em}^2 - \mathrm{FWHM}_{seeing}^2}$.

We defined galaxies as having spatially resolved kinematics if  \mbox{$\mathrm{FWHM}_{em, intr}\geq\mathrm{FWHM}_{seeing}$}. Out of the 144 galaxies in our kinematic sample, 40 (5 of which already defined as ``peculiar cases'', see Sec.~\ref{subsec:kinesample}) have spatially unresolved kinematics. In the analysis presented in Sec.~\ref{sec:results} we excluded those galaxies, along with the ``peculiar cases'',  in order to avoid any possible bias introduced by unreliable kinematic measurements. We verified that this cut did not affect the properties of our kinematic sample, by confirming that the distributions of the galaxy parameters shown in Figure~\ref{fig:properties} still hold.

\subsection{Environment} \label{subsec:environment}
In this work we investigate the possibility of galaxy kinematics being influenced by the environment. In order to quantify ``environment'' we define two metrics, local and global density.   
Local environment relates to smaller physical scales, and describes how galaxies affect one another, while global environment relates to larger physical scales, and dictates how galaxies are affected by residing in their overarching halo for a given period of time. Below we describe in detail how each metric is defined for our sample.

\subsubsection{Local Environment} \label{subsubsec:localdens}
We measure local environment using a Monte-Carlo implementation of Voronoi Tessellation (VMC), a technique described in detail in \citet{Lemaux2017}. 
The advantage of the VMC approach is that high-quality spectroscopic redshifts ($z_{spec}$) are combined with photometric redshift ($z_{phot}$) information to provide a more complete and accurate mapping  of the underlying density field. In this implementation, high-quality $z_{spec}$ information is treated as truth and $z_{phot}$ information enters statistically by sampling the  uncertainties associated with each object lacking a high-quality $z_{spec}$ over many realizations of the density map. More specifically, in each realization a redshift is assigned to each object lacking a high-quality $z_{spec}$ by sampling an asymmetric Gaussian with mean equal to its actual $z_{phot}$  and positive and negative dispersion equal to its effective $\pm$1$\sigma$ uncertainty derived from the full probability distribution function (PDF), respectively.
For a single redshift slice of width $\pm$1500\,\kms~in velocity space, all of the $z_{phot}$ objects which fall within this redshift bin for this iteration are combined with all galaxies with a high-quality $z_{spec}$ that places them in the slice. We then run Voronoi tessellation on this combined sample. This process is run a total of 100 times per redshift slice. The first redshift slice begins at $z=0.55$ and we step forward in steps of  1500\,\kms~until we reach a slice with an upper bound of $z=1.4$.

For each realization of each slice, the area of the Voronoi cell associated with each object is used to define the local density  for that object as the inverse of the cell area multiplied by the square of the angular diameter distance. The resultant density field is then projected onto a two-dimensional grid of 75$\times$75 proper kpc. The final density map is computed by median combining the density maps from the 100 Monte-Carlo realizations. The local overdensity at each pixel (i, j) is calculated as
$log(1 + \delta_{gal})=log(1 + (\Sigma_{i,j} - \tilde{\Sigma})/\tilde{\Sigma})$, where $\tilde{\Sigma}$ is the median density of all pixels where the map is defined well defined.
We use local overdensity rather than local density as our measurement of local environment to mitigate issues of sample selection across different fields of the ORELSE survey and differential bias as a function of redshift.

We show in Figure~\ref{fig:localoverdens} (left panel) a 3D map of the SC1604 LSS color-coded according to the measurement of  $log(1 + \delta_{gal})$ running over $0.55\leq z \leq 1.3$.   We show, also, with black points the location of the galaxies in our kinematic sample. In the right panel of Figure~\ref{fig:localoverdens} we compare the $log(1 + \delta_{gal})$ distribution for our kinematic sample and for the HR-COSMOS sample, which we use for comparison in our analysis presented in Sec.~\ref{sec:results}. The plot highlights that SC1604 covers higher local overdensity than HR-COSMOS, with a shift in the median $log(1 + \delta_{gal})$ of 0.24~dex.

\subsubsection{Global Environment} \label{subsubsec:globaldens}
To quantify the global environment we adopt a phase-space parameterization introduced by \citet{Carlberg1997} that combines the dynamic range of velocities of the galaxies inside a cluster/group and their projected galactocentric distances. This parameter, which we call $\eta$, is defined as: \begin{equation}
\qquad \qquad \qquad \quad\eta= \left( R_{proj}/R_{200}\right) \times \left(  |\Delta v|/\sigma_v \right) ,
\end{equation}
where $R_{proj}$ is the projected distance of each galaxy to the nearest cluster/group center, $R_{200}$ is the radius at which the cluster/group density is 200 times the critical density, $\Delta v$ is the difference between each galaxy velocity and the systemic velocity of the cluster/group, and $\sigma_v$ is the line-of-sight (LOS) velocity dispersion of the cluster/group member galaxies. The systemic velocity and the $\sigma_v$ for each cluster/group are computed using the method described in \citet{Lemaux2012}, while the cluster/group centers are obtained from the $i'$/$z'$-luminosity-weighted center of the members galaxies as described in \citet{Ascaso2014}. The value of $\eta$ for each galaxy is measured with respect to the closest cluster/group. To determine it, we first find all the clusters/groups that are within $\pm$6000\,\kms~in velocity space of a given galaxy, then we compute $R_{proj}/R_{200}$ from the galaxy to all the identified clusters and groups, and we select the one for which $R_{proj}/R_{200}$ is the smallest as the parent cluster/group. If for a given galaxy no clusters/groups within $\pm$6000\,\kms~are found, $\eta$ is computed with respect to all of those clusters/groups in the field and the one with the smallest value is associated with that galaxy.

Quantitatively, following \citet[][and references therein]{Noble2013} we define:
\begin{itemize}
 \item $|\eta|<0.1$ refers to galaxies that are in the virialized cluster core;
 \item $0.1<|\eta|<0.4$ indicates the so-called ``backsplash'' galaxies that have been past pericenter in earlier times and then have moved out \citep[e.g.,][]{Balogh2000, Gill2005};
 \item $0.4<|\eta|<2$ are for galaxies recently accreted, which populate the infall region;
 \item $|\eta|>2$ indicates galaxies that are not associated with the cluster.
 
 \end{itemize}

\begin{figure*}
	\includegraphics[width=\textwidth]{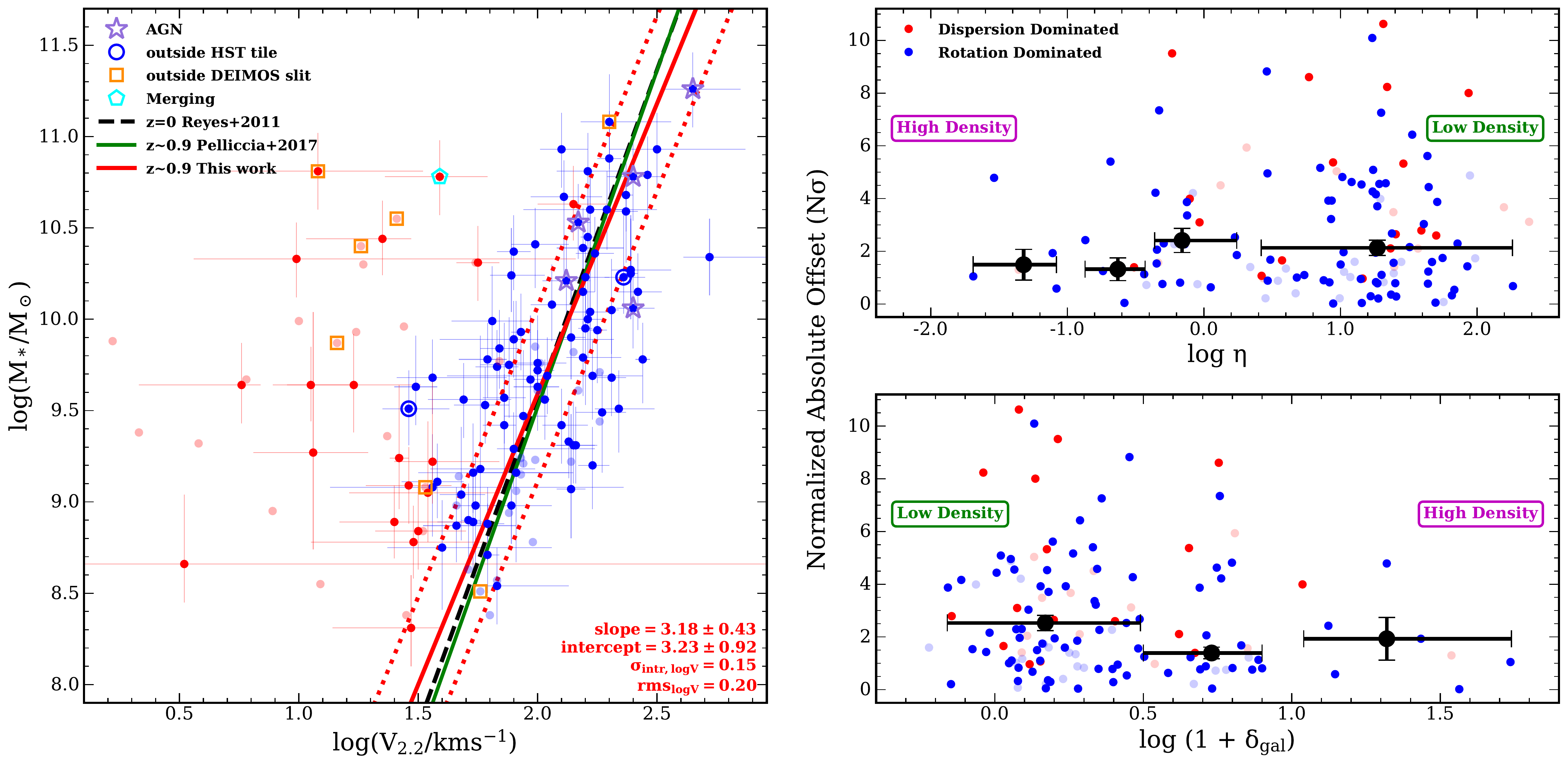}
    \caption{\textit{Left panel: } Stellar-mass Tully-Fisher relation constrained using the rotation dominated sub-sample (blue) of our kinematically reliable sample. In red are shown the measurements for the dispersion dominated galaxies. The light colored points indicate galaxies with spatially unresolved kinematics (see Sec.~\ref{subsubsec:resolved_emiss}), which are not used in our analysis. Open markers refer to peculiar cases described in Sec.~\ref{subsec:kinesample} that are also not used in our analysis, because of unreliable measurements. The fit to the relation and its intrinsic scatter $\sigma_{intr}$ are shown with a red solid and dotted lines. The value of best-fit parameters, $\sigma_{intr}$, and the total scatter $rms$ are presented in the bottom right corner.  We plot as references the relations at $z=0$ (black line) from \citet{Reyes2011}, and at $z\sim0.9$ (green line) for HR-COSMOS from \citet{Pelliccia2017}. \textit{Right panels:} Normalized scatter (described in Sec.~\ref{subsec:smTFR}) of the full kinematically reliable sample around the relation as function of global (top) and local (bottom) environment. Colors are the same as in the left panel. Large points indicate median scatter per environmental bin. The errorbars on the $y$-axis direction represent the uncertainties on the median; the bars on the $x$-axis direction indicate the bin size.  }
    \label{fig:smTFR}
\end{figure*}

\section{Environmental Study} \label{sec:results}
\subsection{Stellar Mass Tully-Fisher Relation} \label{subsec:smTFR}
To constrain the smTFR  for the galaxies in our kinematic sample, we adopt $V_{2.2}$ as our velocity estimator, which correspond to the peak of the rotational velocity for a pure exponential disc rotation curve. \cite{Courteau1997} found that this estimator is the best measure for the TF velocity, providing the smallest internal scatter, minimal TF residuals and the best match to radio (21~cm) line widths for local galaxies.
$V_{2.2}$ is interpolated from the best-fit model rotation curve at the radius $r_{2.2}=2.2\,r_s$, where $r_s$ is the disc exponential scalelength, measured using \textit{HST}/ACS F814W images as described in Sec.~\ref{subsec:kinesample}. The uncertainties on the measurements of $r_s$ are  propagated into the error budget of $V_{2.2}$ in addition to the uncertainties described in Sec.~\ref{subsubsec:kine_uncert}. A typical value of $r_{2.2}$ in our kinematic sample is $\sim 4\,$kpc. For 76\% of this sample the spatial extent of the emission line  \citep[defined by tracing the continuum-subtracted emission line as a function of the spatial position as described in][]{Pelliccia2017} is equal to or larger than $r_{2.2}$. Moreover, for the  ``kinematically reliable'' sample (see below for its definition) only 20\% has $V_{2.2}$ extrapolated at a larger radius than the extent of emission. We plot in Figure~\ref{fig:kinemodel} (right panels) examples of the best-fit model rotation curves showing the interpolation at $V_{2.2}$. We show the same plot for the entire ``kinematically reliable'' sample in  Appendix~\ref{app:figures}. Also, measurements of $V_{2.2}$ and $\sigma_0$, along with other galaxy parameters measurements, are provided in Appendix~\ref{app:tab}. 

From now on, we use in our analysis only the sample of 94 galaxies derived from the whole kinematic sample presented in Sec.~\ref{subsec:kinesample}, which excludes the galaxies with spatially unresolved emission (Sec.~\ref{subsubsec:resolved_emiss}) and the ``peculiar'' cases (Sec.~\ref{subsec:kinesample}). This sample is referred to hereafter as our ``kinematically reliable'' sample. We show, however, sometime as reference (e.g., Figure~\ref{fig:smTFR}) the measurements for the excluded galaxies.
We find that 82\% (77 galaxies) of our kinematically reliable sample is rotation dominated, while 18\% (17 galaxies) is dispersion dominated, with ``rotation'' and ``dispersion'' dominated galaxies being those with $V_{2.2}/\sigma_0>1$ and $V_{2.2}/\sigma_0<1$, respectively.
These values are consistent with those found by \citet{Pelliccia2017} for the HR-COSMOS sample at similar redshift (see Figure~\ref{fig:properties} left panel). 

The smTFR for SC1604 galaxies is presented in the left panel of Figure~\ref{fig:smTFR} and is described by the following functional form:
\begin{equation} \label{eq:smTF}
\\ \\  \qquad logM_\ast= slope \times(logV_{2.2}-logV_{2.2, 0}) + intercept \, ,
\end{equation}
where $logV_{2.2, 0}$ is chosen to be equal to 2.0~dex to minimize the correlation between the uncertainties on $slope$ and $intecept$ \citep{Tremaine2002}. The relation is constrained using only the rotation dominated galaxies of our kinematically reliable sample, as the TF relation is known to be valid only for rotating galaxies, while for the analysis presented later the dispersion dominated sample is also included. In Figure~\ref{fig:smTFR} the rotation dominated galaxies are shown in solid blue colors, while the dispersion dominated ones are in solid red. The points in lighter colors are the measurements for the galaxies with spatially unresolved kinematics. The ``peculiar'' cases are also shown in the left panel with open markers.
The relation is obtained in the same way as presented in \citet{Pelliccia2017}, by fitting an inverse linear regression using the software \mbox{MPFITEXY} \citep*{William2010}, which adopts a least-squares approach and accounts for the uncertainties in both coordinates. The \mbox{1$\sigma$} errors on $slope$ and $intecept$ were determined by taking the dispersion of the distribution of 100 bootstrapped estimations of the same parameters. The smTFR best-fit parameters are shown in the bottom-right corner of Figure~\ref{fig:smTFR}, where the reader can find also the value of the intrinsic scatter $\sigma_{intr}$ and the total scatter $rms$ on the velocity variable. As comparison we show in Figure~\ref{fig:smTFR} the smTFR constrained by \citet{Reyes2011} at $z=0$ and the one obtained for the HR-COSMOS sample \citep{Pelliccia2017} at similar redshift, but for galaxies residing in lower local overdensities (see Figure~\ref{fig:localoverdens} right panel). These comparisons highlight that: i) the smTFR presented in this paper confirms once again the non-evolution of the relation with redshift up to $z\sim1.2$ as already found by some works \citep{Conselice2005, Miller2011, Pelliccia2017}, although recently rejected by others \citep{Tiley2016, Turner2017}; ii) the best-fit relation is consistent with the one for the HR-COSMOS sample, which is in general at less dense environment with respect to SC1604 galaxies, suggesting a non-dependence of the smTFR on environment. We find, though, that $\sigma_{intr}$ on $logV_{2.2}$ is slightly larger (0.15~dex) for SC1604 than for HR-COSMOS (0.11~dex). Although this difference in $\sigma_{intr}$ is very small, it may be a sign of the environment affecting galaxy kinematics. 

To better investigate this possible effect, excluding any bias that may arise from the comparison of different samples, we perform an internal analysis within the SC1604 sample by measuring the scatter around the relation for the entire kinematically reliable sample (including rotation and dispersion dominated galaxies) against two metrics of environment: local, described by the quantity $log(1 + \delta_{gal})$ (Sec.~\ref{subsubsec:localdens}); global, expressed by the parameter $\eta$ (Sec.~\ref{subsubsec:globaldens}). Using both metrics of environment allows us to discriminate, in case of positive environmental sign, between different physical processes taking place. 
We, therefore, define the scatter around the smTFR as the \mbox{1$\sigma$} error-weighted (combined error on $logM_\ast$ and $logV_{2.2}$) shortest distance of each individual data point to the relation. More details about how this scatter is computed is presented in Appendix D of \citet{Pelliccia2017}. In the right panels of Figure~\ref{fig:smTFR} we show the distribution of the scatter as a function of $\eta$ (top) and $log(1 + \delta_{gal})$ (bottom). We compute the median of the scatter per environment bin in order to investigate the presence of any trend. The global environment is binned according to the definition for the values of $\eta$ described in Sec.~\ref{subsubsec:globaldens}, while the local environment is binned such that $log(1 + \delta_{gal})<0.5$, $0.5<log(1 + \delta_{gal})<1$ and $log(1 + \delta_{gal})>1$ describe galaxies in low, intermediate and high local overdensity, respectively. The normalized median absolute deviation \citep[$\sigma_{NMAD}$,][]{Hoaglin1983} is used to estimate the uncertainty on the medians as  $\sigma_{NMAD}/\sqrt{n-1}$ \citep[$n$\,=\,bin size, see][and reference therein]{Lemaux2017vuds}.
We observe that the median \mbox{1$\sigma$} error-normalized offsets from the relation have values oscillating between 1.3$\sigma$ to 2.5$\sigma$ with no clear trend with both the environment metrics.  To determine if the observed fluctuations of the median values as a function of the environment are significant, we perform a \textit{Kolmogorov-Smirnov} (K-S) test between all the distributions of the scatter in each environment bin (both local and global), and we rejected the null hypothesis that the two considered distributions  are drawn by the same sample if the \mbox{p-value$<0.05$}. The results from the nine permutations of the K-S test (six for the global densities and three for the local ones) provide  \mbox{p-values} between 0.2 and 0.9, which tell us that we cannot exclude the possibility that all the sub-samples are drawn for the same distribution, and therefore that the fluctuations observed in the median scatters cannot be attributed to the environment. We have, moreover, repeated the same analysis by using the normalized scatter around the relation for only the rotation dominated galaxies, and we found that the conclusion is invariant.  This result is consistent with past studies \citep[e.g.,][]{Bosch2013, PerezMartinez2017} that have found no evidence of environmental effect on the smTFR.

We are aware of the issue recently brought up by some authors \citep[e.g.,][]{Turner2017}  that galaxies at higher redshifts ($z\geq1$) exhibit higher velocity dispersions than their counterpart at lower $z$, contributing to the dynamical support of the galaxy. For such galaxies it is likely that the gas velocity dispersion should be combined with the rotation velocity in the determination of the smTF relation. For this reason, we repeat this analysis using as velocity estimator the circular velocity $V_{circ}$, which combines the contribution of the rotation and the dispersion of the gas by adopting an asymmetric drift correction. Following \citet{Meurer1996}, we assume that  i) the galaxy kinematics is axisymmetric, ii) the velocity dispersion is isotropic, iii) the velocity dispersion and the scaleheight of the galaxy disc are constant with the radius $R$, and iv) the gas surface density follows an exponential profile. The value of $V_{circ}(R)$ corrected for asymmetric drift is thus given by:
\begin{equation} \label{eq:vcirc1}
\\   \qquad V_{circ}(R)= \sqrt{V^2_{rot}(R) + \sigma_0^2\times \frac{R}{r_s}} \, .
\end{equation}
\noindent This formula for $V_{circ}(R)$ is consistent with past works \citep[e.g.,][]{Lelli2014}, although others \citep[e.g.,][]{Burkert2010, Burkert2016} adopt an asymmetric drift correction with the dispersion component two times larger than the one in Eq.~\ref{eq:vcirc1}. As before, we measure $V_{circ}$ at the radius $R=r_{2.2}$.

 We re-fit the smTFR using the full kinematically reliable sample (including rotation and dispersion dominated galaxies)  and find that the best-fit value of $slope$ and $intercept$ are consistent within the errors with the values showed in Figure~\ref{fig:smTFR} (left panel). Moreover, we confirm no trend in the normalized scatter around the relation with the environment.

\begin{figure}
	\includegraphics[width=\columnwidth]{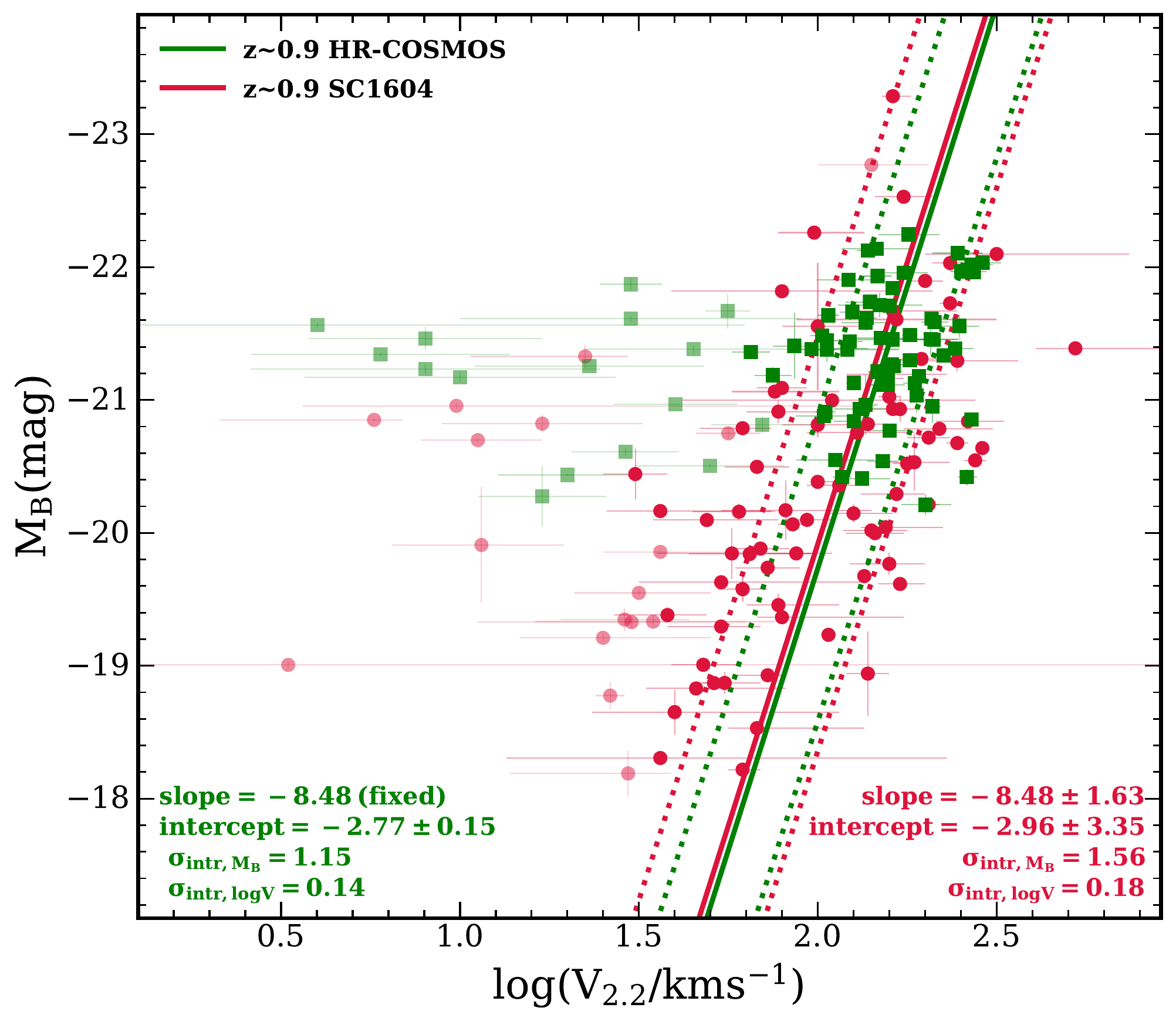}
    \caption{B-band Tully-Fisher relation constrained using the rotation dominated sub-sample (solid red) of our kinematically reliable sample compared to the relation for the rotation dominated galaxies in HR-COSMOS (solid green). In light colors are shown the measurements for the dispersion dominated galaxies of both samples. The fitted relations are expressed by the form $M_B= slope \times(logV_{2.2}-logV_{2.2, 0}) + intercept \, $,
where $logV_{2.2, 0}= 2.0~dex$, and are shown with solid red and green lines for SC1604 and HR-COSMOS respectively.  The intrinsic scatters $\sigma_{intr}$ around the relations are shown with dotted lines. The value of the best-fit parameters and $\sigma_{intr}$ for $M_B$ and $logV_{2.2}$ are presented in the bottom left and right corner for HR-COSMOS and SC1604, respectively.  }
    \label{fig:BbandTFR}
\end{figure}

\begin{figure*}
	\includegraphics[width=0.56\textwidth]{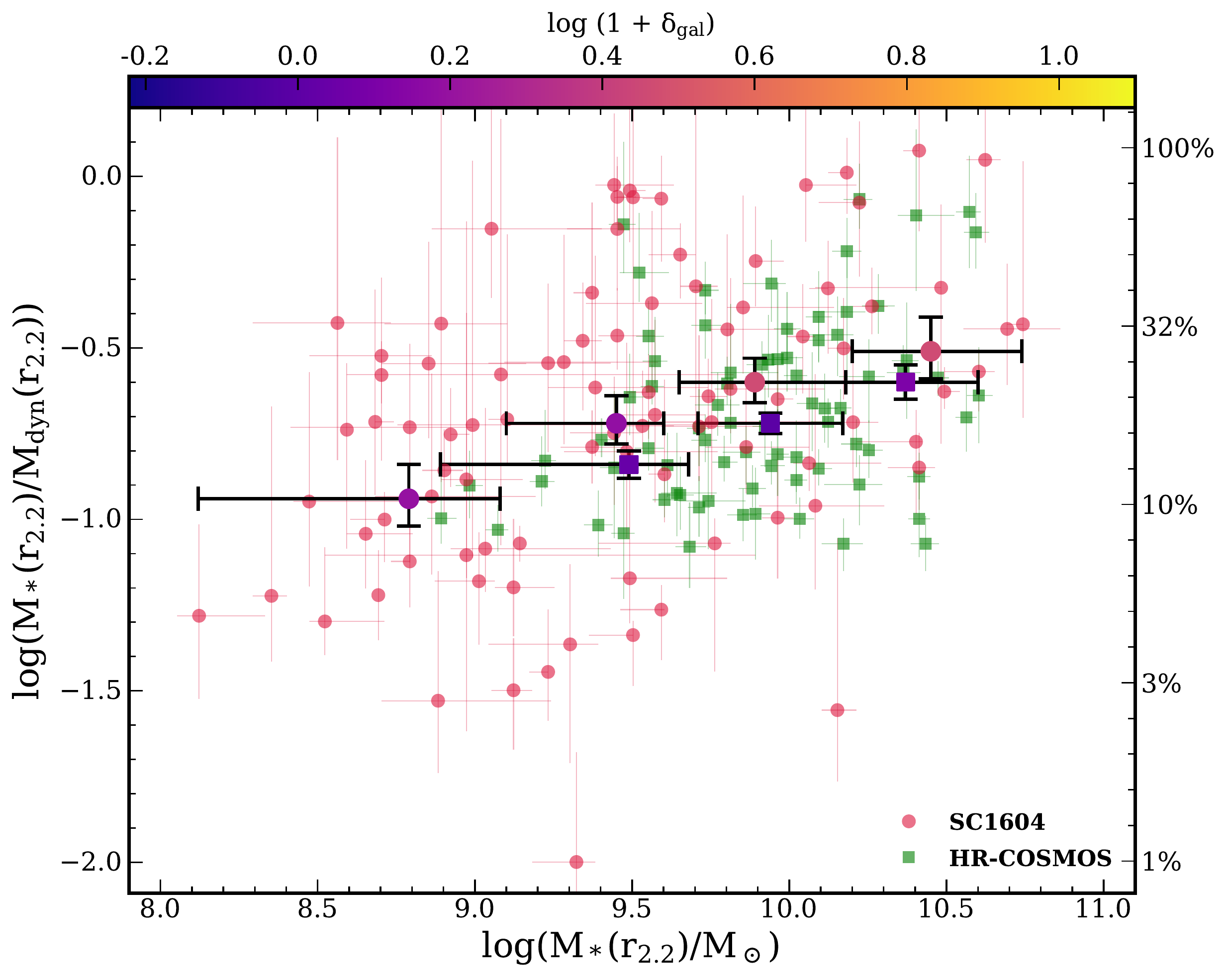}
		\includegraphics[width=0.43\textwidth]{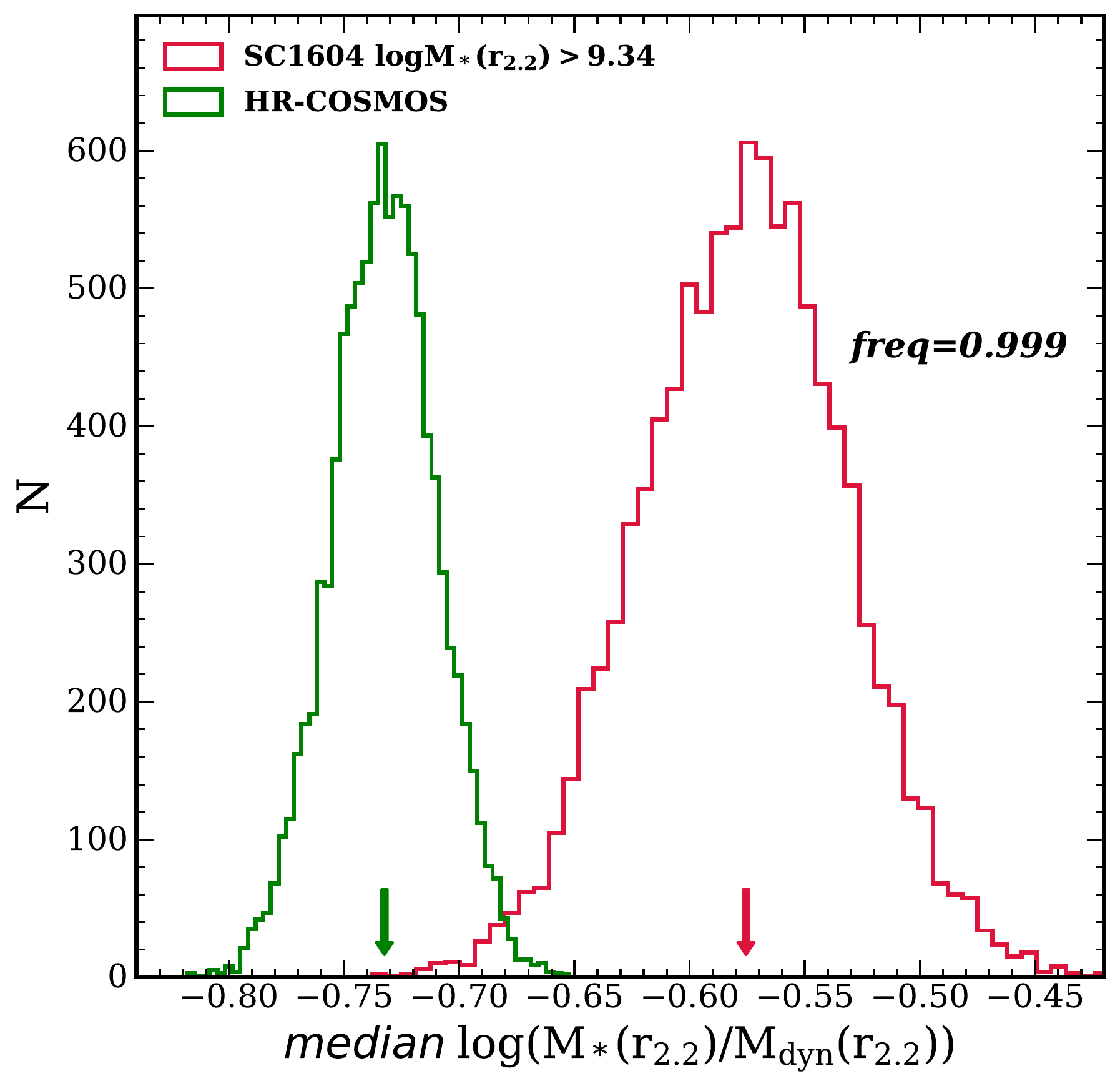}
    \caption{\textit{Left panel:} Stellar-to-dynamical mass ratio versus stellar mass within $r_{2.2}$ in logarithmic unit. The light red points and light green squares show the measurements for SC1604 and HR-COSMOS, respectively. The large markers represent the median $log(M_\ast/M_{dyn})$ per stellar mass bin (see Table~\ref{table:ratios}) and are color-coded according to the median value of local overdensity in each bin. The errorbars on the $y$-axis direction represent the uncertainties on the median; the bars on the $x$-axis direction indicate the bin size. \textit{Right panel: } Comparison of the distributions of 10000 Monte-Carlo realizations of the median $log(M_\ast/M_{dyn})$ for SC1604 cut at $M_\ast > 10^{9.34} M_\odot$ (red) and HR-COSMOS (green). The arrows point to the median value of each distribution, and $freq$ represent the relative frequency that, for 100000 random draws from both distributions, the median $log(M_\ast/M_{dyn})$ for SC1604 is larger than for HR-COSMOS.}
    \label{fig:stellardynamic_ratio}
\end{figure*}
\subsection{B-band Tully-Fisher Relation} \label{subsec:BbandTFR}
It has been proposed by some works that the effect of the cosmic time or the environment on galaxies may not affect the smTFR because the changes in velocity and stellar mass happen on the same timescale along the relation. However, galaxy luminosity, i.e, rest-frame B band, can change on shorter timescales, since it is sensitive to a galaxy's recent star-formation history. For this reason, the B-band TFR is thought to be evolving with time \citep[e.g.,][]{Portinari2007, Miller2011} and environment \citep[e.g.,][]{Milvang-Jensen2003, Bamford2005}. In a recent paper \citet{PerezMartinez2017} performed a kinematic analysis for six XMM2235-2557 cluster galaxies and three field galaxies at $z\sim 1-1.4$ and found that, for a given velocity, cluster galaxies are 1.6~mag more luminous in rest-frame B band ($M_B$) than their field counterparts. 

Here, we constrain the B-band TFR for our kinematically reliable sample in SC1604 and compare it to the relation for HR-COSMOS galaxies in Figure~\ref{fig:BbandTFR}. The best-fit relation for both samples was obtained using the same technique described in Sec.~\ref{subsec:smTFR} for the smTFR. Since the HR-COSMOS sample has a smaller range of B band luminosities, the constrained slope would have large uncertainties; therefore, we decide to fix it to the value obtained for SC1604. The best-fit parameters, as well as the intrinsic scatter in $M_B$ and $logV_{2.2}$, are shown in the bottom left and right corners of Figure~\ref{fig:BbandTFR} for HR-COSMOS and SC1604, respectively. We find that the B-band TFR is largely consistent between the two samples, showing no evolution of the $intercept$ due to the environment. As done in Sec.~\ref{subsec:stellarmass} for the other rest-frame colors, we investigated the differences in $M_B$ between the two samples induced by the different SED fitting codes, the different templates, and the k-correction due to different filter used, and find a maximum difference of $\Delta M_B=0.03$~mag, which is very small and does not bias our result.  

As observed for the smTFR, SC1604 galaxies also exhibit 0.41~mag larger intrinsic scatter in $M_B$ than HR-COSMOS ones, which may hint at some environmental effect. To verify that, we analyze, as done in Sec.~\ref{subsec:smTFR} for the smTFR, the 1$\sigma$ error-weighted scatter of our kinematically reliable sample (including rotation and dispersion dominated galaxies) around the B-band TFR against the measurements of the local and global overdensities. We observe median offset from the relation ranging from $\sim3\sigma$ to $4\sigma$. Although these values are larger compared to those measured for the smTFR, again no clear trend with both environment metrics is observed. We repeated the same analysis using only the normalized scatter of the rotation dominated galaxies around the relation, and we found that the result does not change.
Therefore, the internal investigation on our SC1604 kinematic sample shows a non-dependence of smTFR and B-band TFR on the environment. However, larger samples, both in the low and high overdensity environments,  are necessary to investigate if any more subtle dependence exists.

\subsection{Stellar-to-Dynamical Mass Ratio} \label{subsec:stellar-to-dynamical}
One of the greatest powers of galaxy kinematic measurements is that they provide information about the enclosed dynamical mass ($M_{dyn}$), i.e., the total mass of the galaxy including the contribution from stars, gas and dark matter ($M_{dyn} = M_\ast + M_{gas}+ M_{DM}$). Indeed, the maximum circular velocity ($V_{circ}$, see Eq.~\ref{eq:vcirc1}), measured at a certain radius $R$, provides the measure of the galaxy $M_{dyn}$ within $R$ following the formula:
\begin{equation} \label{eq:Mdyn}
\\ \\  \qquad M_{dyn}(R)= \dfrac{R\times V_{circ}^2(R)}{G} \, ,
\end{equation}
where $G$ is Newton's gravitational constant. Since we use the value of $V_{circ}$ at $r_{2.2}$, we also measure $M_{dyn}$ at the same radius. 
The values of $M_{dyn}(r_{2.2})$ for our kinematically reliable sample range between $9.8\times 10^{8}\,M_{\odot}$ and $5.1\times 10^{11}\,M_{\odot}$ with a median value of $1.5\times 10^{10}\,M_{\odot}$.

To make a fair comparison with the galaxy stellar mass, knowing that this is measured at larger radii then $M_{dyn}(r_{2.2})$ (i.e., corrected for $r_\infty$), we apply a correction to $M_\ast$ by estimating that 65\% of the total light is contained within $r_{2.2}$ for a galaxy described by an exponential profile and assuming a constant stellar mass-to-light ratio as function of radius, as done by \citet{Pelliccia2017}. 
We compute the stellar-to-dynamical mass ratio ($M_\ast/M_{dyn}$) within the radius $r_{2.2}$ for our kinematically reliable sample, and find a median value of 0.21, which confirms that the contribution of the stellar component to the galaxy total mass is small, while gas+dark matter make up  most of the total mass, in line with other works \citep[e.g.,][]{Stott2016,Pelliccia2017}. We note that four galaxies exhibit values of $M_\ast/M_{dyn}$ larger than 1 at the $\gtrsim 1.5\sigma$ level, which, if true, is unphysical. 
We inspected these four galaxies and found that: one galaxy (ID=LFC\_SC1\_05297) is blended with another galaxy in the ground-based image used for the SED fitting, and therefore the measured $M_\ast$  is most likely affected by the contribution of the two galaxies; one galaxy (ID=LFC\_SC1\_03266) shows a reduced $\chi^2$ from the fitting performed with FAST to derive $M_\ast$ measurements (see Sec.~\ref{subsec:stellarmass}) equal to 49, which is considerably higher than the median reduced $\chi^2$ of the entire sample ($\sim$1.1); therefore, the measurement of $M_\ast$ for this galaxy is likely not accurate; the last 2 galaxies (ID=LFC\_SC1\_05250, ID=COS\_SC1\_02200) are the only ones to show a combination of large offset between PA$_{galaxy}$ and PA$_{slit}$ (|$\Delta$PA| $\sim40$\degr) and being almost face-on (inclination $\sim30$\degr), which together provide the highest uncertainties  in the determination of the velocity; therefore, the measurements of the $M_{dyn}$ may be incorrect. We decided to exclude these galaxies from the following analysis in order not to bias our results.

\begin{table}
\caption{Median stellar-to-dynamical ratio for the two comparing samples: ORELSE-SC1604 and HR-COSMOS}             
\label{table:ratios}              
\begin{tabular}{c c c c} 
\multicolumn{4}{c}{\textbf{ORELSE-SC1604}} \\
\hline\hline     
\noalign{\smallskip}   
N$^a$ &  $\widetilde{log(M_{\ast}(r_{2.2})/M_\odot)}^b$& $\widetilde{log(1+\delta_{gal})}^c$ & $\widetilde{log(M_\ast/M_{dyn})}^d$ \\ 
\hline         
\noalign{\smallskip}   
 27   &   8.79  & 0.19 &   $-0.94_{-0.08}^{+0.10}$ \\   
 \noalign{\smallskip}   
30   &    9.45  & 0.20 &   $-0.72_{-0.06}^{+0.08}$  \\
 \noalign{\smallskip} 
21   &    9.89  & 0.46 &   $-0.60_{-0.06}^{+0.07}$ \\
 \noalign{\smallskip} 
12   &    10.45  & 0.46 &  $-0.51_{-0.08}^{+0.10}$ \\
 \noalign{\bigskip  } 
\multicolumn{4}{c}{\textbf{HR-COSMOS}} \\
\hline\hline     
\noalign{\smallskip}   
N$^a$ &  $\widetilde{log(M_{\ast}(r_{2.2})/M_\odot)}^b$& $\widetilde{log(1+\delta_{gal})}^c$ & $\widetilde{log(M_\ast/M_{dyn})}^d$ \\ 
\hline         
\noalign{\smallskip}   
 21   &   9.49  & 0.03 &   $-0.84_{-0.04}^{+0.04}$\\   
 \noalign{\smallskip}   
36   &   9.94  &  -0.01 &   $-0.72_{-0.03}^{+0.03}$\\
 \noalign{\smallskip} 
19   &    10.37  & 0.10 &   $-0.60_{-0.05}^{+0.05}$\\
\hline
\end{tabular}
{\footnotesize$^{(a)}$ Number of galaxies in each $M_\ast$ bin\\
 $^{(b)}$ Median $M_\ast$ within $r_{2.2}$ per bin in logarithmic unit\\
 $^{(c)}$ Median value of local overdensity per bin\\
 $^{(d)}$ Median $M_\ast/M_{dyn}$ within $r_{2.2}$ per bin in logarithmic unit}
\end{table}
To investigate any dependence of $M_\ast/M_{dyn}$ with environment we decide to make a comparison with the that measured in the HR-COSMOS, a sample which probes lower local overdensities (see Figure~\ref{fig:localoverdens} right panel). We attempted to internally trace changes in $M_\ast/M_{dyn}$ as a function of the local and global environment measurements available for SC1604, but the number of galaxies per environment bin (especially in higher dense environments) were not large enough to provide statistically significant results.
The environmental analysis is performed by computing the median per stellar mass bin of the $M_\ast/M_{dyn}$ ratios for both SC1604 and HR-COSMOS samples and comparing it for sub-samples of galaxies characterized by similar  median $M_\ast(r_{2.2})$. The measurements for SC1604 are divided into four bins of $M_\ast(r_{2.2})$, while for HR-COSMOS, given its smaller range of stellar masses, we divide the measurements into three $M_\ast(r_{2.2})$ bins. In Table~\ref{table:ratios} we report the number of galaxies and the median values of $M_\ast(r_{2.2})$, local environment and $M_\ast/M_{dyn}$ ratio in each mass bin for the two comparing samples.
 The $M_\ast(r_{2.2})$ range in each bin is chosen to be a compromise between having a significantly large number of galaxies and similar median $M_\ast(r_{2.2})$ in the two samples.
The median  $M_\ast/M_{dyn}$ in each bin is computed using a Monte-Carlo approach, where 10000 realizations of the median are computed after randomly perturbing the single $M_\ast/M_{dyn}$ values according to their errors. From the distribution of the realizations, we determine the average $M_\ast/M_{dyn}$ and its 1$\sigma$ uncertainty as the median and the 16th and 84th percentiles of the distribution, respectively.

\begin{figure*}
	\includegraphics[width=\textwidth]{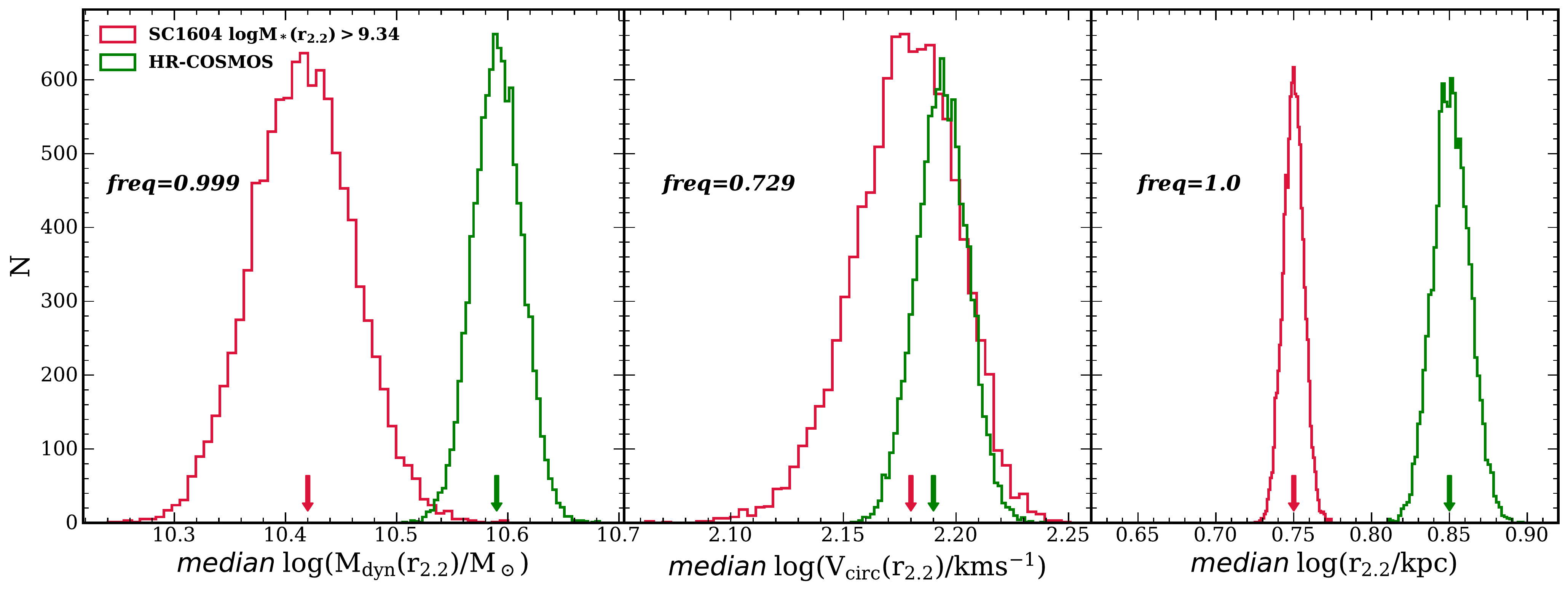}
    \caption{Comparison of the distributions of 10000 Monte-Carlo realizations of the median $logM_{dyn}(r_{2.2})$ (left panel), $logV_{circ}(r_{2.2})$ (middle panel) and $logr_{2.2}$ (right panel) for SC1604 cut at $M_\ast > 10^{9.34} M_\odot$ (red) and HR-COSMOS (green). The arrows point to the median value of each distribution, and $freq$ represent the relative frequency that, for 100000 random draws from both distributions, the drawn value for SC1604 is smaller than for HR-COSMOS.}
    \label{fig:diagnostic}
\end{figure*}

The stellar-to-dynamical mass ratio as a function of the stellar mass within $r_{2.2}$ for both samples is shown in Figure~\ref{fig:stellardynamic_ratio} (left panel), where the light red points and the light green squares represent the single measurements for SC1604 and HR-COSMOS, respectively. The larger markers show the median $M_\ast/M_{dyn}$ per $M_\ast(r_{2.2})$ bin and are color-coded according to the median $log(1+\delta_{gal})$ measured in each bin. Once again, it is clear, even for the single stellar mass bins, that SC1604 galaxies reside always at higher local overdensities than HR-COSMOS ones, having average values of $log(1+\delta_{gal})$ $\sim1.5-3$ times higher (see Table~\ref{table:ratios}). The first result that emerges from the left panel of Figure~\ref{fig:stellardynamic_ratio} is that for both samples stellar mass is only a small contribution (maximum $\sim$30\%) to the galaxy total mass within $r_{2.2}$, which is, therefore, dominated by the gas+dark matter mass at $z\sim 0.9$. We also observe a decrease of $M_\ast/M_{dyn}$ towards lower stellar masses, meaning that the contribution of $M_\ast$ to the total mass within $r_{2.2}$ drops from $\sim$30\% to $\sim$10\% from high to low stellar mass galaxies.  This is in agreement with the observations showing that in general low mass galaxies have higher gas and/or dark matter fraction than high mass galaxies \citep[e.g.,][]{Geha2006, Blanton2009, Battaglia2013}. Moreover, we find that $M_\ast/M_{dyn}$ for SC1604 galaxies is systematically higher than the ratio for the HR-COSMOS sample. If we neglect the first mass bin in SC1604 and we make a one-to-one comparison between the value of the median $M_\ast/M_{dyn}$ in the remaining three bins of SC1604 and in the three bins of HR-COSMOS, we find that $M_\ast/M_{dyn}$, from higher to lower $M_\ast(r_{2.2})$,  is $\sim$19\%,  $\sim$24\%, $\sim$24\% smaller for HR-COSMOS within 1.1$\sigma$, 2$\sigma$ and 2$\sigma$, respectively. Knowing that the statistical significance of this comparison is low in every given bin, to verify that the observed offset towards higher $M_\ast/M_{dyn}$ is real and characteristic of the population of galaxies in denser environment, we compare the median $M_\ast/M_{dyn}$ within $r_{2.2}$ for the SC1604 sample cut at  $logM_\ast(r_{2.2})>9.34$ (corresponding to $logM_\ast>9.53$ ), which has $\widetilde{log(1+\delta_{gal})}=0.35$, and for the HR-COSMOS sample with $\widetilde{log(1+\delta_{gal})}=0.03$. The stellar mass cut in SC1604 is applied to account for the lack of lower $M_\ast(r_{2.2})$ in HR-COSMOS and is imposed in an attempt to make the two samples comparable.
A K-S test performed on the stellar mass distributions of the original HR-COSMOS sample and the cut SC1604 sample returns a p-value of 0.21, which does not allow us to exclude the possibility that they are drawn from the same underlying distribution.

\begin{table*}
\caption{Comparison of median parameters between SC1604 (with stellar mass cut) and HR-COSMOS }             
\label{table:MdynVr_compare}              
\begin{tabular}{c c c c c c c } 
\hline\hline     
\noalign{\smallskip}   
& N &  $\widetilde{log\left(\dfrac{M_{\ast}(r_{2.2}))}{M_\odot}\right)}$& $\widetilde{log\left( 1+\delta_{gal}\right) }$ & $\widetilde{log\left( \dfrac{M_{dyn}(r_{2.2})}{M_\odot}\right) }$ &  $\widetilde{log\left(\dfrac{V_{circ}(r_{2.2})}{km\,s^{-1}}\right) }$ & $\widetilde{log\left( \dfrac{r_{2.2}}{kpc}\right) }$\\ 
\hline         
\noalign{\smallskip}   
ORELSE ($M_{\ast} (r_{2.2})>10^{9.34}M_\odot$) & 54   &   9.78  & 0.35 &   10.42$\pm$0.05 & 2.18$\pm$0.02 & 0.75$\pm$0.01 \\   
 \noalign{\smallskip}   
HR-COSMOS & 76   &   9.93  & 0.03 &   10.59$\pm$0.02& 2.19$\pm$0.01 & 0.85$\pm$0.01 \\
 \noalign{\smallskip} 
 \hline
\end{tabular} \\
\end{table*}
Using this new stellar mass cut SC1604 sample, we then computed the median of the ratio $M_\ast/M_{dyn}$ of this sample relative to the original HR-COSMOS sample. The final median of each sample was estimated using the same Monte-Carlo approach as described above, and we show the distribution of the individual medians for each of the 10000 realizations for both samples in the right panel of Figure~\ref{fig:stellardynamic_ratio}. The median $log(M_\ast/M_{dyn})$ for HR-COSMOS is equal to $-0.73\pm 0.02$, while for the SC1604 sample with $logM_\ast(r_{2.2})>9.34$ the median ratio is $-0.58\pm 0.04$, telling that SC1604 is characterized by $M_\ast/M_{dyn}$ ratio  in median $\sim$1.4 times larger than HR-COSMOS with a significance of 3.9$\sigma$. We double-check the significance of the observed offset by  randomly drawing a value from both distributions of the median shown in Figure~\ref{fig:stellardynamic_ratio} (right panel) 100000 times,  finding that in 99.9\% of the cases we obtain a larger median $log(M_\ast/M_{dyn})$ for SC1604 than for HR-COSMOS.
This result appears to be somewhat different than what found by \cite{Darvish2015}, who analyzed the $M_\ast/M_{dyn}$ ratio for a sample of 28 star-forming galaxies in a large filament at $z\sim 0.53$  in the COSMOS field, and compared them with a sample of 30 field galaxies. They found no differences in the $M_\ast/M_{dyn}$ ratio for the two samples. We note, however, that our results may not be comparable to the one presented by \citet{Darvish2015}, since their samples are at lower redshift with respect to both SC1604 and HR-COSMOS, the distribution of the galaxy stellar mass has a narrower range of $M_\ast= \sim10^9-10^{10.5} M_\odot$, and filament environments differ from group/cluster environments. Moreover, we measured $M_\ast/M_{dyn}$ at  $r_{2.2}$ that is the radius at which we constrained $M_{dyn}$ (as described above); conversely, \citet{Darvish2015} measured $M_{dyn}$ using the galaxy velocity dispersion and its half-light radius.

Given the fact that we compare the stellar-to-dynamical mass ratio for fixed stellar mass, the observed differences of $M_\ast/M_{dyn}$ in the two samples is driven by a difference in their $M_{dyn}(r_{2.2})$. Specifically,  SC1604 galaxies have in median $\sim$30\% smaller $M_{dyn}(r_{2.2})$ than HR-COSMOS galaxies, as shown in the left panel of Figure~\ref{fig:diagnostic}, where we compare the distributions of the Monte-Carlo realizations of the median $logM_{dyn}(r_{2.2})$ for the two samples. Indeed, we find that for 100000 random draws from these two distributions, 99.9\% of the times $M_{dyn}(r_{2.2})$ for SC1604 is smaller than for HR-COSMOS. To understand the causes of smaller dynamical masses for galaxies in denser environments, we investigate  the environmental dependence of the galaxy parameters that define  $M_{dyn}(r_{2.2})$ in Eq.~\ref{eq:Mdyn}: the circular velocity $V_{circ}(r_{2.2})$ described by Eq.~\ref{eq:vcirc1} and the radius $r_{2.2}$.
We show in Figure~\ref{fig:diagnostic} (middle and right panels) the comparison of the Monte-Carlo realizations of the median values of these two parameters for SC1604 sample cut at $logM_\ast(r_{2.2})>9.34$ and HR-COSMOS. Their final median values are, also, reported in Table~\ref{table:MdynVr_compare}. 

We find that the distributions of the medians values of $logV_{circ}(r_{2.2})$ for the two samples largely overlap, with the HR-COSMOS one being narrower and shifted to slighter higher values ($\widetilde{logV_{circ}(r_{2.2})}= 2.19\pm 0.01$) than SC1604 ($\widetilde{logV_{circ}(r_{2.2})}= 2.18\pm 0.02$), which instead shows a broader distribution. We are confident that this comparison is fair since the kinematic measurements for both samples have been performed using the same technique.  Conversely, we find that the distributions of the median values of  $logr_{2.2}$ for the two samples are completely disparate, with no overlap, and with the HR-COSMOS median $logr_{2.2}$ being 0.1~dex larger than  for SC1604. We note that the formal errors on $r_{2.2}$ for HR-COSMOS are not available to us; therefore, in the Monte-Carlo iterations we adopt uncertainties similar to the ones measured for SC1604, by randomly sampling from $r_{2.2}$ error distribution of SC1604 galaxies. We also repeat the measurement by adopting uncertainties twice as large as the SC1604 ones, and we find that the result does not change.
Moreover, aware of the fact that $r_{2.2}$ for the two samples was measured using different softwares  \citep[see Sec.~\ref{subsec:kinesample} and][]{Pelliccia2017}, we verify that the difference observed is not a result of different measurement techniques. To that end, we retrieve the archival  \textit{HST}/ACS F814W images for the HR-COSMOS galaxies, which were observed using similar strategy adopted for SC1604 images (e.i., one \textit{HST} orbit exposure), and we run SExtractor in the same way as done for SC1604 as described in Sec.~\ref{subsec:kinesample}. We find that, comparing the original and the re-measured $r_s$ (from which we derive $r_{2.2}$, see Sec.~\ref{subsec:smTFR}), the SExtractor measurements are in median 10\% larger than the value used in \citet{Pelliccia2017}, confirming that difference in size between SC1604 and HR-COSMOS galaxies is indeed real, and, in fact, it may be larger than we estimate here.

This investigation led us to affirm that the cause of smaller dynamical masses in denser environments are mainly due to the smaller size of the galaxies in this environment. This result is consistent with what found by  \citet{Kuchner2017}, who constrained the mass-size relation for galaxies in cluster environment at $z=0.44$, and compared the relation for star-forming cluster members to the one obtained by \citet{vanderWel2014} for star-forming galaxies in the field. They found that the mass-size relation for cluster galaxies is in general shifted towards smaller values of the galaxy size, with a shift equal to $\sim 0.1-0.2$~dex  for stellar masses similar to our sample.

\subsection{Specific Angular Momentum} \label{subsec:specific_angular_momentum}
In the $\Lambda$CDM cosmology, dark matter haloes acquire rotation from the tidal torques \citep{Hoyle1951}. This rotation is quantified by the dimensionless spin parameter:
\begin{equation} \label{eq:lambda}
\\ \\  \qquad \lambda= \dfrac{J_{DM}\,|E|^{1/2}}{G\, M_{DM}^{5/2}} \, ,
\end{equation}
where $J$ is the angular momentum, $G$ is the gravitational constant, $E$ is the energy and $M$ the mass of the system \citep{Peebles1969}. From the tidal torque theory and N-body simulations $\lambda$ is predicted to follow a log-normal distribution, with expectation value $\langle\lambda\rangle=0.035$ and a $1\sigma$ log dispersion of 0.23~dex \citep{Maccio2008}, relatively insensitive to cosmological parameters, time, galaxy mass and environment \citep[e.g.,][]{BarnesEfstathiou1987, Zurek1988, SteinmetzBartelmann1995, ColeLacey1996, Maccio2007, Bryan2013}.

By inverting Eq.~\ref{eq:lambda} and defining the specific angular momentum of the dark matter halo as $j_{DM}=J_{DM}/M_{DM}$, we can express it in terms of $\lambda$ and $M_{DM}$:
\begin{equation} \label{eq:jDM}
\\ \\  \qquad j_{DM}\propto \lambda\,M_{DM}^{2/3} \, .
\end{equation}
If we assume that initially the baryons are well mixed with the dark matter of the parent halo and have the same angular momentum, according to the standard theory of disc galaxy formation,  baryons should retain their specific angular momentum as they collapse into the center of the halo \citep{FallEfstathiou1980, Mo1998}. However several processes (e.g., mergers) can reduce/increase the galaxy specific angular momentum. We introduce, therefore, the fraction $f_j$ of the specific angular momentum retained by the baryons during the galaxy formation, defined as the ratio between the stellar and dark matter specific angular momentum ($f_j = j_\ast/j_{DM}$, assuming that $ j_\ast$ follows the baryonic specific angular momentum). In addition, considering that only a fraction ($f_\ast$) of the cosmological baryon fraction $f_b$ (=0.17) is converted into stars we define $f_\ast = M_\ast/(f_b\times M_{DM})$. From Eq.~\ref{eq:jDM} we can, then, predict the stellar specific angular momentum to be:
\begin{equation} \label{eq:jstar_pred}
 \frac{j_\ast}{km\,s^{-1} kpc} = 9.07\cdot 10^3 \lambda\, f_j \,(f_b f_\ast)^{-2/3}\left( \frac{H(z)}{H_0}\right)^{-1/3} \left( \frac{M_\ast}{10^{11} M_\odot}\right)^{2/3}   \, ,
\end{equation}
where $H(z)=H_0\,[\Omega_{\Lambda, 0} + \Omega_{M, 0} \times (1+z)^3)]^{1/2}$. This expression is equivalent to the one derived by \citet{Harrison2017} and similar to other derivations by e.g., \citet{RomanowskyFall2012}, and  \citet{Burkert2016}.

\begin{figure}
	\includegraphics[width=\columnwidth]{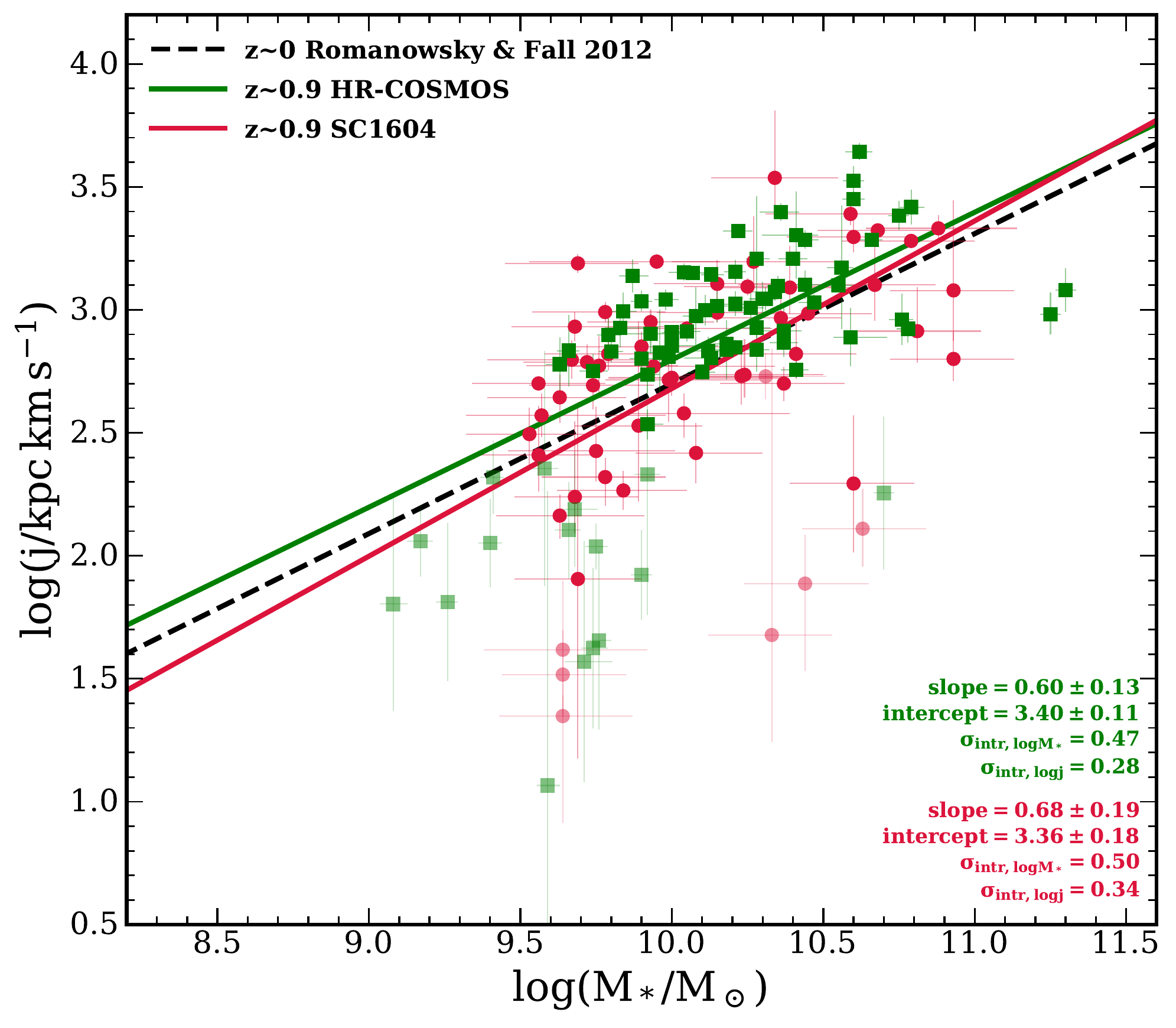}
    \caption{Relationship between  $j_\ast$ and $M_\ast$ for our SC1604 sample with mass cut (see text, Sec.~\ref{subsec:specific_angular_momentum}) in solid red color compared to the relation for HR-COSMOS in solid green. In light colors are shown the measurements for the dispersion dominated galaxies of both samples. The fitted relations are expressed by the form $logj_\ast= slope \times(logM_\ast-logM_{\ast, 0}) + intercept \, $, where $logM_{\ast, 0}= 9.8~dex$, and are shown with solid red and green lines for SC1604 and HR-COSMOS, respectively. The value of the best-fit parameters and $\sigma_{intr}$ for both $logM_\ast$ and $logj_\ast$ are presented in the bottom right corner in red color for SC1604  and in green color for HR-COSMOS. We show as reference the relation derived by \citet{RomanowskyFall2012} for spiral galaxies at $z=0$.}
    \label{fig:jstarMstar}
\end{figure}
The measurements presented in this paper, allow us to measure the stellar specific angular momentum $j_\ast$. For spiral galaxies approximated to be axisymmetric, infinitely thin discs with an exponential surface density profile, and assuming that the ionized gas traces the stellar disc, we adopt:
\begin{equation} \label{eq:jstar_obs}
\\ \\  \qquad j_\ast =  2\, r_s\, V_{2.2} \, .
\end{equation}
This value of $j_\ast$ is an approximation based on global measurements, due to the lack of detailed kinematic maps. It is widely used in literature and provides remarkably good results when compared to the ``true'' value \citep{ObreschkowGlazebrook2014}. On a $j_\ast - M_\ast$ diagram galaxies follow a linear relation with similar slope $\sim 2/3$ \citep[e.g.,][]{Fall1983, RomanowskyFall2012}, which is consistent with the $M_\ast$ dependence predicted in  Eq.~\ref{eq:jstar_pred}, and have different normalization according to the galaxy morphology \citep[see e.g.,][]{RomanowskyFall2012}.
We present the comparison of $j_\ast$ measurements as a function of $M_\ast$ for SC1604 and HR-COSMOS in Figure~\ref{fig:jstarMstar}. We apply to SC1604 the same stellar mass cut discussed in Sec.~\ref{subsec:stellar-to-dynamical} in order to avoid bias in the comparison due to the lack of lower $M_\ast$ in HR-COSMOS.  We fit a relation to the full kinematically reliable sample (including rotation and dispersion dominated galaxies) using the same technique adopted for the smTFR  (Sec.~\ref{subsec:smTFR}) and the B-band TFR (Sec.~\ref{subsec:BbandTFR}), obtaining a best-fit slope of $0.68\pm0.19$ for SC1604 galaxies and  $0.60\pm0.13$ for \mbox{HR-COSMOS} galaxies, both consistent with value found by \citet{RomanowskyFall2012} at $z=0$ and with $j_\ast \propto M_\ast^{2/3}$  predicted from  Eq.~\ref{eq:jstar_pred}. When we fit the relation with fixed slope $=2/3$ to both samples, we find best-fit values of the intercept equal to $3.45\pm0.04$ an $3.35\pm0.05$ for HR-COSMOS and SC1604, respectively. This offset of 0.1~dex in the normalization of the relation is suggestive, and possibly indicates a signature of galaxy transformation from low to high density environments. 

Following \citet{Burkert2016}, we combine Eq.~\ref{eq:jstar_obs} and Eq.~\ref{eq:jstar_pred} to estimate the fraction $f_j$ of specific angular momentum retained by the baryons during the formation of the galaxy. To that end, we remove from  Eq.~\ref{eq:jstar_pred} the dependence on $f_\ast$ by using the fitting function proposed by \citet{Dutton2010}  for late-type galaxies:
\begin{equation} \label{eq:fstar_dutton}
 f_\ast = 0.29 \times \left( \dfrac{M_\ast}{5\times10^{10} M_\odot}\right) ^{0.5} \times  \left[ 1+\left( \dfrac{M_\ast}{5\times10^{10} M_\odot}\right) \right] ^{-0.5} \, .
\end{equation}
This function was largely used by previous works \citep[e.g.,][]{RomanowskyFall2012, Burkert2016, Harrison2017}. We note that a recent study by \citet{Posti2018} showed that the dependence of $f_j$ on the galaxy stellar mass varies considerably according to the chosen $f_\ast$. They found that the function proposed by \citet{Dutton2010} provides $f_j$ values approximately constant with $M_\ast$, while other functions \citep[e.g.,][]{Behroozi2013} provide values of $f_j$ that are strongly dependent on the stellar mass, showing a sharp decrease at large $M_\ast$ ($\gtrsim 10^{10.5} M_\odot$).  Testing different $f_\ast$ is outside of the scope of this paper, however, we plan to do it in a future work.

\noindent Assuming for simplicity that all the galaxies analyzed here have the same dark matter spin parameter $\lambda=0.035$, we find that galaxies in the SC1604  have a median value of $f_j = 0.71\pm 0.06$, while for HR-COSMOS galaxies in median $f_j = 0.91\pm 0.04$. This is telling us that galaxies in higher density environments have lost $\sim30\%$ of their original angular momentum, while galaxies in less dense environments have lost only $\sim10\%$ of it. 
This difference persists when we compute the median $f_j$ for only rotation dominated galaxies, obtaining that purely rotating galaxies in low-density environments are able to retain all their original angular momentum ($\tilde{f_j}=1.04 \pm 0.04$), while rotating galaxies in higher density have lost 20\% of their angular momentum ($\tilde{f_j}=0.80 \pm 0.06$). Moreover, we find that the dispersion dominated galaxies in both environments have lost most ($\sim90\%$) of their angular momentum, having  $\tilde{f_j}=0.16 \pm 0.03$  in HR-COSMOS and $\tilde{f_j}=0.07 \pm 0.02$ in SC1604.

One of the main processes thought to influence a galaxy's specific angular momentum is galaxy-galaxy mergers.  \citet{Lagos2018}, using the EAGLE simulations, investigated the impact of galaxy minor (mass ratios $0.1-0.3$) and major (mass ratios $\geq0.3$) mergers  on galaxy's specific angular momentum. 
They found that, for galaxies with $M_\ast\geq10^{9.5}M_\odot$ and in the redshift range $0\leq z\leq2.5$, a single major merger is able to reduce the specific angular momentum, measured within the half-stellar mass radius $r_{50}$, by $\sim20\%$, while one minor merger contributes only with 2\% of $j_\ast$ loss. They were also able to discriminate between the effect on $j_\ast$ due to gas-poor and gas rich mergers, finding that gas-poor mergers are more effective in reducing $j_\ast$, causing a loss of $\sim40\%$  (major mergers) and $\sim20\%$  (minor mergers), while the effect of gas-rich mergers is negligible. Their definition of gas-poor and gas-rich merger is based on the measurements of the combined gas fraction ($f_{gas,\,merger}$) of the two merging galaxies with respect to the combined stellar masses, and requires that $f_{gas,\,merger}\leq0.2$ for gas-poor mergers and $f_{gas,\,merger}\geq0.5$ for gas rich merger. However, in a following paper, using slightly different parameter to describe the specific angular momentum, \citet{Lagos2018b} have shown that for galaxies with $M_\ast>10^{10}M_\odot$ the reduction of the spin parameter $\lambda_{r_{50}}$ due to major mergers is mostly constant at $\sim40\%$ for a range of $f_{gas, merger}\simeq0.02-0.5$, with a rapid decrease to $\sim0\%$ of $\lambda_{r_{50}}$ loss to $f_{gas, merger}\simeq0.8$. 
\citet{Tomczak2017} have demonstrated with a simple semi-empirical model that  galaxy-galaxy merging is a relevant process in dense regions, showing that a larger fraction of mergers is required to reproduce the galaxy stellar mass function in dense environments with respect to the low density environments. We want, therefore, to investigate whether this larger fraction of mergers can account for the reduction of $f_j$ observed in SC1604 compared to HR-COSMOS. 

To this end, we divide both samples in three local overdensity bins (low, intermediate and high overdensities, equal to the ones adopted in Sec.~\ref{subsec:smTFR} and in the right bottom panel of Figure~\ref{fig:smTFR}), and we keep for this analysis only the HR-COSMOS galaxies in the low overdensity bin (63 galaxies) and the SC1604 galaxies in the intermediate and high overdensity bins (20 galaxies). The idea is to understand whether, by correcting $f_j$ observed in SC1604 galaxies in intermediate and high densities for the decrease due to the expected number of mergers they have undergone to reach these environments, we are able to recover the typical  $f_j$ observed in HR-COSMOS galaxies at low densities. In other words, we want to understand if mergers can be considered as the main (or the only) process that caused the decrease of angular momentum in SC1604. 

We make use, therefore, of the semi-empirical model from \citet{Tomczak2017} to estimate how many more mergers galaxies undergo in intermediate/high densities with respect to the galaxies in lower densities. We provide, here, a brief description of the semi-empirical model; for a full discussion we refer the reader to \citet{Tomczak2017}. The model begins with a sample of $\approx 10^6$ galaxies that are simulated to match the $z=5$ universe in terms of their distributions of stellar masses and star-formation rates. These galaxies progress forward in discrete time intervals of 100\,Myr until $z=0.8$ (the median redshift of the ORELSE sample) where in each time-step prescriptions for star-formation, quenching, and galaxy-galaxy merging are enforced. The first two of these prescriptions are informed by empirical relations whereas the latter is allowed to vary. Realizations of this model are tested by comparing the final ($z=0.8$) stellar mass distribution to the observed galaxy stellar mass function in the three environmental density bins used for this analysis. Galaxy-galaxy merging is treated as a free parameter in that multiple realizations of the model are generated with a variable bulk merger count. This bulk merger count is defined as the fraction of the initial sample of $\approx 10^6$ galaxies that merge by the end of the simulation at $z=0.8$, which ranges between 0\% (i.e., no merging) to 95\% (i.e., only 5\% of galaxies remaining). With each time-step galaxy pairs are selected randomly to be merged in accordance with the adopted bulk merger count, with the only constraint that minor mergers (mass ratios $<1:4$) occur 3 times more frequently than major mergers (mass ratios $1:4-1:1$) \citep{Lotz2011}.
For each galaxy we record the total number of major and minor mergers at $z=0.9$ (the median redshift of SC1604) of the most massive progenitor that matches the SC1604 stellar mass cut.

 From this simulation we find that galaxies at intermediate and high density undergo $3.8-4$ times more major mergers ($3-3.5$ times more minor mergers) than galaxies in low density, with on average 0.18 and 0.2 extra major mergers (0.56 and 0.60 extra minor mergers) per galaxy with respect to the low density environment, respectively. If we scale the values of the decrease in $j_\ast$ per major/minor merger from \citet{Lagos2018} by the number of extra mergers each galaxy in intermediate/high density undergoes with respect to the low density, we can use them as a correction for the SC1604 galaxies in order to recover the fraction ($f_{j, corr}$) of retained angular momentum as it would have been if galaxies had not undergone these extra mergers. 
We perform this exercise by adopting a Monte-Carlo approach as follows. 

We begin by computing, for each of the realizations, a new value of $f_j$ for each of the galaxies in the intermediate/high density environment by perturbing the measured values by their errors. For each realization, we define a correction factor to $f_j$ per major and minor merging event by sampling from a Gaussian distribution with mean equal to 0.82 and 0.98, respectively, and dispersion equal to the 1$\sigma$ uncertainties from \citet{Lagos2018}. The mean values of these Gaussians correspond to the median fraction of $j_\ast$ retained per merger as estimated in the simulations of \citet{Lagos2018} for all types of mergers. These two new correction factors per merger are then scaled by the excess number of mergers that galaxies experience in intermediate/high density environments as estimated from the simulation presented in \citet{Tomczak2017}. In order to determine the excess number of mergers per galaxy in each realization, we draw from the distribution of the number of minor and major mergers experienced by simulated galaxies with stellar masses that are within $\pm0.1$~dex from the $M_\ast$ value of each galaxy in our intermediate/high density SC1604 sample. To these numbers we subtract values coming from a similar sampling of similarly massive simulated galaxies in low density environments. The differences of these two pairs of values set the excess number of major and minor mergers each galaxy experiences for that realization. From the ensemble of corrected $f_j$ values ($f_{j, corr}$) computed in this way from each realization, we computed a median $f_{j, corr}$. This process is performed a total of 10000 times, and, from the distribution of the median $f_{j, corr}$ values computed from all realizations we derive the final median corrected value and its uncertainties. After having performed this exercise, we find that for the intermediate/high density SC1604 galaxies, which had an original $f_j=0.76\pm0.10$, the median $f_{j, corr}$ was $0.80_{-0.10}^{+0.12}$. This corrected value is consistent with the original value and is insufficient to recover the $f_j$ value measured for the galaxies in HR-COSMOS in low density environments. 

However, the adopted value of $j_\ast$ loss per merger from \citet{Lagos2018} is an overly conservative value,  since we are not able to discriminate between gas-poor a gas-rich mergers for our sample, and it represents an average value between the effect of gas-poor and gas-rich mergers, which are, respectively, highly effective and not effective at reducing $j_\ast$.  Although we cannot make any definitive statement about the amount of gas fraction in the SC1604 galaxies,  we have reasons to believe that in high density environments the principal type of merger occurring is gas-poor. For example, it ha been observed that spiral galaxies in cluster environments at z=0 show a deficiency of neutral gas with respect to their field counterparts \citep{BoselliGavazzi2006}. Moreover, \citet{Lin2010}  investigated the environments of wet (blue galaxy pairs), mixed (blue-red pairs), and dry (red galaxy pairs) mergers at $0.75 <z< 1.2$ in the Deep Extragalactic Evolutionary Probe 2 \citep[DEEP2,][]{Davis2003, Newman2013_deep2} survey, showing that high local overdensity regions are the preferred environment in which dry and mixed mergers occur, and, indeed, that the combined incidence of such mergers exceeds those of wet mergers in their highest density environments, i.e., galaxy groups. This disparity is only likely to be exacerbated in cluster environments. 
Because of these lines of evidence, we repeat the above analysis using the values of the median fraction of $j_\ast$ retained per major ($0.64\pm 0.09$) and minor ($0.84\pm 0.03$) merger that \citet{Lagos2018} found for gas-poor mergers, we find that the median $f_{j, corr}$ for the SC1604 galaxies in intermediate/high density is equal to $0.91\pm0.15$. This value is consistent within the uncertainties with the median $f_j$ measured for HR-COSMOS in the low density bin ($f_j=0.96\pm0.04$). However, this $f_{j, corr}$ is also consistent within 1$\sigma$-error with the median value of $f_j$ for SC1604 galaxies before the correction for angular momentum loss due to mergers. This consistency with both values is a consequence of largeness of the uncertainties on the individual $f_j$ values, the small sample of galaxies (N=20) in the SC1604 sample at intermediate/high density, and a very conservative approach in the determination of the error budget. We note, however, that, using this approach, if we draw 10000 times from the $f_{j, corr}$ and $f_j$ distributions of the intermediate/high density SC1604 sample and from the $f_j$ distribution of the low-density HR-COSMOS sample, we find that $
62\%$ of the time $f_{j, corr}$ has a value closer to that of the low-density HR-COSMOS sample. Thus, under such a scenario, it is at least plausible to recover nearly all of the angular momentum lost by the SC1604 intermediate/high density sample, and, if correct, gas-poor mergers would be, indeed, the principal cause of the observed reduction of the specific angular momentum in dense environments. However, the large number of assumptions taken in the exercise, the sample size, and the large uncertainties does not allow us to confirm this statement. A larger sample of galaxies in intermediate/high density environments as well as gas fraction measurements would allow for a full and, perhaps definitive investigation, of this scenario.

\section{Conclusions}
We presented here an investigation of the environmental effect on the kinematics of a sample of star-forming galaxies at $z\sim 0.9$, which are part of the ORELSE survey. ORELSE is a large photometric and spectroscopic campaign dedicated to map out and characterize galaxy properties across a full range of environments in 15 fields containing LSSs in a redshift range of $0.6 < z < 1.3$. The sample in this paper is taken from the field SC1604, which is dominated by the known SC1604 supercluster at $z\sim 0.9$. Galaxy samples from two serendipitous clusters discovered along the line of sight at $z=0.60$ and $z=1.18$ are also included in the sample.
We constrained the rotation velocity for our kinematic sample using high-resolution semi-analytical models, and we measured the environment, both local and global.
We constructed the stellar-mass/B-band Tully-Fisher relation, we measured the stellar-to-dynamical mass ratio and the stellar specific angular momentum, and
we investigated their dependence on the environment. Our main results are summarized below.

\begin{itemize} \itemsep5pt
\item[$-$] We constrained the smTFR for SC1604 rotation dominated galaxies, and we compared it with the relation obtained for the HR-COSMOS sample \citep{Pelliccia2017}, which in general is at lower local overdensities than SC1604. We found the two relations to be consistent within the uncertainties. However, we found that the intrinsic scatter $\sigma_{intr}$ on the velocity is slightly larger (0.11~dex vs 0.15~dex) for SC1604 compared to HR-COSMOS. To verify whether this difference in $\sigma_{intr}$ may be an environmental effect, we performed an investigation internal to the SC1604 sample, by analyzing the 1$\sigma$-error normalized scatter around the smTFR for the entire kinematically reliable sample (i.e., including rotation and dispersion dominated galaxies) against two metrics of environments: local and global. We found that the median \mbox{1$\sigma$} error-normalized offsets from the relation have values oscillating between 1.3$\sigma$ to 2.5$\sigma$ with no clear trend with either of the environment metrics. This result does not change if we repeat this analysis using only the 1$\sigma$-error normalized scatter for the rotation dominated galaxies.

\item[$-$] Since changes in velocity and stellar mass may happen on the same timescale along the smTFR, we repeated the same analysis on the B-band TFR, because the rest-frame B-band luminosity is more sensitive to recent episodes of star-formation, and thus subject to changes on shorter timescales. Once again, SC1604 galaxies exhibit 0.41~mag larger $\sigma_{intr}$ in $M_B$ than HR-COSMOS ones; however, no trend with local nor global environment is observed for the  1$\sigma$-error normalized scatter around the B-band TFR. We concluded, therefore, that we did not find evidence that the environment affects the smTFR and the B-band-TFR. This result is consistent with past works using global environment \citep[e.g.,][]{Ziegler2003, Nakamura2006, Jaffe2011} and local environment \citep[e.g.,][]{Pelliccia2017} measurements, although those studies were characterized by small galaxy samples or a small range of densities.

\item[$-$] We measured $M_\ast/M_{dyn}$ ratios within $r_{2.2}$ for the SC1604 sample and compared them to the ones measured for HR-COSMOS. 
After applying a stellar mass cut to the SC1604 sample to account for the lack of lower $M_\ast(r_{2.2})$ in HR-COSMOS, we found that SC1604 galaxies, which on average reside in local overdensities twice higher than HR-COSMOS galaxies, are characterized by $M_\ast/M_{dyn}$ ratio  in median $\sim$1.4 times larger than HR-COSMOS with a significance of 3.9$\sigma$.

\item[$-$] The observed difference in $M_\ast/M_{dyn}$ at fixed stellar mass between the two samples is associated with a difference in their $M_{dyn}(r_{2.2})$, which is $\sim$30\% smaller in SC1604. To understand what causes smaller dynamical masses for galaxies in denser environment we investigated  the environmental dependence of the galaxy parameters that define  $M_{dyn}(r_{2.2})$: circular velocity and radius $r_{2.2}$. We found that in median the values of circular velocity are consistent between the two samples, while $r_{2.2}$ is in general 1.3 times larger for HR-COSMOS. This result is consistent with other works \citep[e.g.,][]{Maltby2010, CebrianTrujillo2014, Kuchner2017} that found that galaxies in high density environments are smaller than their counterparts in less dense environments. \citet{Kuchner2017} proposed ram pressure stripping in combination with a gradual gas starvation as the possible responsibles for the effect of the environments on the galaxy sizes. However, it has been also proposed \citep[e.g.,][]{Bekki1998, Querejeta2015} that major mergers can produce S0 galaxy remnants, which are characterized by more a concentrated bulge and a faded disk. This process would effectively make galaxies more compact.

\item[$-$] We took advantage of the measurements performed in this paper to constrain the stellar specific angular momentum $j_\ast$, which is considered a more fundamental quantity to investigate galaxy formation and evolution. We compared $j_\ast$ for SC1604 galaxies and HR-COSMOS in order to investigate the effect of the local environment on it. We found that both samples follow a $j_\ast - M_\ast$ relation with slope consistent with the relation at $z=0$ from \citet{RomanowskyFall2012}. However, the relation for SC1604 galaxies has a lower normalization, which implies a sample with a larger contribution from lower $j_\ast$ galaxies.

\item[$-$] By comparing measured $j_\ast$ values with those predicted by theory for cases under which galaxies do not lose their initial angular momentum, we were able to estimate the fraction $f_j$ of the original angular momentum that galaxies retained throughout their evolution. We found that SC1604 galaxies, in general, show $\sim20\%$ lower $j_\ast$ than those of HR-COSMOS. Galaxy-galaxy mergers are a process that can be responsible for this loss of angular momentum in galaxies and we attempted to investigate this scenario. Adopting literature estimates of the excess rate of galaxy-galaxy mergers in intermediate/high-density environments \citep{Tomczak2017} and the average amount of $j_\ast$ loss per merger event from the EAGLE simulations \citep{Lagos2018},  we showed that gas-poor mergers could account completely for the observed loss of $j_\ast$. However, because of the small number of SC1604 galaxies at intermediate/high densities (N=20), this result was not statistically significant.
A larger sample of galaxies in intermediate/high density environments, bolstered by measurements of gas fractions in these galaxies, are needed to confirm or deny this scenario. As such, we were unable to definitively confirm that mergers are the only process responsible for $j_\ast$ loss, which leaves room for other processes, such as ram pressure stripping or tidal interactions, to also contribute to the angular momentum loss of group and cluster galaxies throughout their evolutionary history.
\end{itemize}

This study highlighted the potential of the ORELSE survey in proving the possibility to investigates galaxy kinematics as a function of a wide range of environments. In this paper we focus our analysis on one of the 15 ORELSE fields; however, we plan to extend this study to more fields in order to collect a larger sample of galaxies, especially in the most dense environments.

\section*{Acknowledgements}
We thank Olga Cucciati for helping with the 3D rendering.
This material is based upon work supported by the National Science Foundation under Grant No. 1411943 and the National Aeronautics and Space Administration under NASA Grant Number NNX15AK92G.
A portion of this work made use of the Peloton computing cluster operated by the Division of Mathematical and Physical Sciences at the University of California, Davis. This study is based, in part, on data collected at the Subaru Telescope and obtained from the SMOKA, which is operated by the Astronomy Data Center, National Astronomical Observatory of Japan. This work is based, in part, on observations made with the Spitzer Space Telescope, which is operated by the Jet Propulsion Laboratory, California Institute of Technology under a contract with NASA. UKIRT is supported by NASA and operated under an agreement among the University of Hawaii, the University of Arizona, and Lockheed Martin Advanced Technology Center; operations are enabled through the cooperation of the East Asian Observatory. When the data reported here were acquired, UKIRT was operated by the Joint Astronomy Centre on behalf of the Science and Technology Facilities Council of the U.K. This study is also based, in part, on observations obtained with WIRCam, a joint project of CFHT, Taiwan, Korea, Canada, France, and the Canada-France-Hawaii Telescope which is operated by the National Research Council (NRC) of Canada, the Institut National des Sciences de l'Univers of the Centre National de la Recherche Scientifique of France, and the University of Hawai'i. The spectroscopic observations used in this work were obtained at the W.M. Keck Observatory, which is operated as a scientific partnership among the California Institute of Technology, the University of California, and the National Aeronautics and Space Administration. The Observatory was made possible by the generous financial support of the W.M. Keck Foundation. We wish to thank the indigenous Hawaiian
community for allowing us to be guests on their sacred mountain, a privilege, without which, this work would not have been possible. We are most fortunate to be able to conduct observations from this site.




\bibliographystyle{mnras}
\bibliography{ORELSE_smTF} 



\appendix

\section{Additional Figures} \label{app:figures}

\begin{figure*}
	\includegraphics[width=\textwidth]{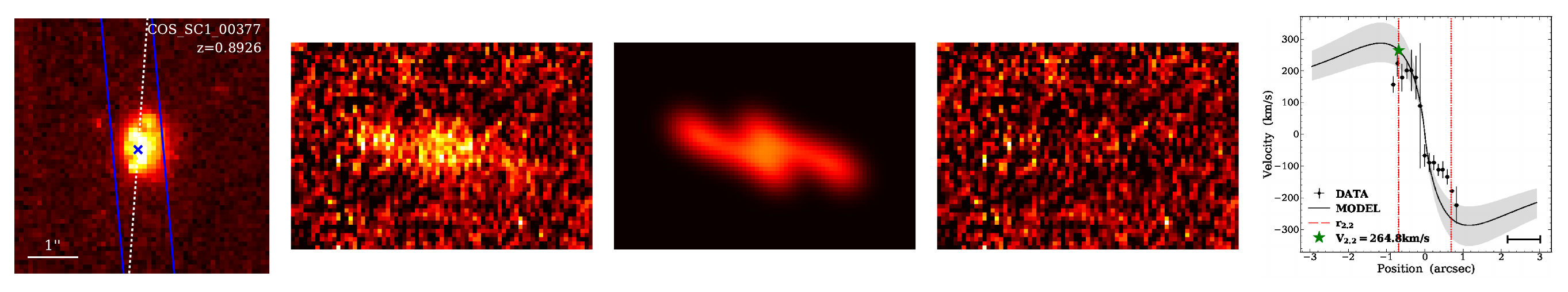}
\medskip
\includegraphics[width=\textwidth]{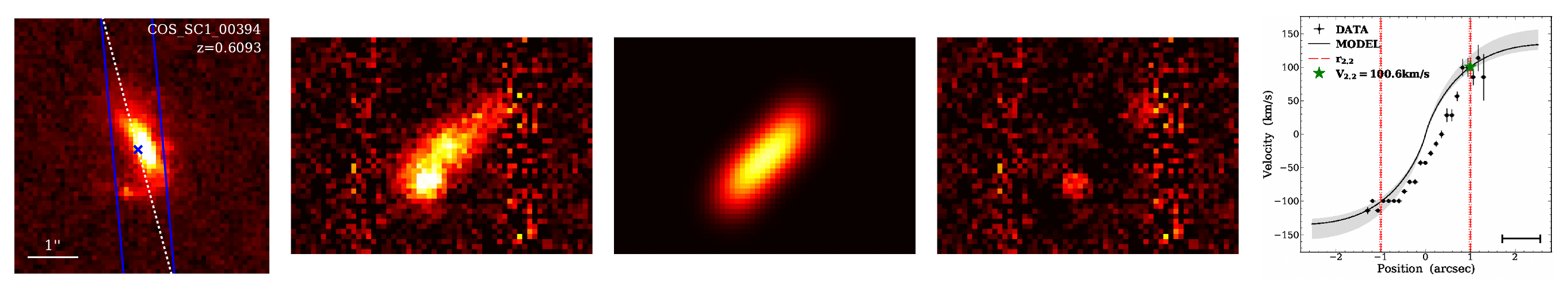}
\medskip
\includegraphics[width=\textwidth]{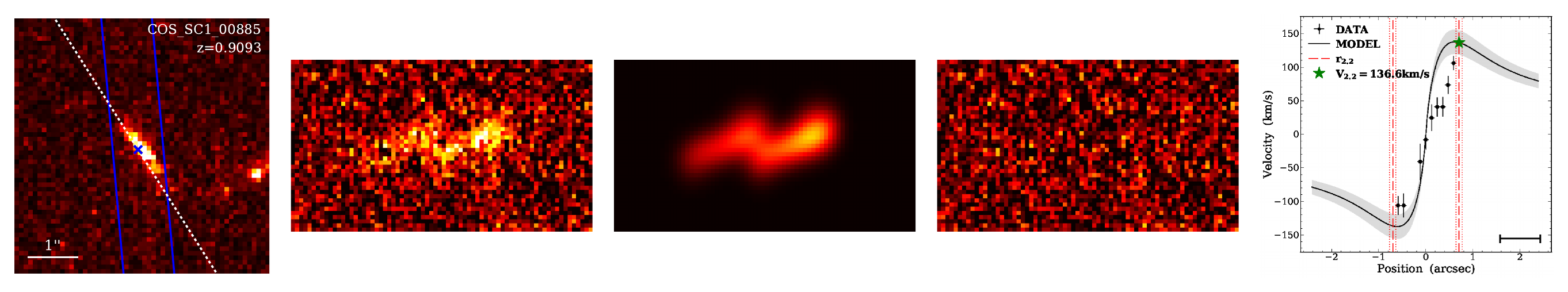}
\medskip
\includegraphics[width=\textwidth]{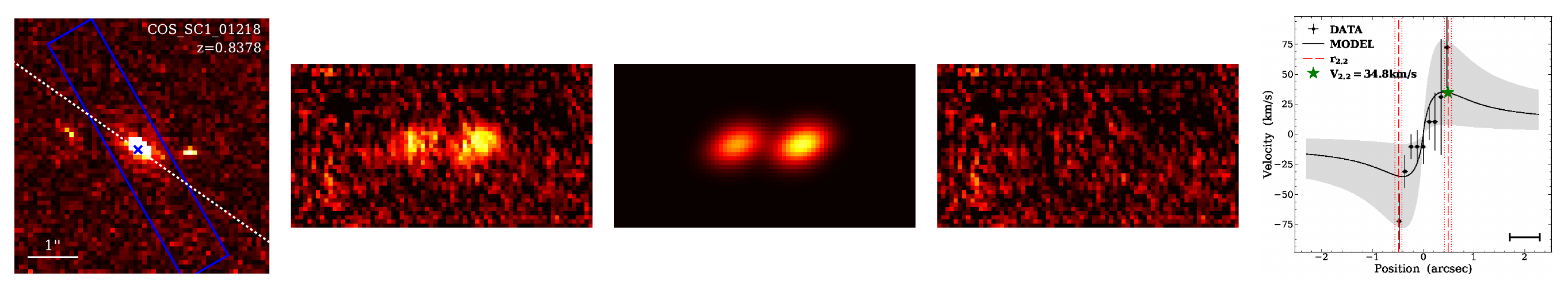}
\medskip
\includegraphics[width=\textwidth]{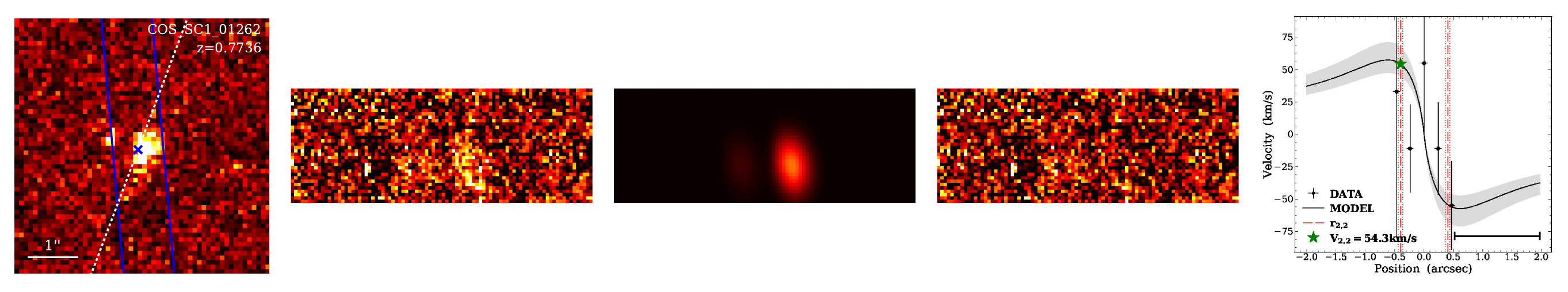}
\medskip
\includegraphics[width=\textwidth]{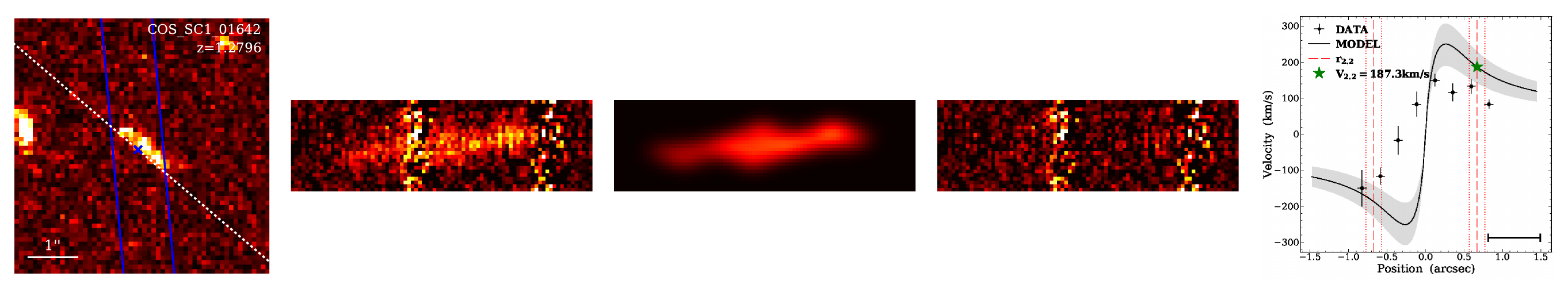}
    \caption{Kinematic modeling for the entire ``kinematically reliable'' sample (Sec.~\ref{subsec:smTFR}) \textit{Each row:} From left to right: \textit{HST}/ACS F814W postage stamp (5\arcsec$\times$5\arcsec) with superimposed in blue the DEIMOS slit and in white the orientation of the galaxy PA;  continuum-subtracted 2D spectrum centered at the emission line; best-fit kinematic model; residual image between the 2D spectrum and the best-fit model on the same intensity scale as the 2D spectrum; high-resolution rotation curve model (black line), corrected for the inclination, with 1$\sigma$ uncertainty (shaded area), compared to the observed rotation curve (black points). The red dashed and dotted lines indicate the radius $r_{2.2}$ (see Sec.~\ref{subsec:smTFR}) and its uncertainty, respectively. The horizontal black bar on the bottom right corner represents the DEIMOS spatial PSF. }
    \label{fig:appendix_kinemodel}
\end{figure*}

\begin{figure*}
\includegraphics[width=\textwidth]{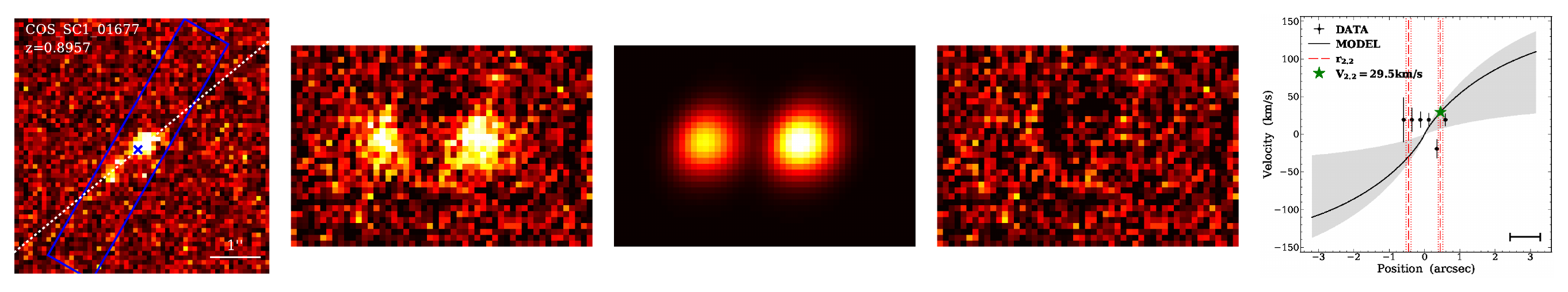}
\includegraphics[width=\textwidth]{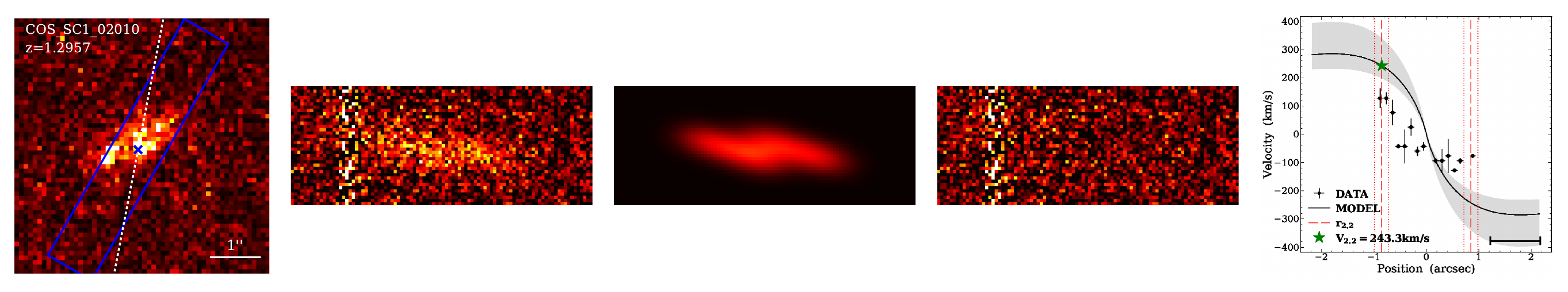}
\includegraphics[width=\textwidth]{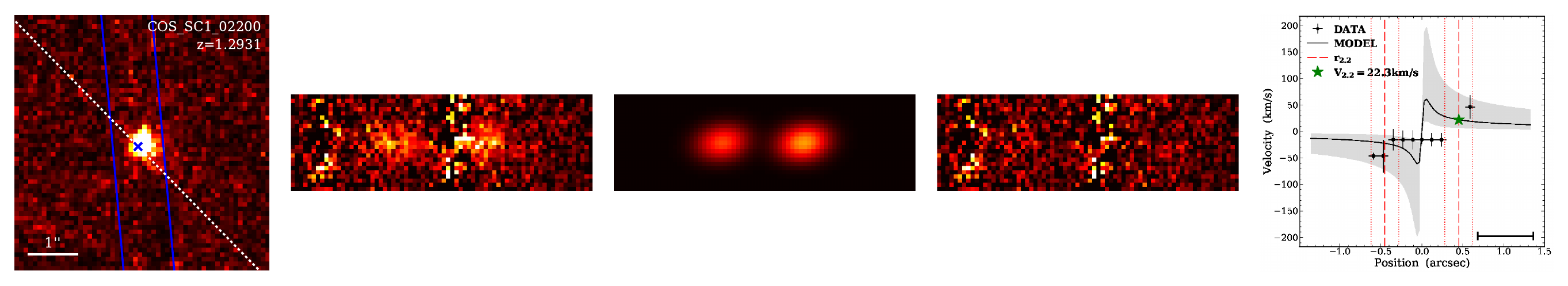}
\includegraphics[width=\textwidth]{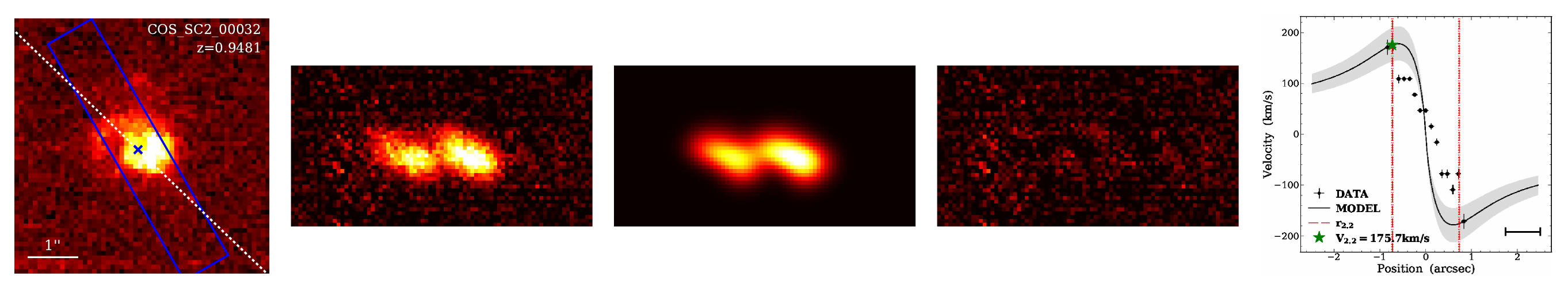}
\includegraphics[width=\textwidth]{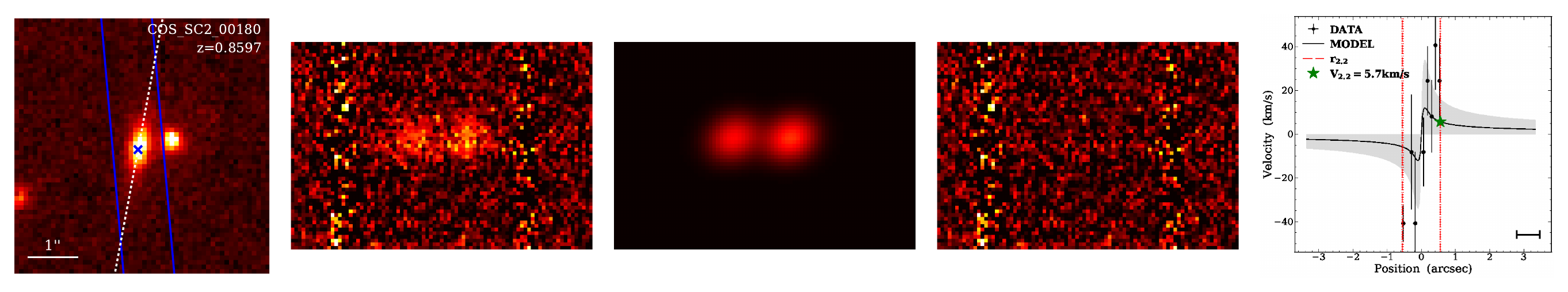}
\includegraphics[width=\textwidth]{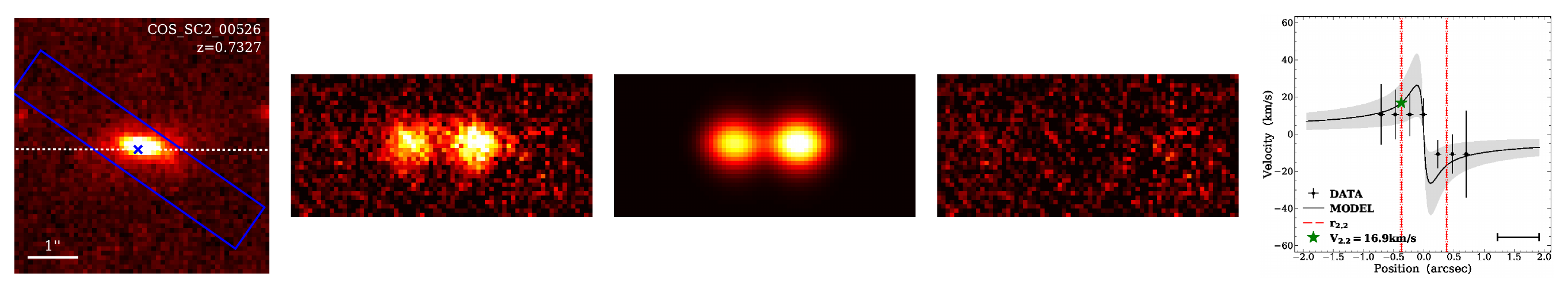}
\includegraphics[width=\textwidth]{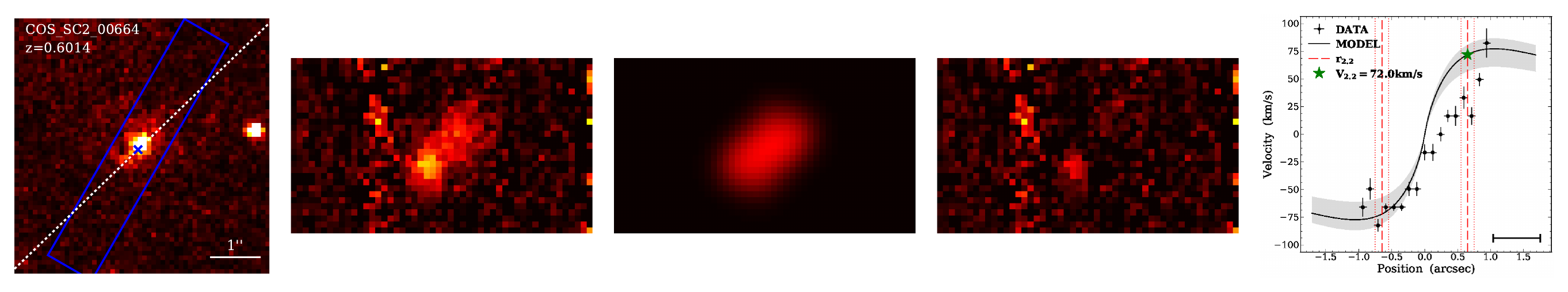}
\contcaption{}
\label{fig:appendix_kinemodel}
\end{figure*}

\begin{figure*}
\includegraphics[width=\textwidth]{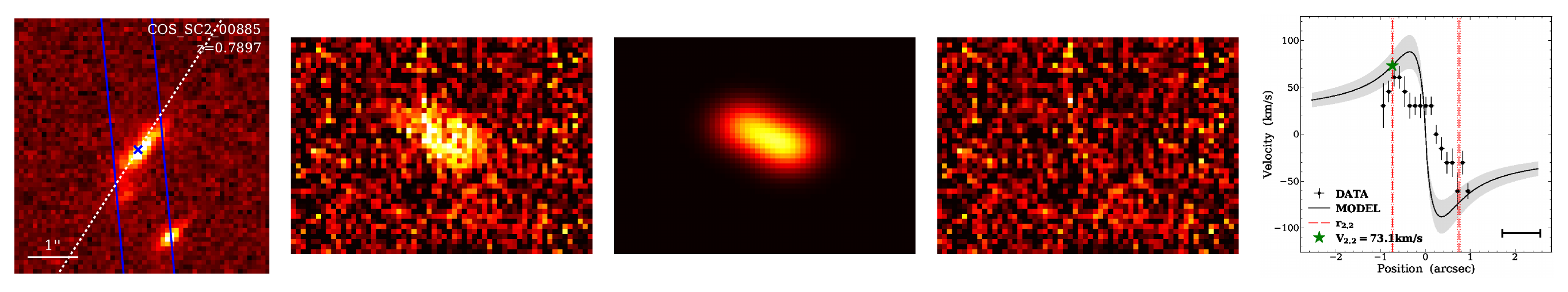}
\includegraphics[width=\textwidth]{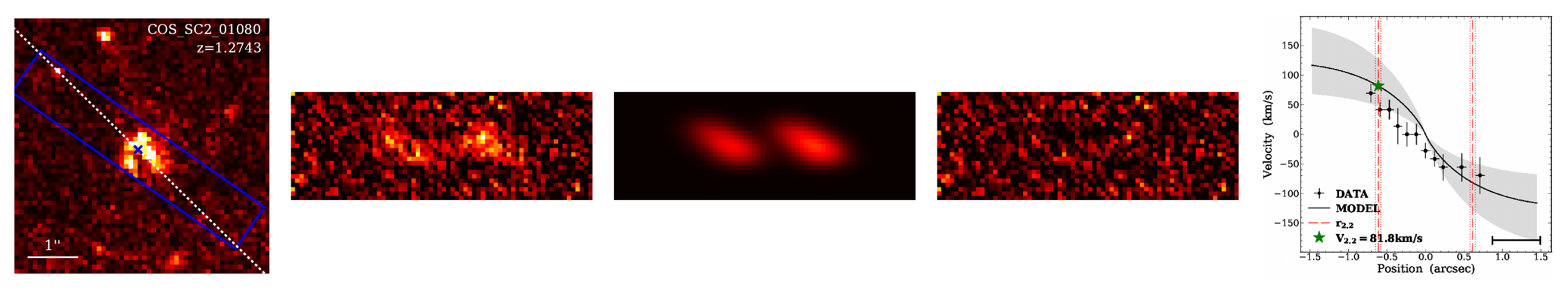}
\includegraphics[width=\textwidth]{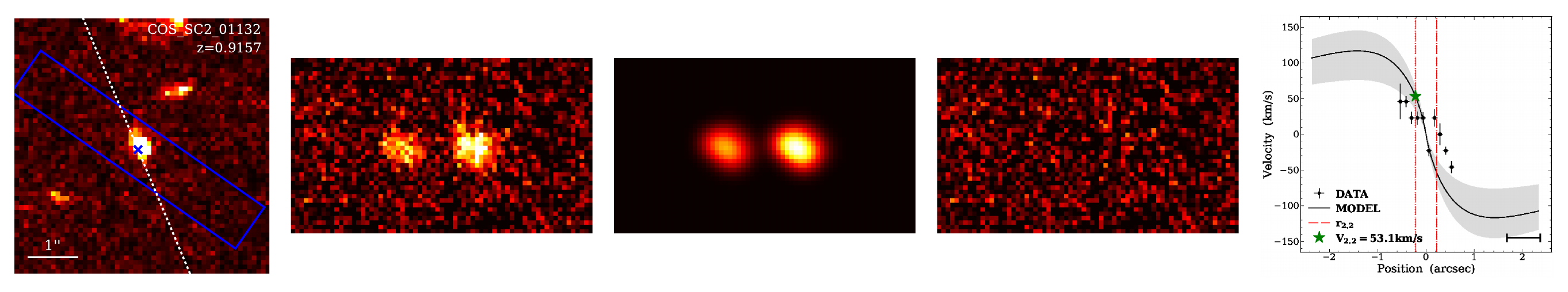}
\includegraphics[width=\textwidth]{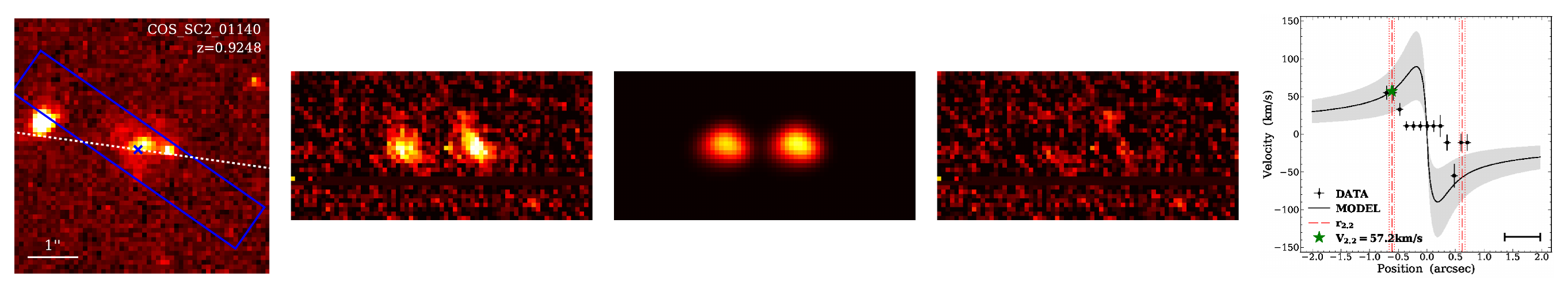}
\includegraphics[width=\textwidth]{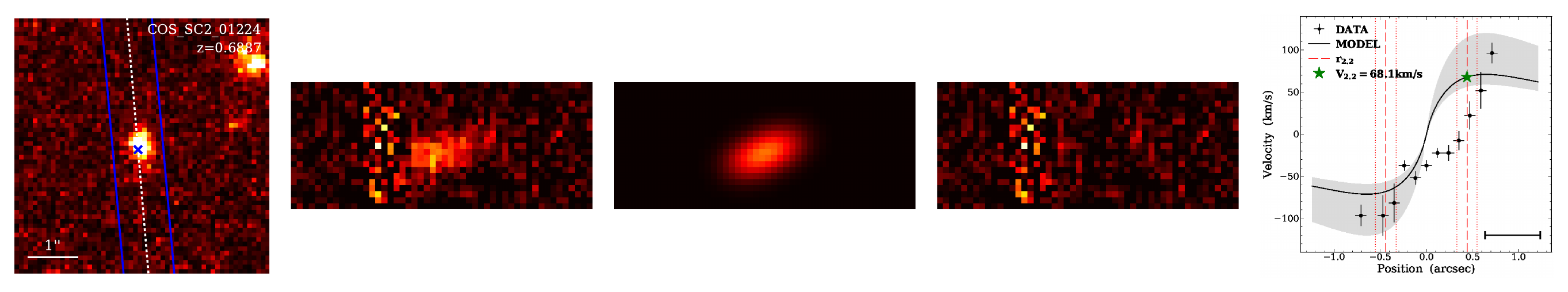}
\includegraphics[width=\textwidth]{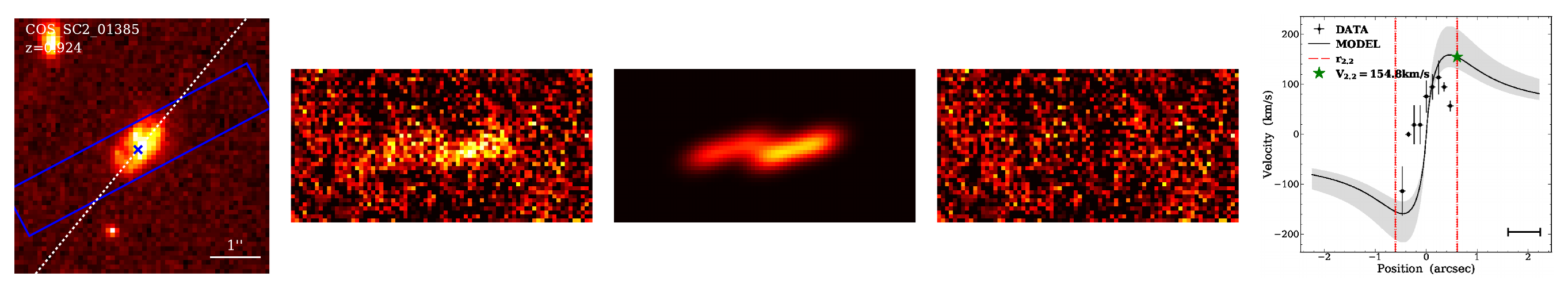}
\includegraphics[width=\textwidth]{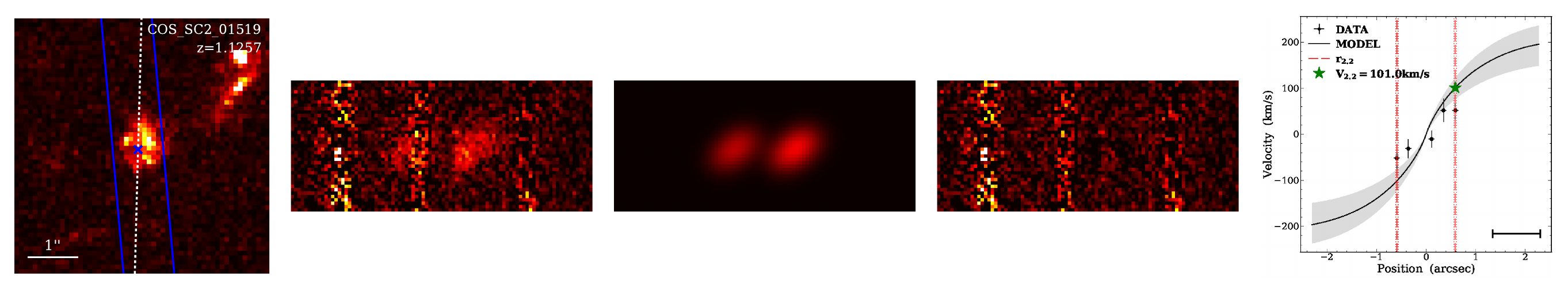}
\contcaption{}
\label{fig:appendix_kinemodel}
\end{figure*}

\begin{figure*}
\includegraphics[width=\textwidth]{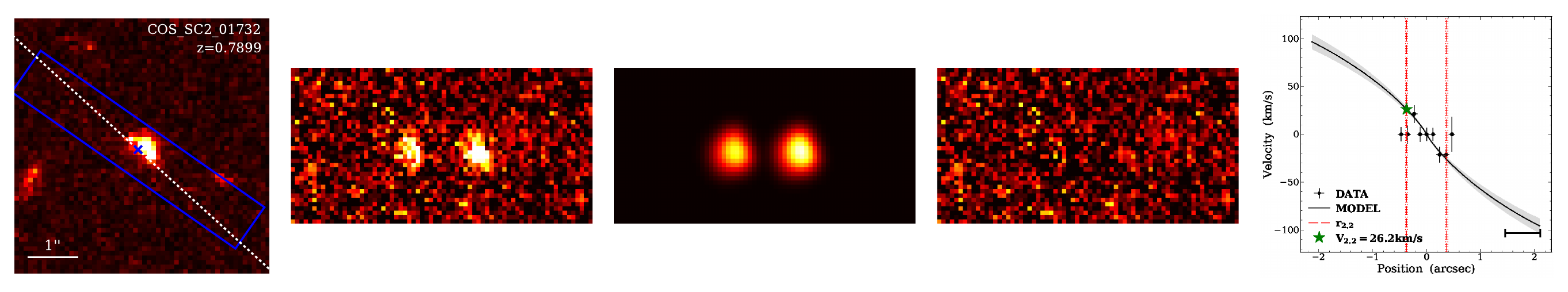}
\includegraphics[width=\textwidth]{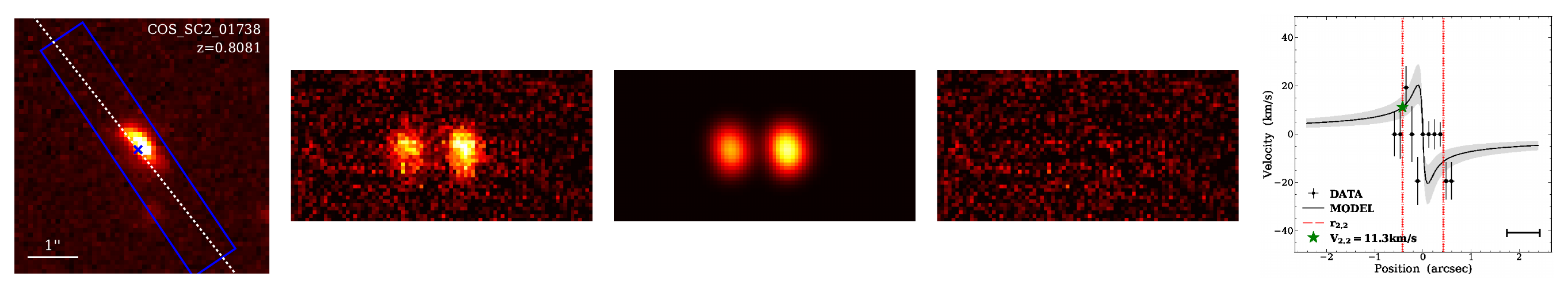}
\includegraphics[width=\textwidth]{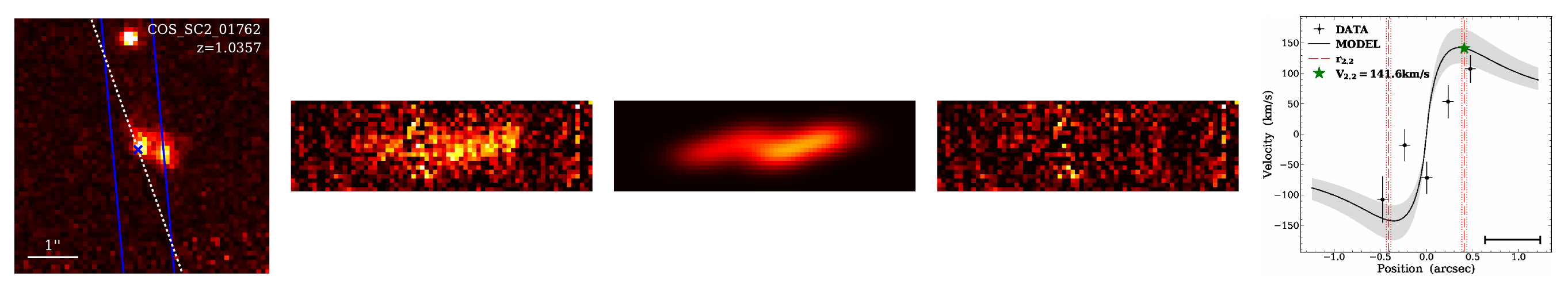}
\includegraphics[width=\textwidth]{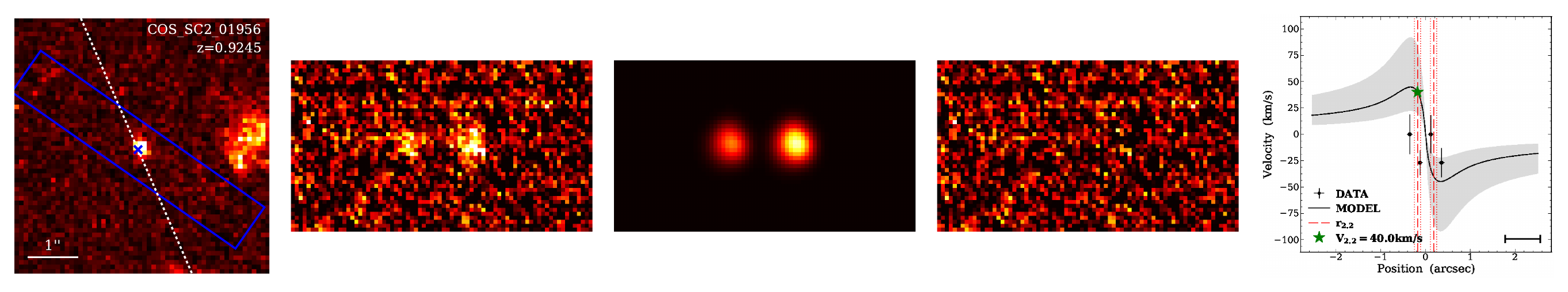}
\includegraphics[width=\textwidth]{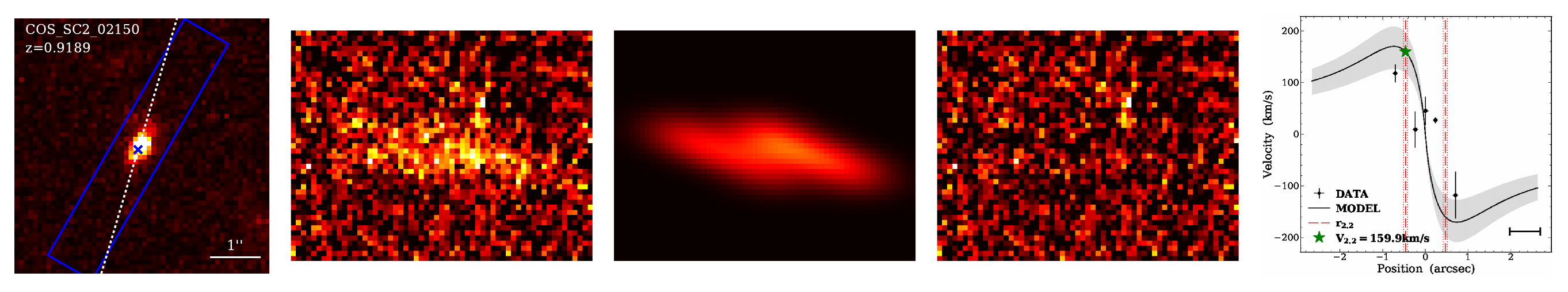}
\includegraphics[width=\textwidth]{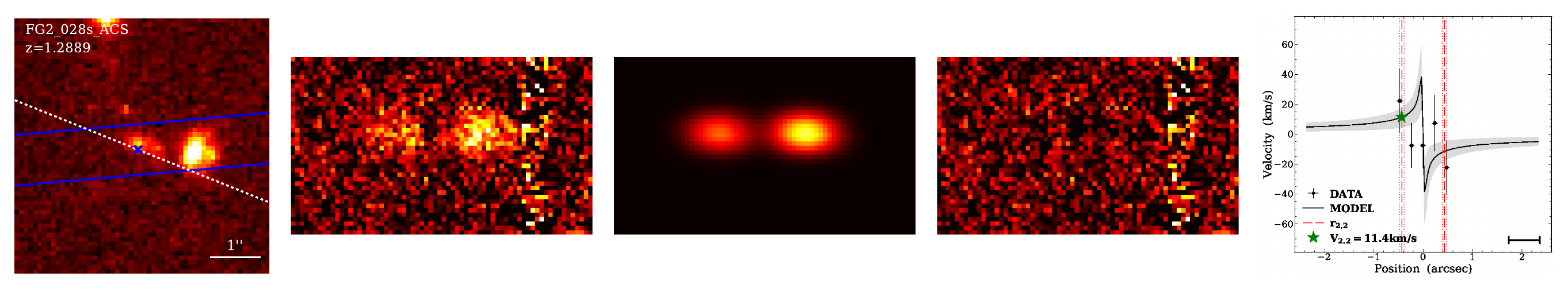}
\includegraphics[width=\textwidth]{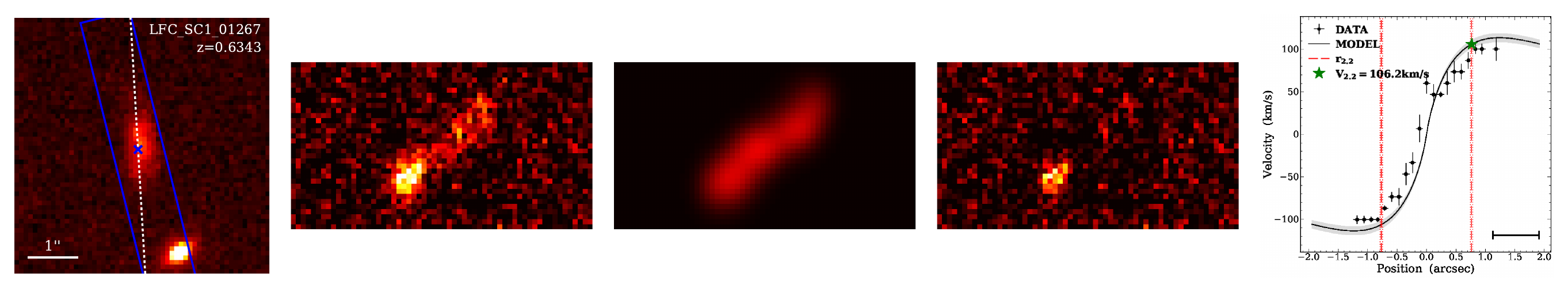}
\contcaption{}
\label{fig:appendix_kinemodel}
\end{figure*}

\begin{figure*}
\includegraphics[width=\textwidth]{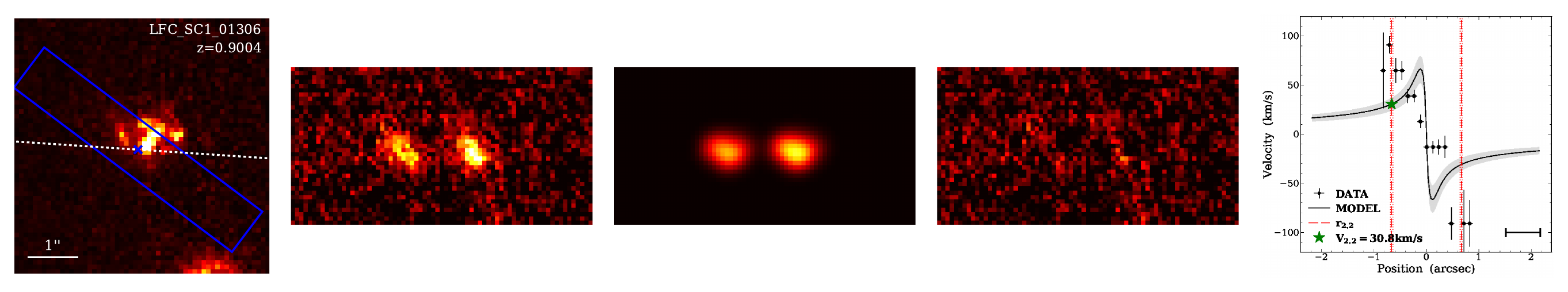}
\includegraphics[width=\textwidth]{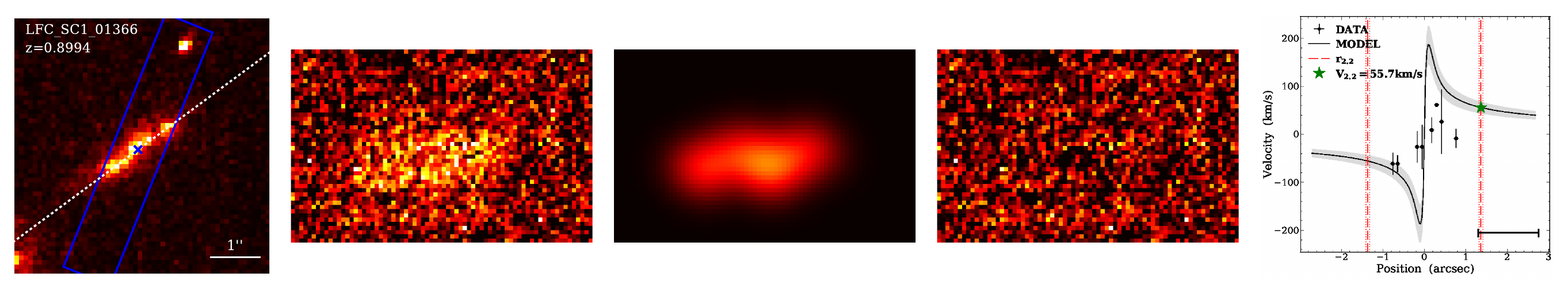}
\includegraphics[width=\textwidth]{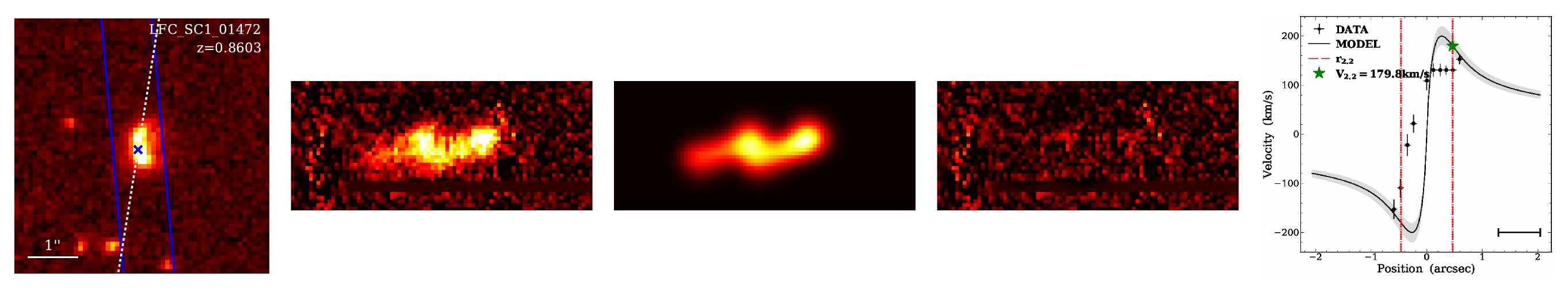}
\includegraphics[width=\textwidth]{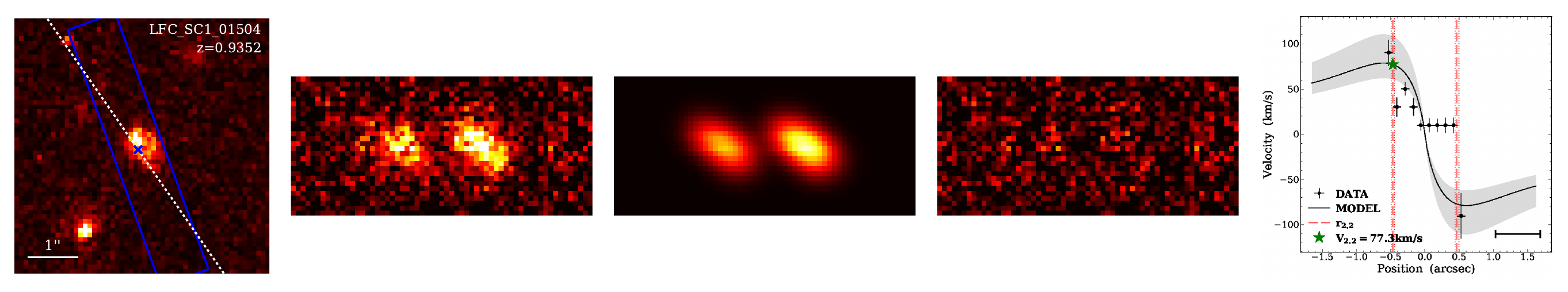}
\includegraphics[width=\textwidth]{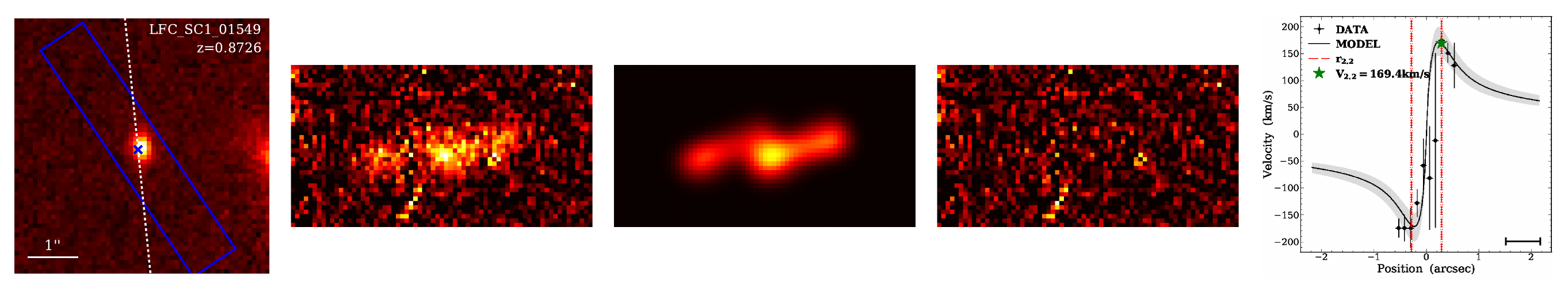}
\includegraphics[width=\textwidth]{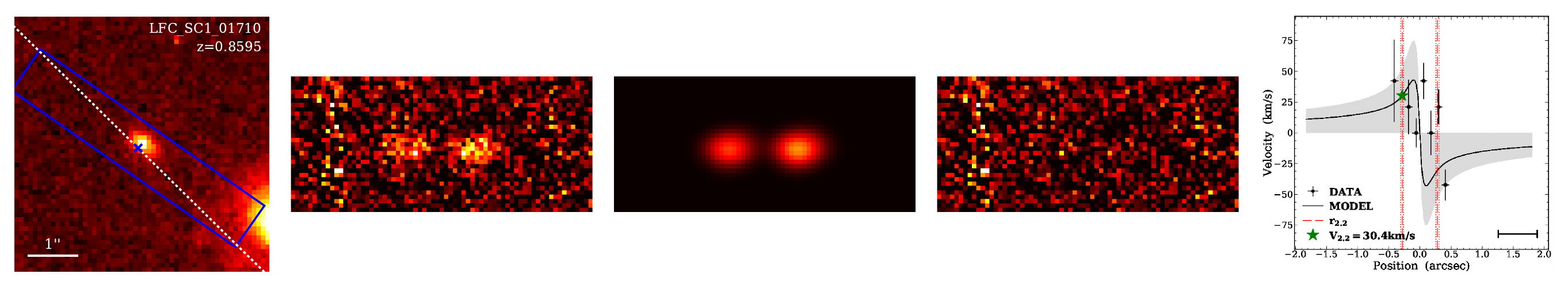}
\includegraphics[width=\textwidth]{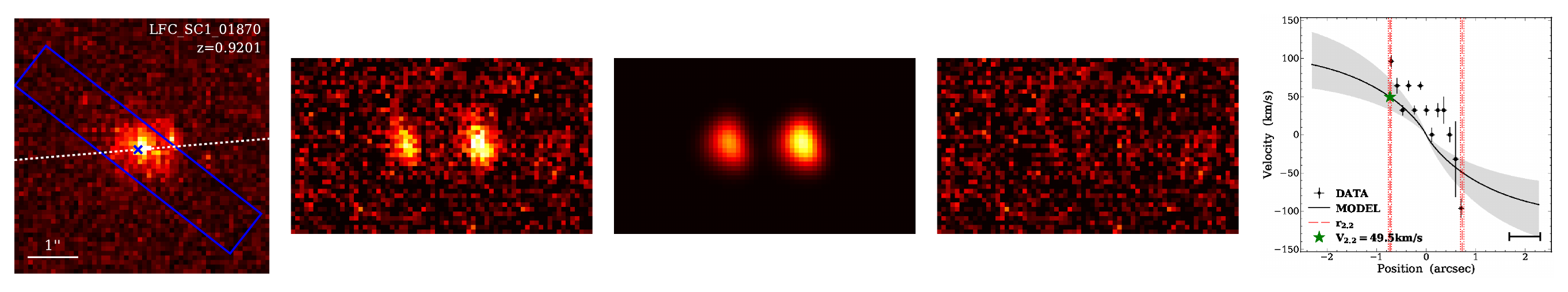}
\contcaption{}
\label{fig:appendix_kinemodel}
\end{figure*}

\begin{figure*}
\includegraphics[width=\textwidth]{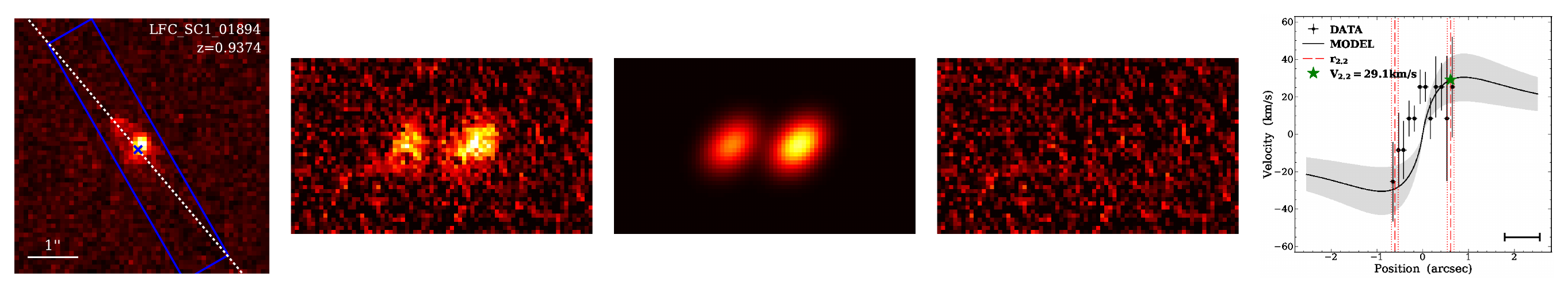}
\includegraphics[width=\textwidth]{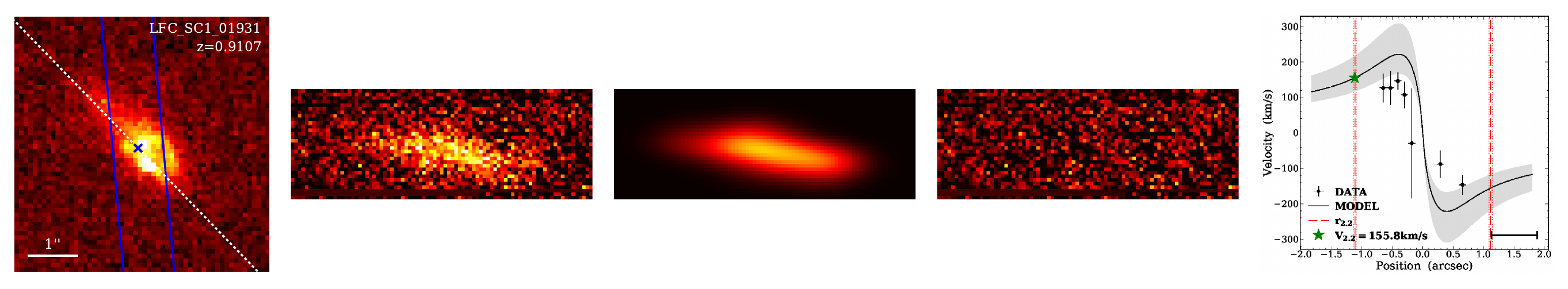}
\includegraphics[width=\textwidth]{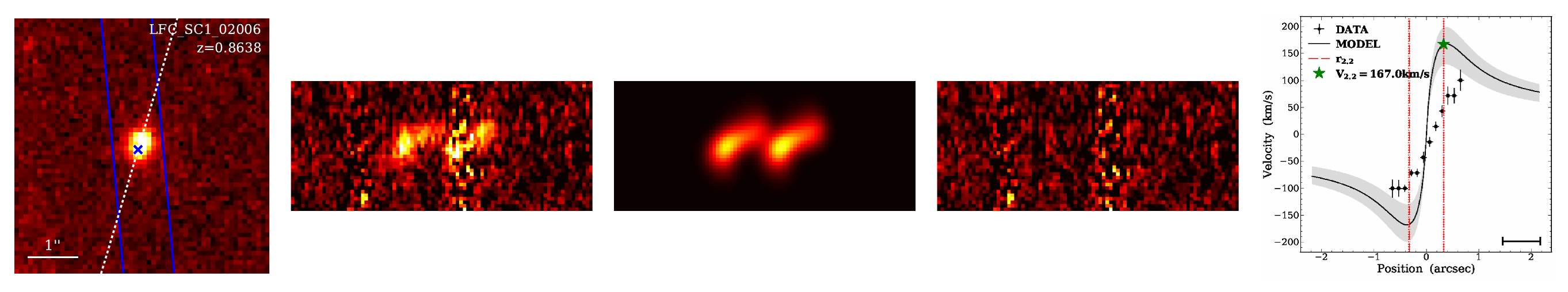}
\includegraphics[width=\textwidth]{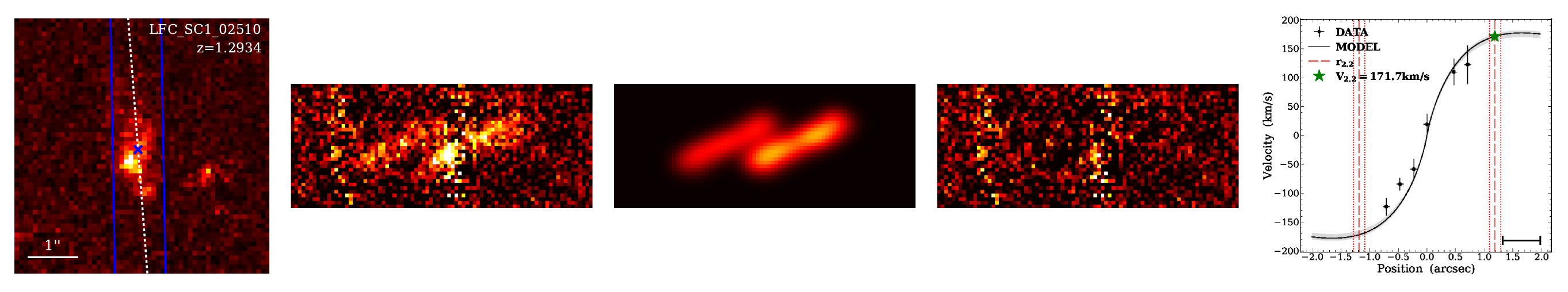}
\includegraphics[width=\textwidth]{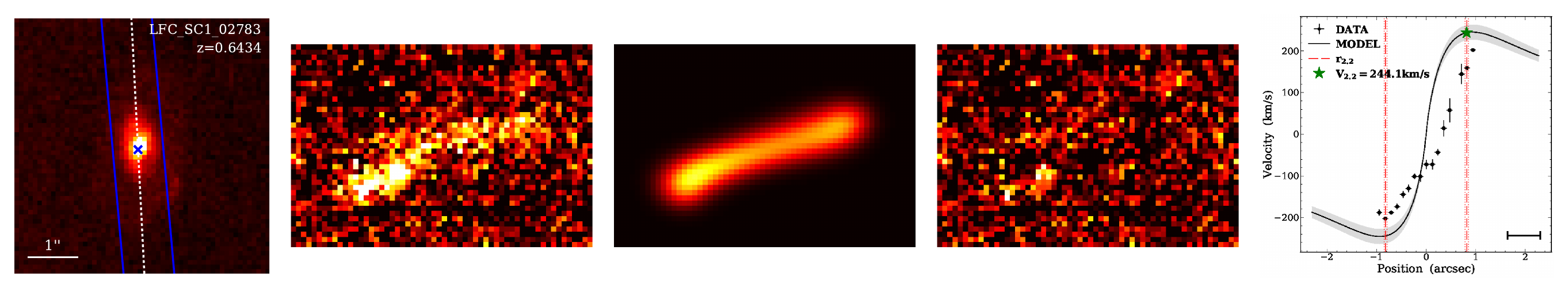}
\includegraphics[width=\textwidth]{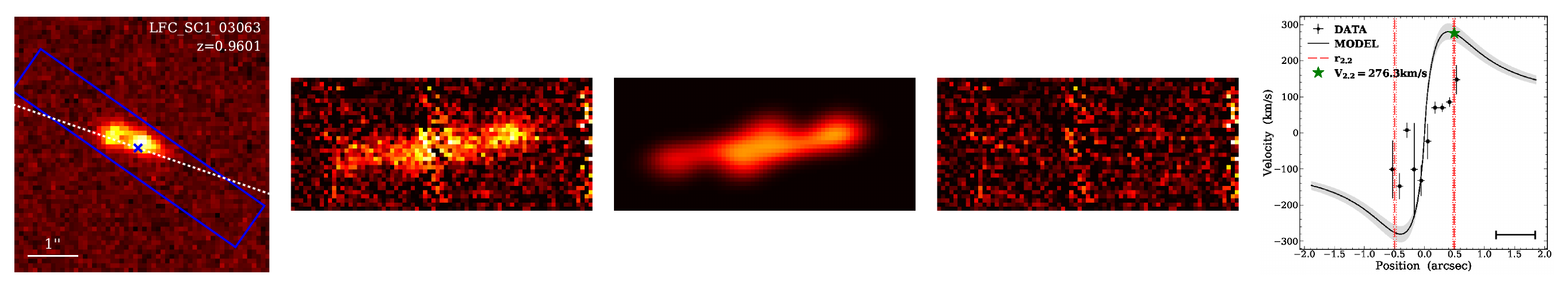}
\includegraphics[width=\textwidth]{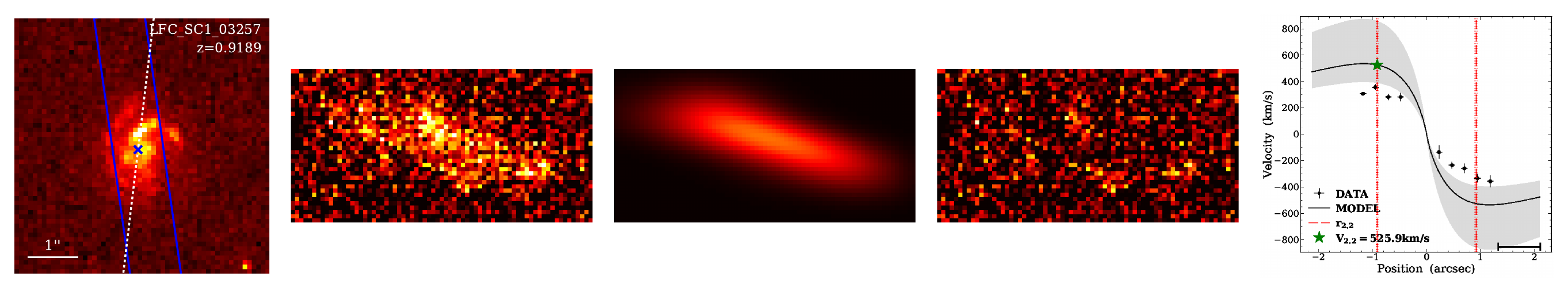}
\contcaption{}
\label{fig:appendix_kinemodel}
\end{figure*}

\begin{figure*}
\includegraphics[width=\textwidth]{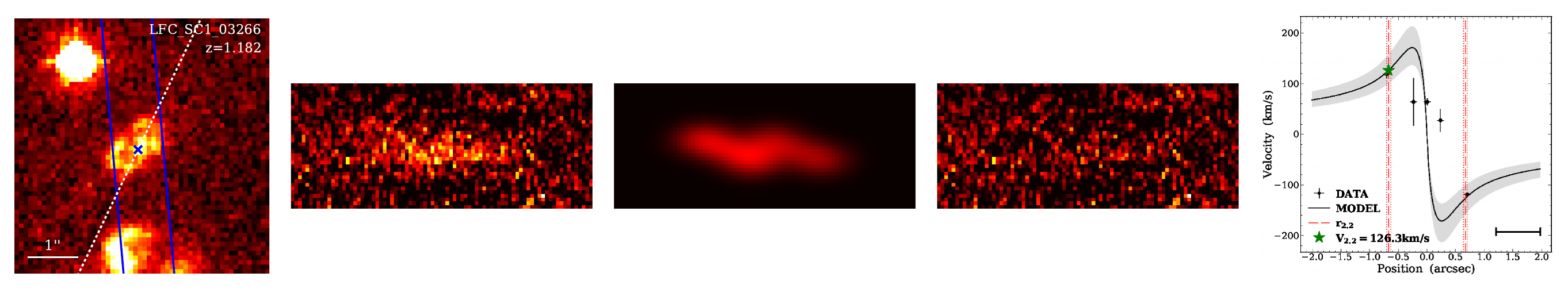}
\includegraphics[width=\textwidth]{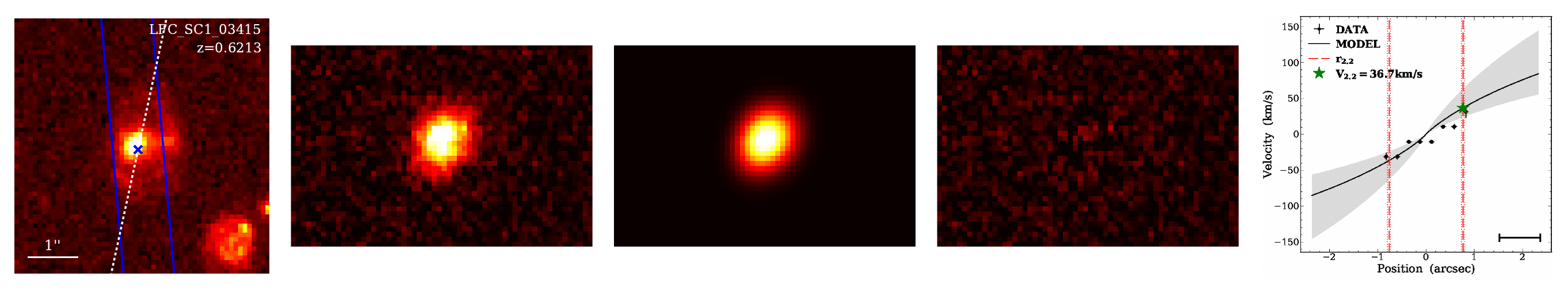}
\includegraphics[width=\textwidth]{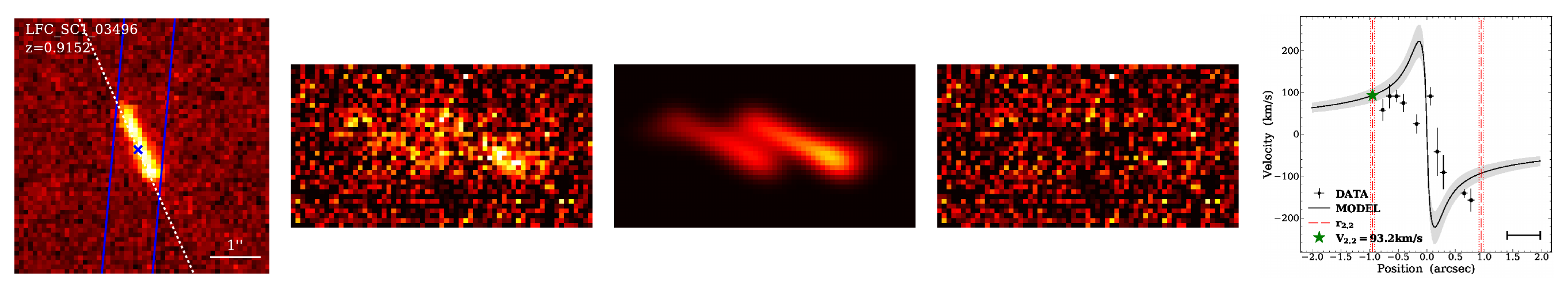}
\includegraphics[width=\textwidth]{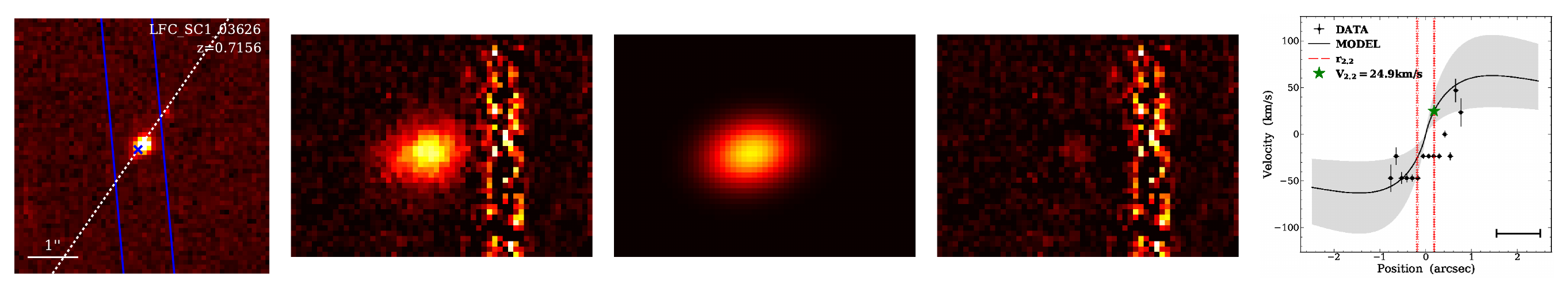}
\includegraphics[width=\textwidth]{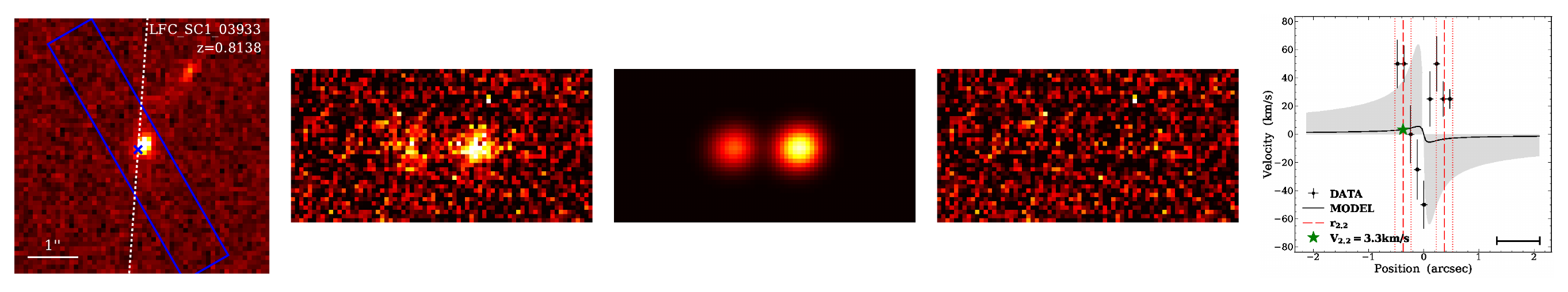}
\includegraphics[width=\textwidth]{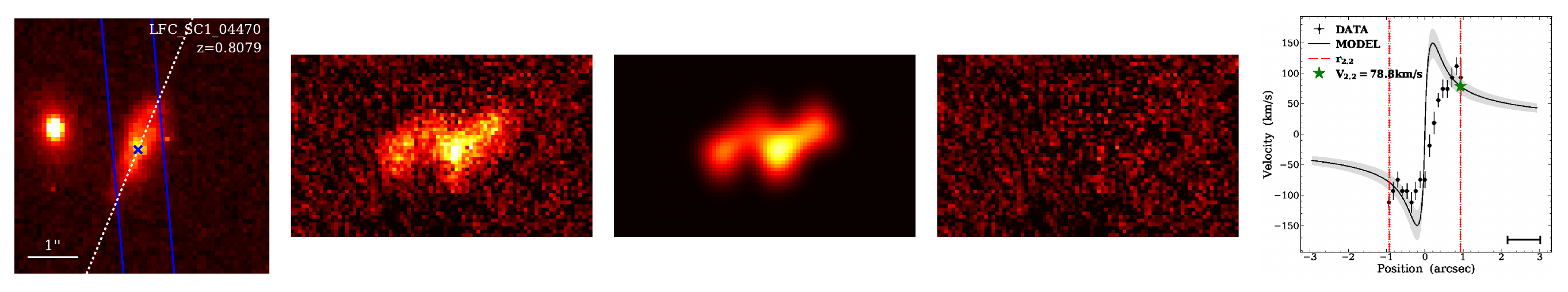}
\includegraphics[width=\textwidth]{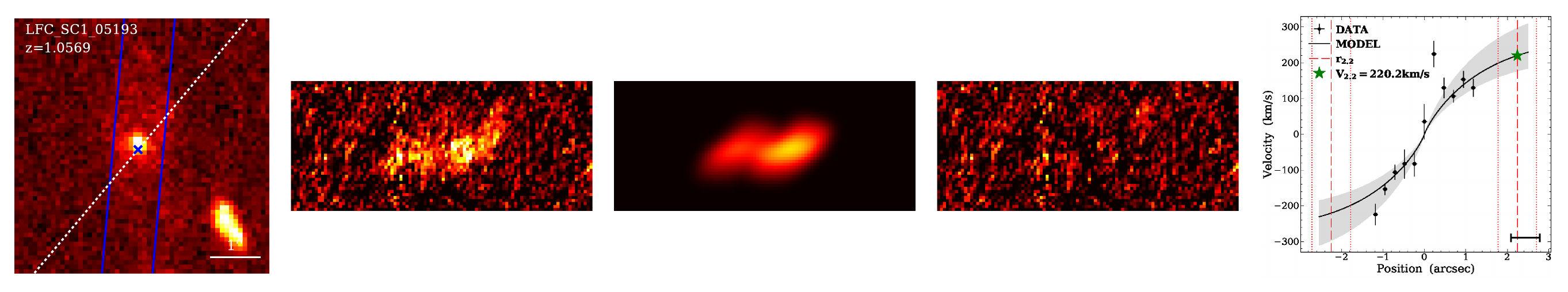}
\contcaption{}
\label{fig:appendix_kinemodel}
\end{figure*}

\begin{figure*}
\includegraphics[width=\textwidth]{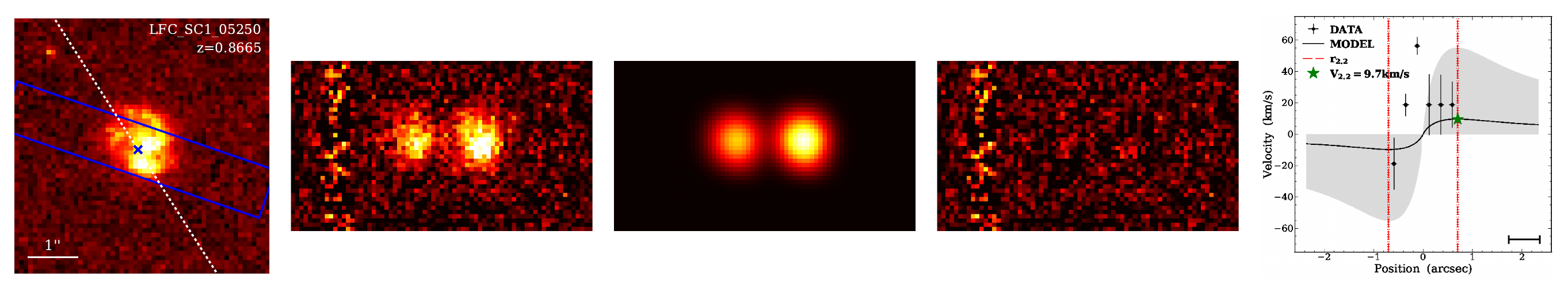}
\includegraphics[width=\textwidth]{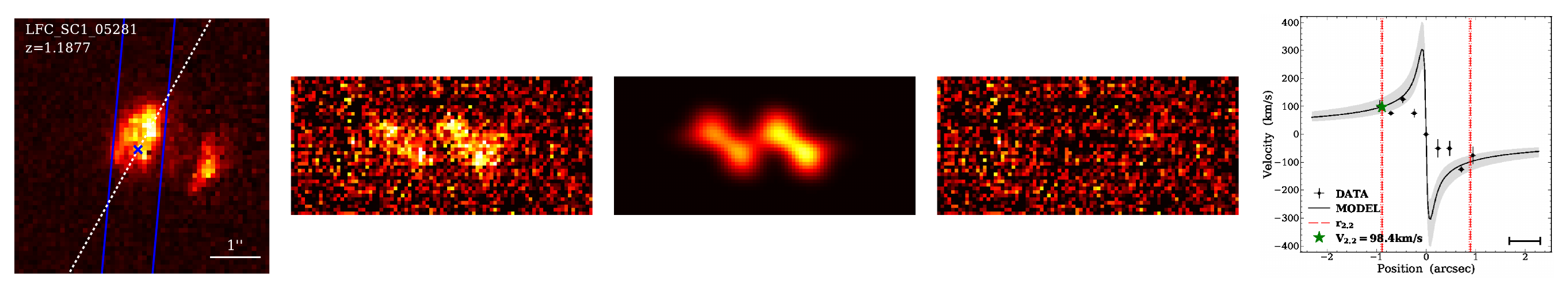}
\includegraphics[width=\textwidth]{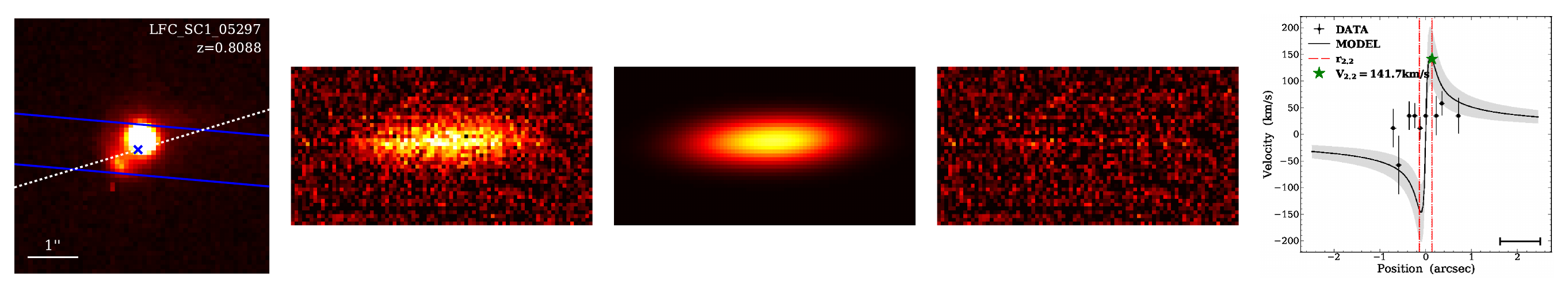}
\includegraphics[width=\textwidth]{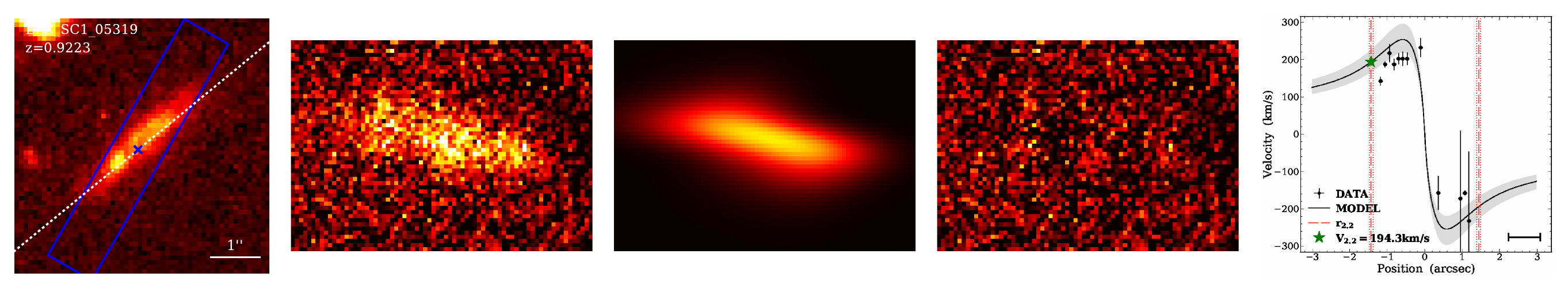}
\includegraphics[width=\textwidth]{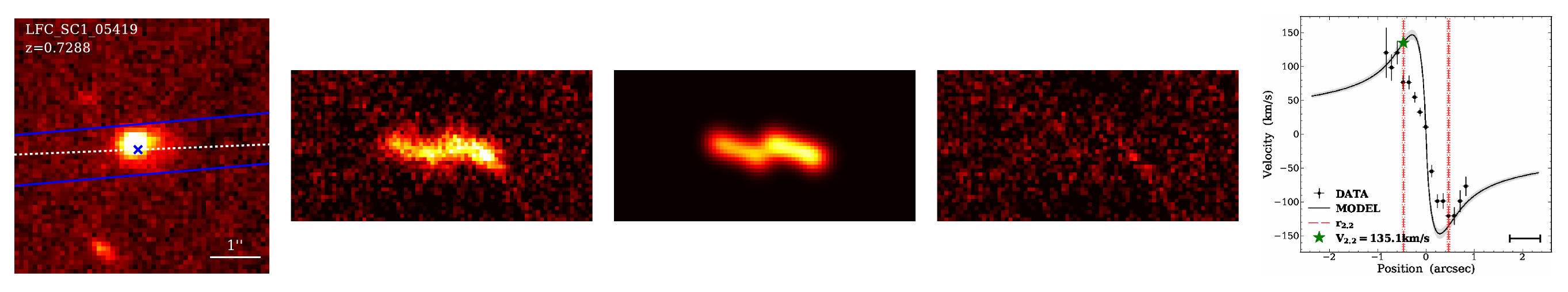}
\includegraphics[width=\textwidth]{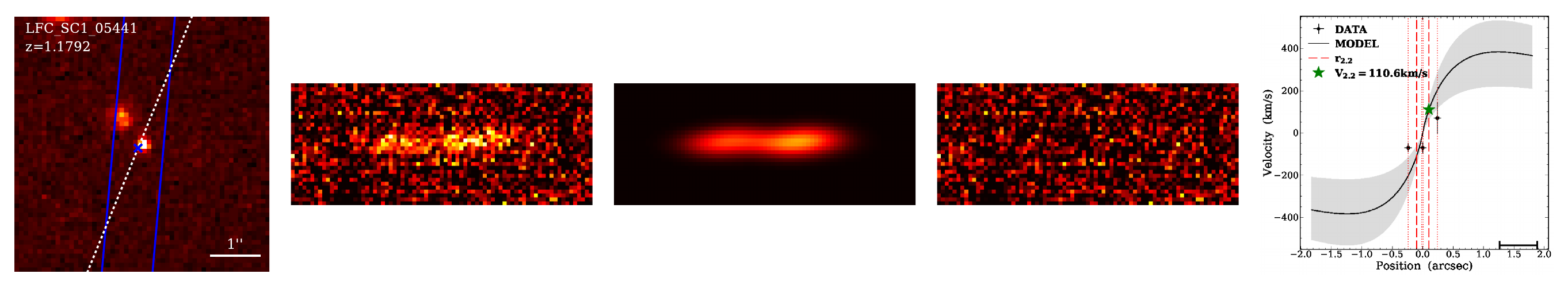}
\includegraphics[width=\textwidth]{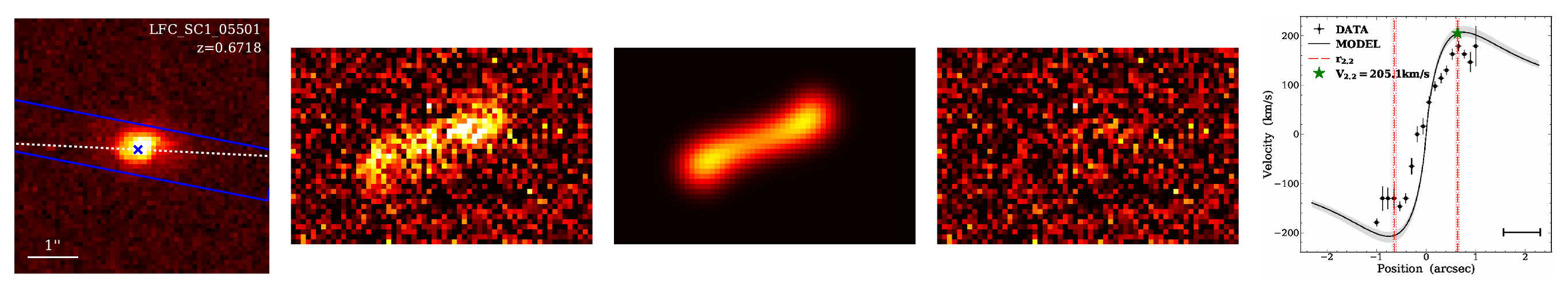}
\contcaption{}
\label{fig:appendix_kinemodel}
\end{figure*}

\begin{figure*}
\includegraphics[width=\textwidth]{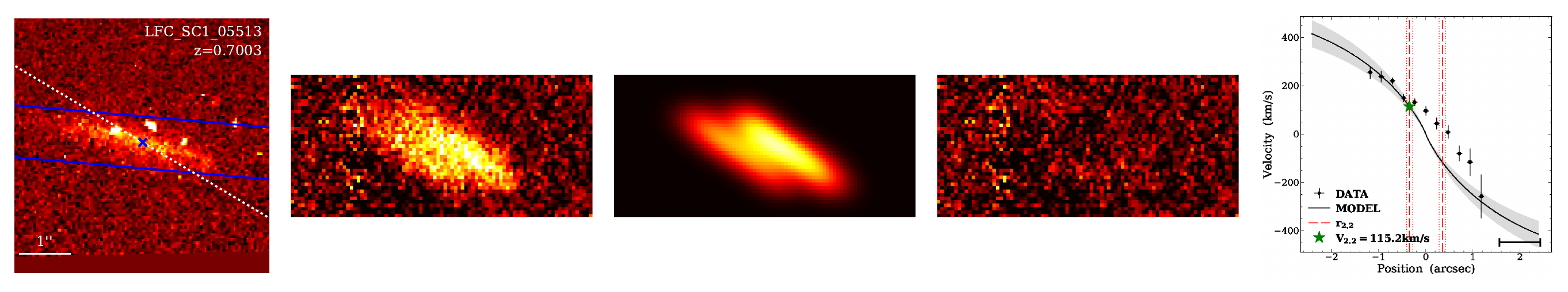}
\includegraphics[width=\textwidth]{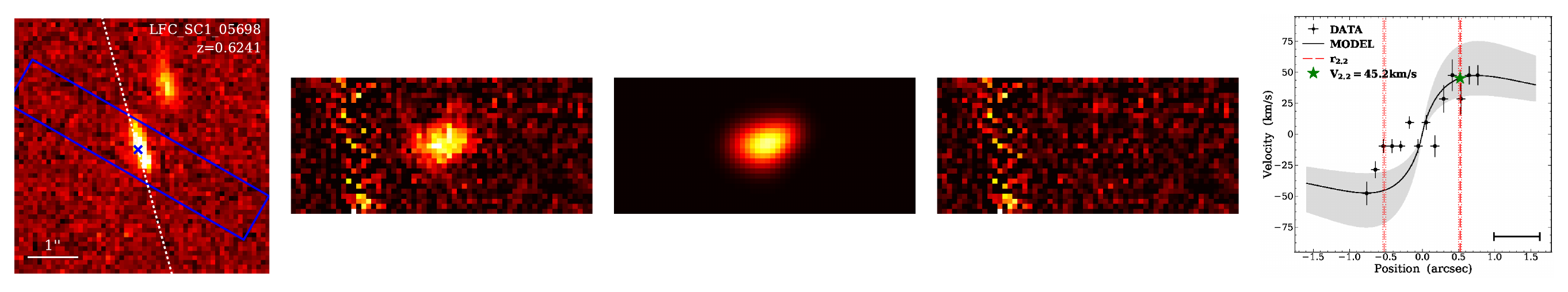}
\includegraphics[width=\textwidth]{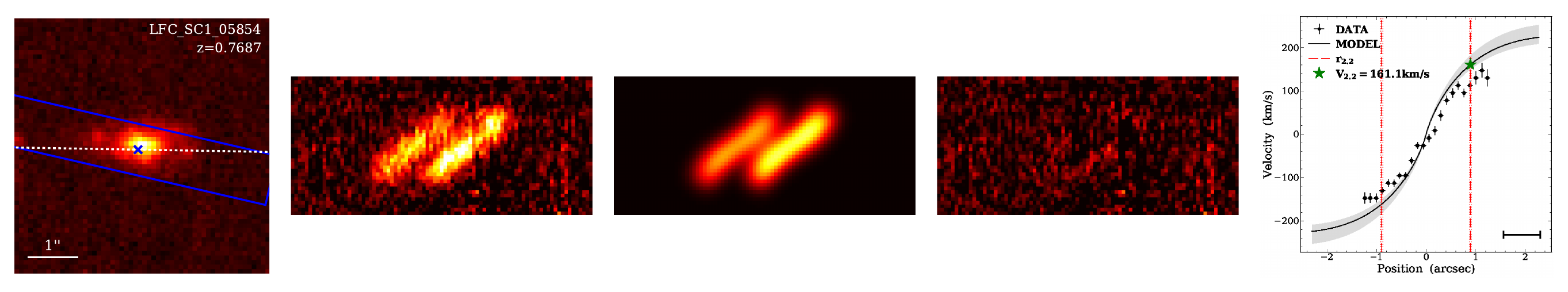}
\includegraphics[width=\textwidth]{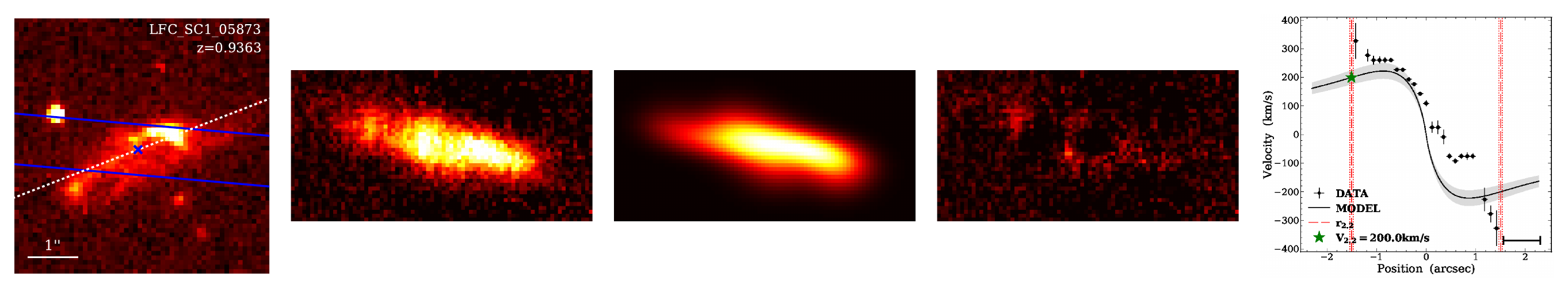}
\includegraphics[width=\textwidth]{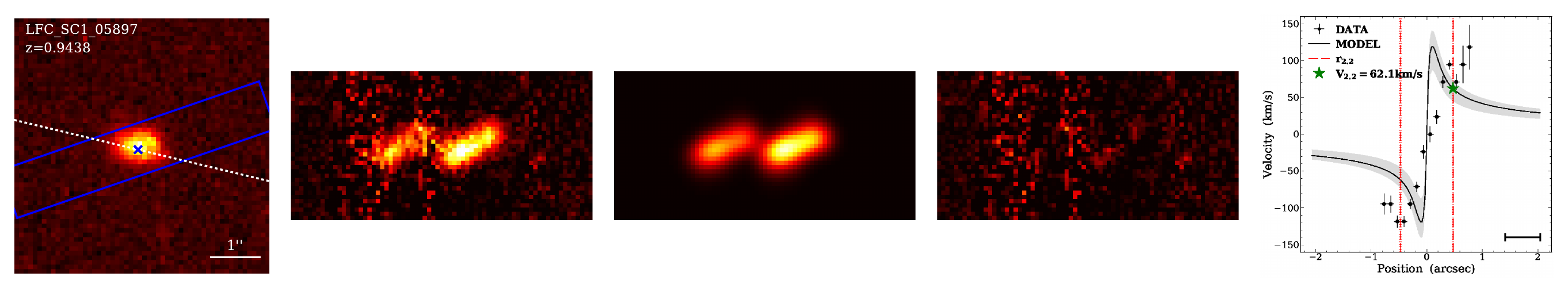}
\includegraphics[width=\textwidth]{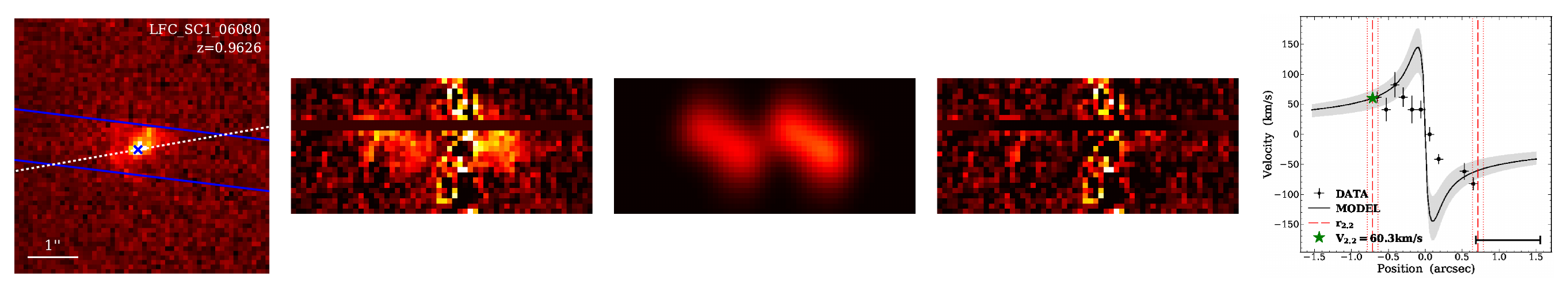}
\includegraphics[width=\textwidth]{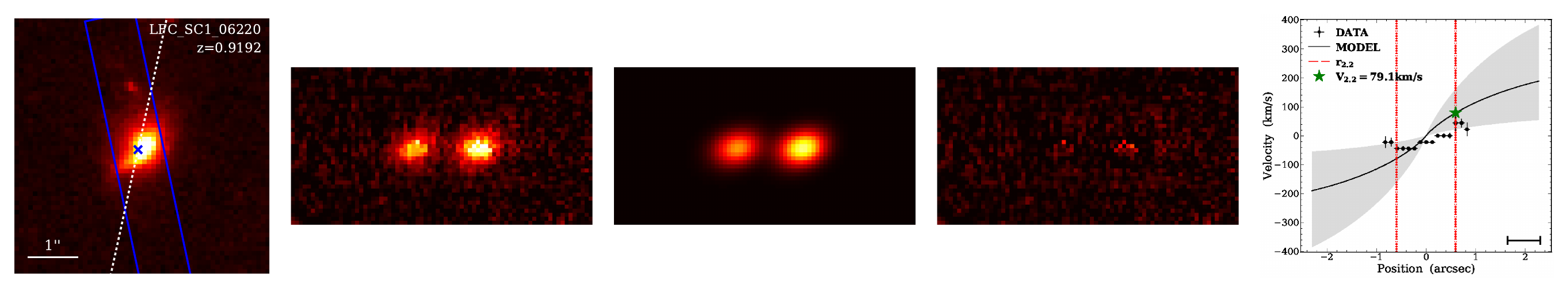}
\contcaption{}
\label{fig:appendix_kinemodel}
\end{figure*}

\begin{figure*}
\includegraphics[width=\textwidth]{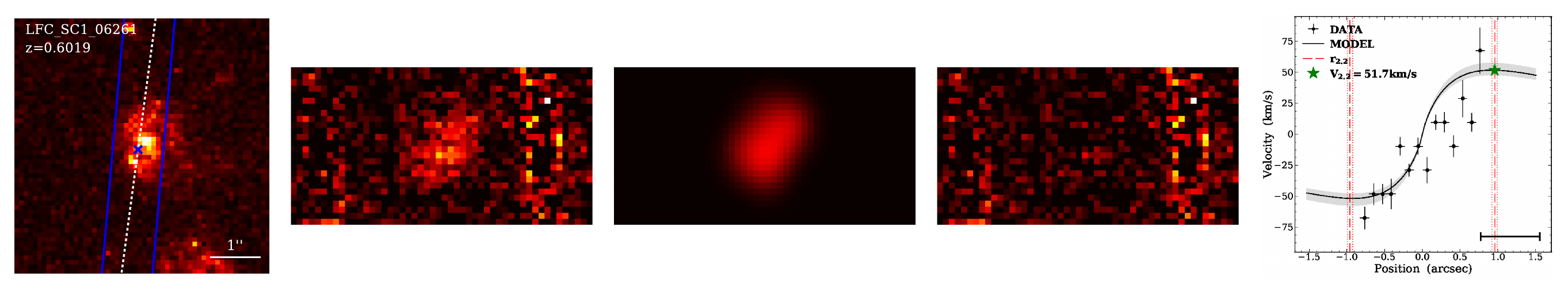}
\includegraphics[width=\textwidth]{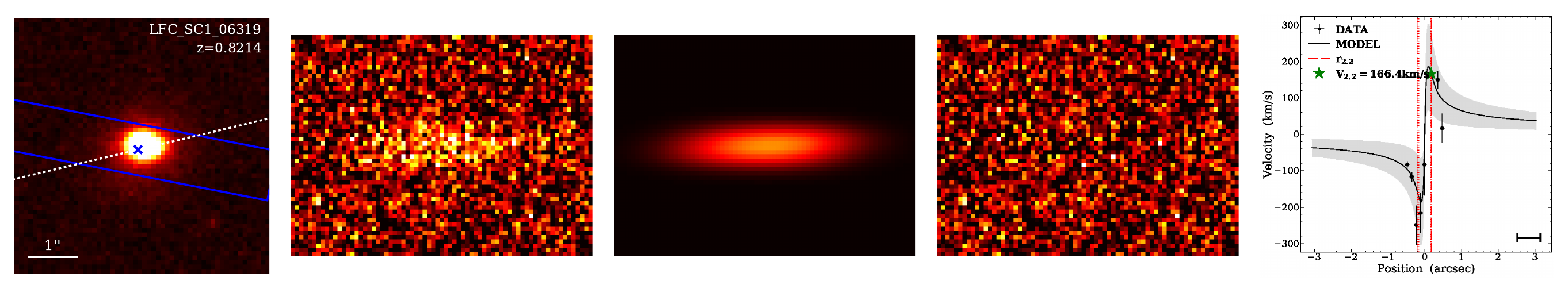}
\includegraphics[width=\textwidth]{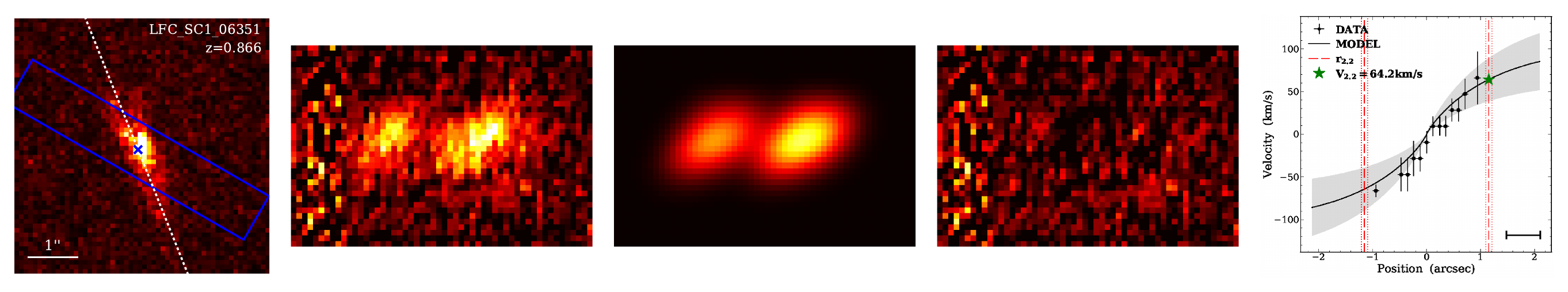}
\includegraphics[width=\textwidth]{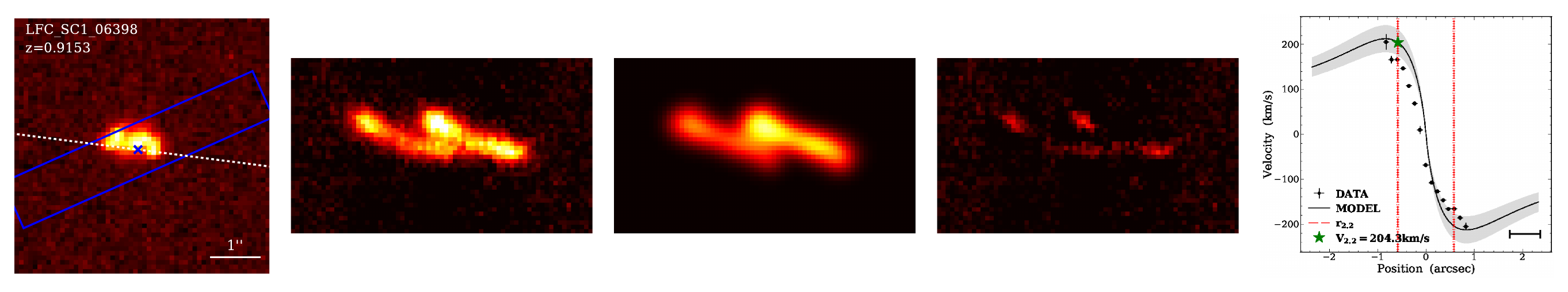}
\includegraphics[width=\textwidth]{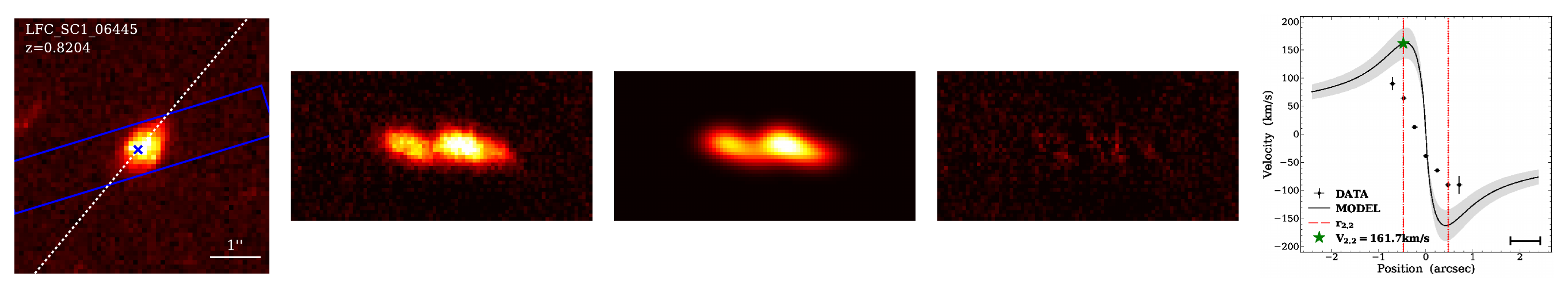}
\includegraphics[width=\textwidth]{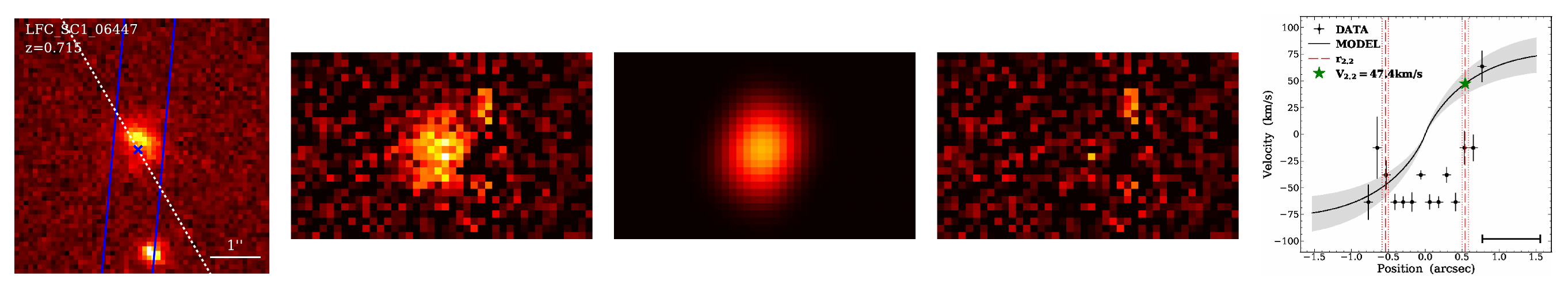}
\includegraphics[width=\textwidth]{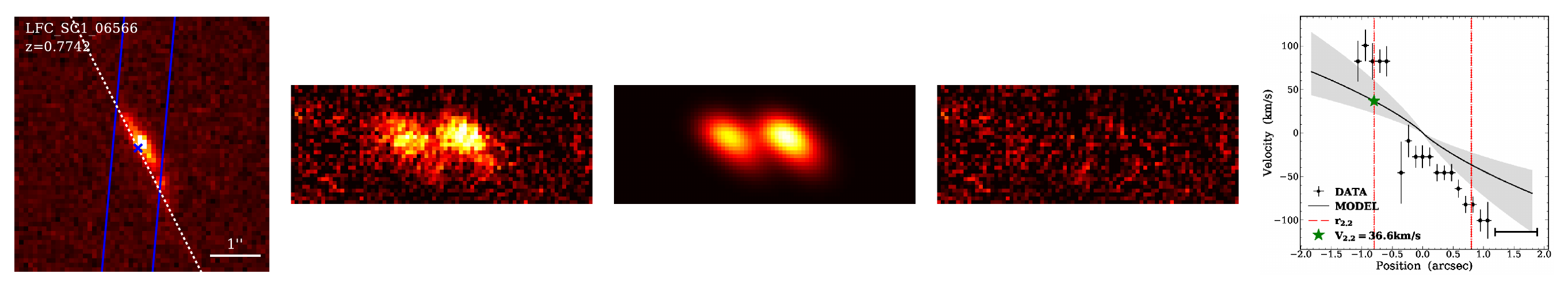}
\contcaption{}
\label{fig:appendix_kinemodel}
\end{figure*}

\begin{figure*}
\includegraphics[width=\textwidth]{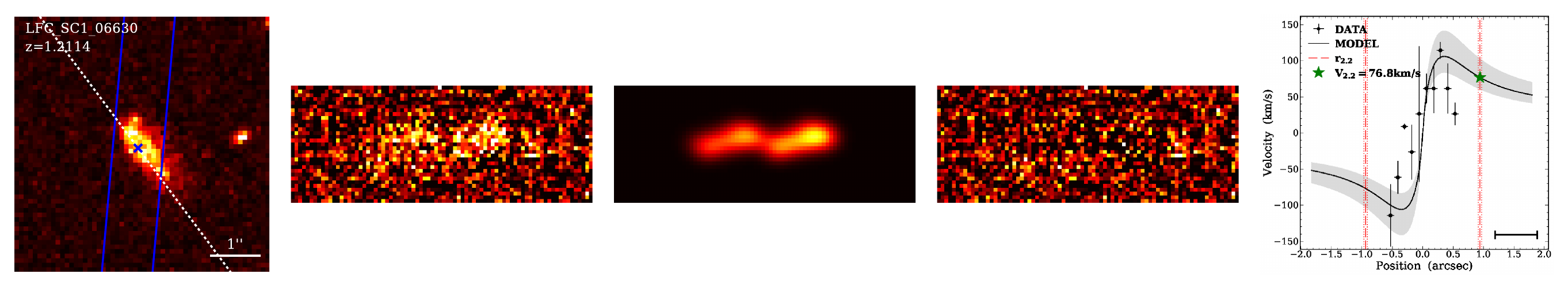}
\includegraphics[width=\textwidth]{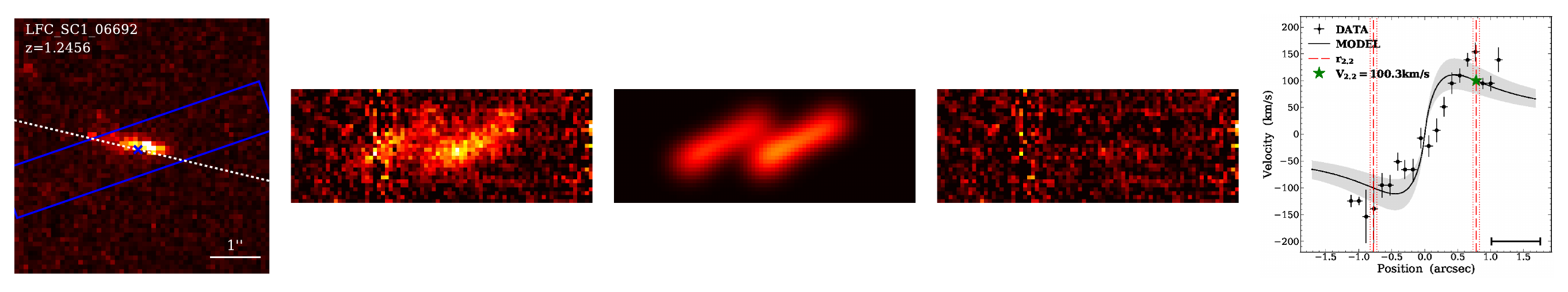}
\includegraphics[width=\textwidth]{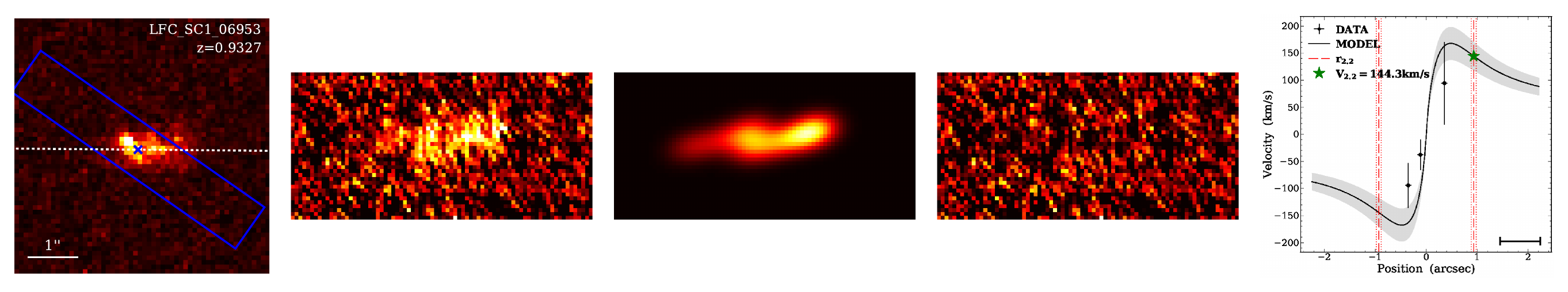}
\includegraphics[width=\textwidth]{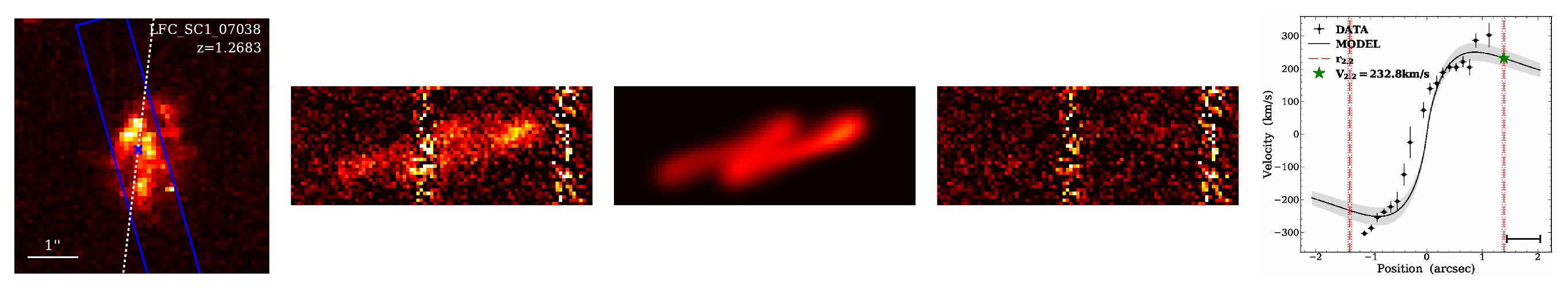}
\includegraphics[width=\textwidth]{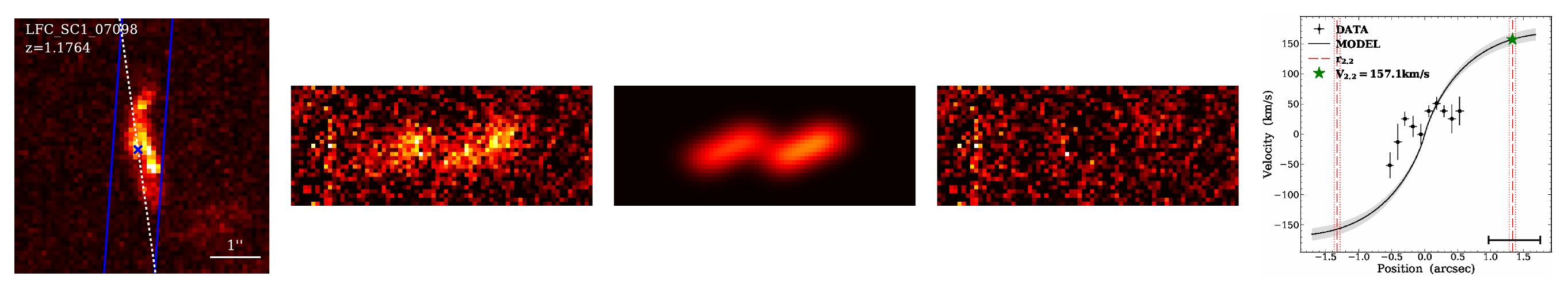}
\includegraphics[width=\textwidth]{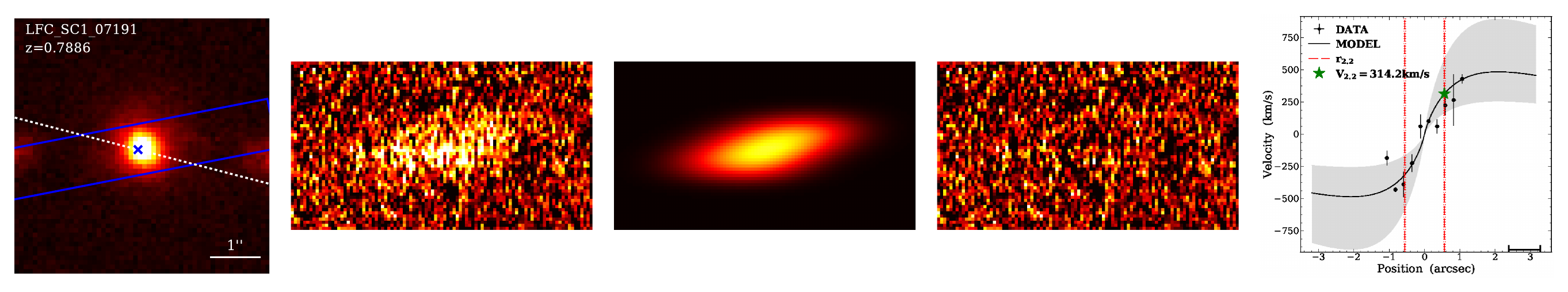}
\includegraphics[width=\textwidth]{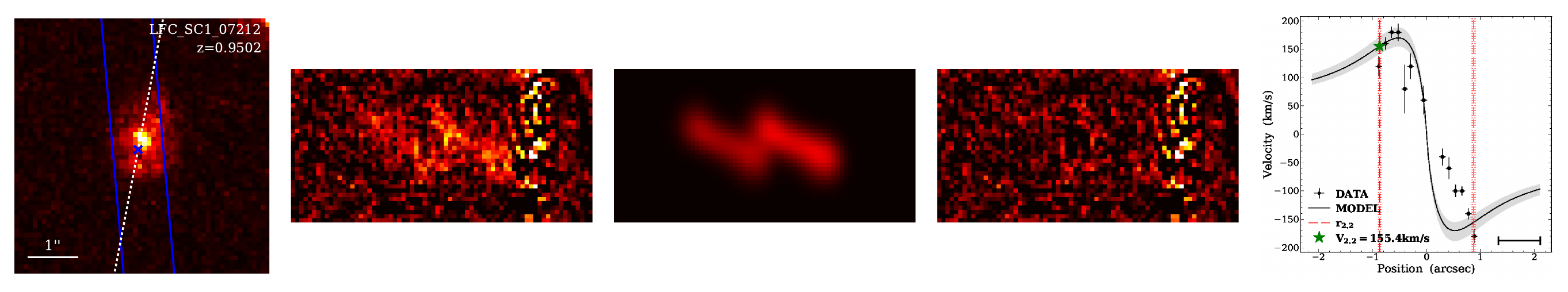}
\contcaption{}
\label{fig:appendix_kinemodel}
\end{figure*}

\begin{figure*}
\includegraphics[width=\textwidth]{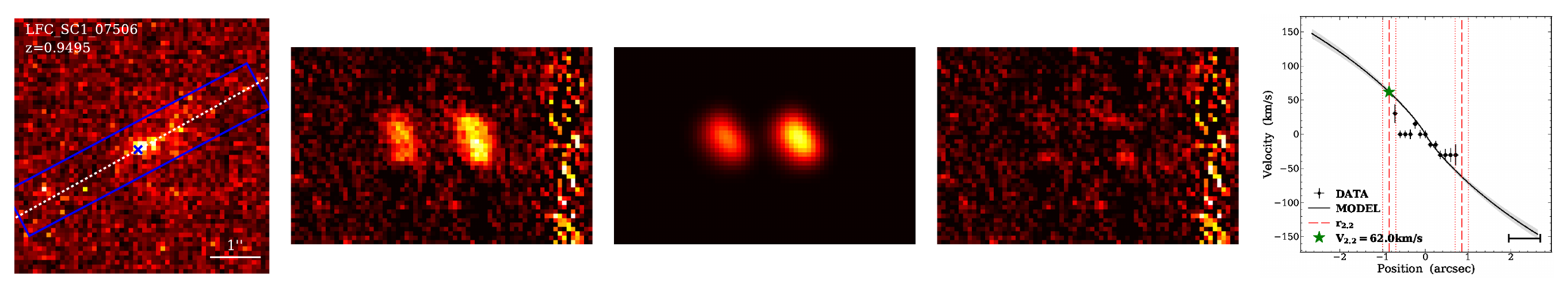}
\includegraphics[width=\textwidth]{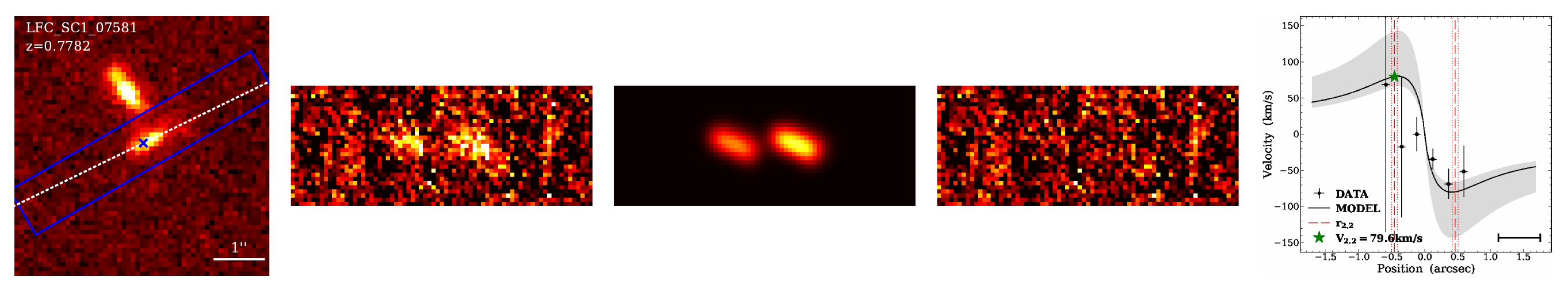}
\includegraphics[width=\textwidth]{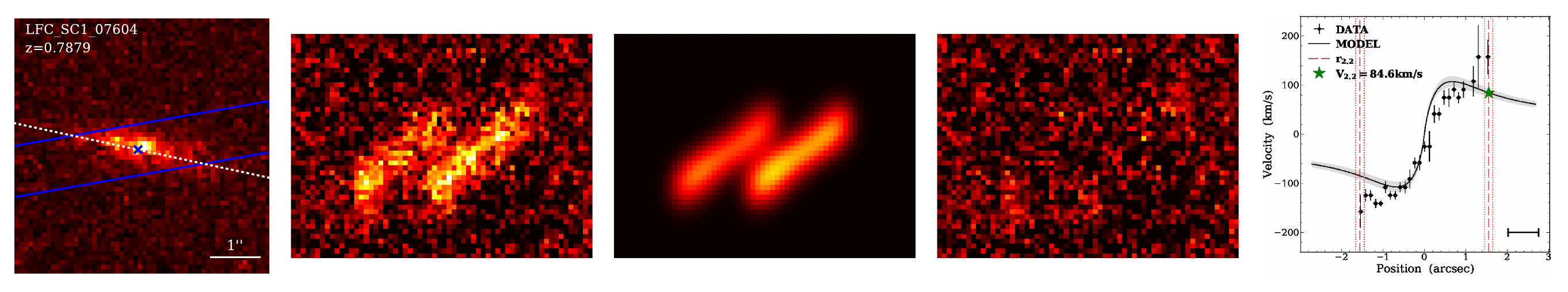}
\includegraphics[width=\textwidth]{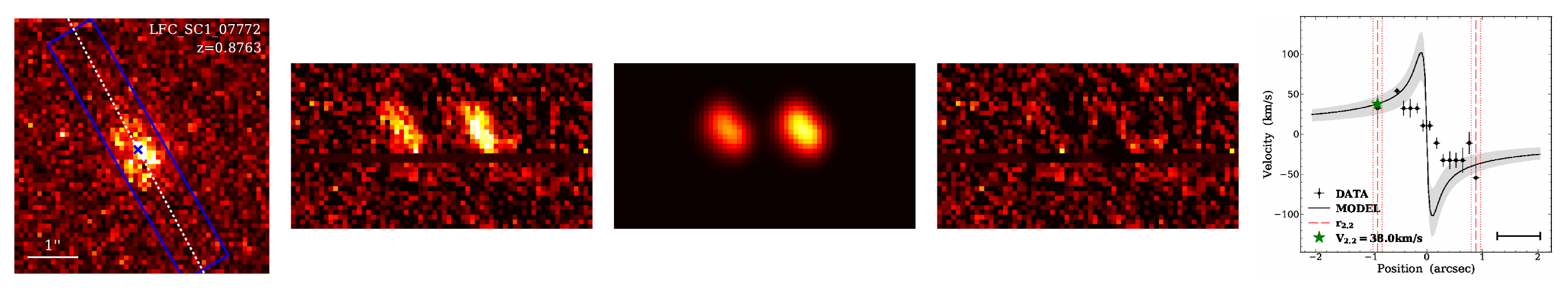}
\includegraphics[width=\textwidth]{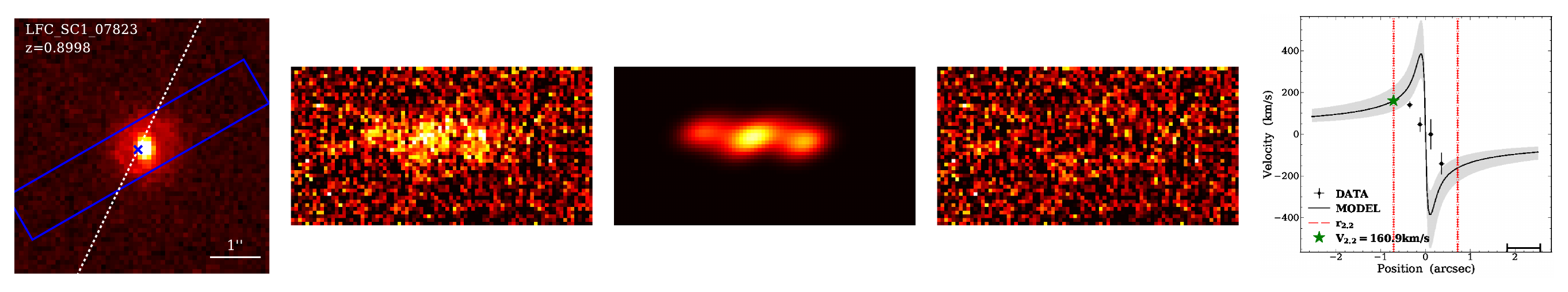}
\includegraphics[width=\textwidth]{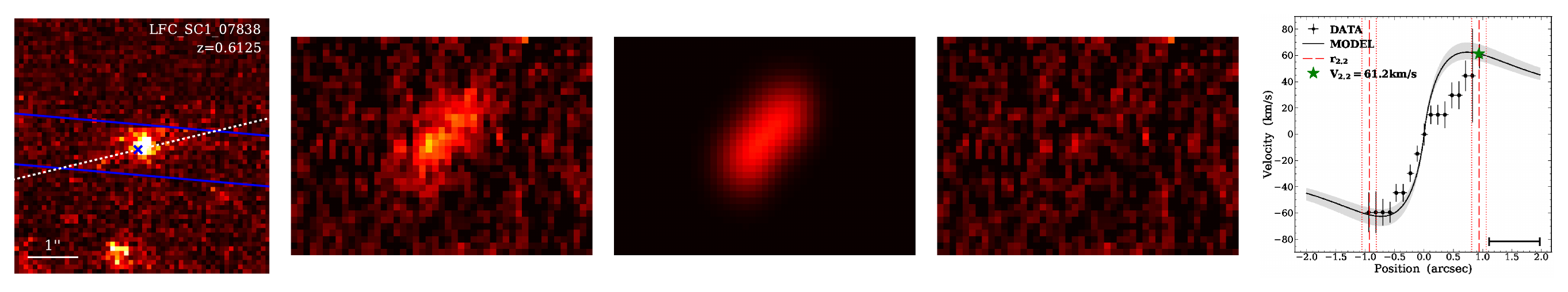}
\includegraphics[width=\textwidth]{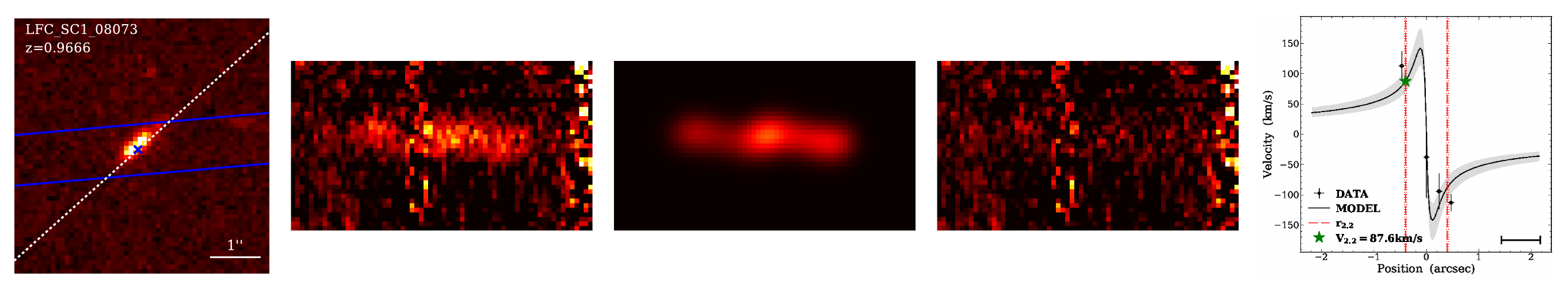}
\contcaption{}
\label{fig:appendix_kinemodel}
\end{figure*}

\begin{figure*}
\includegraphics[width=\textwidth]{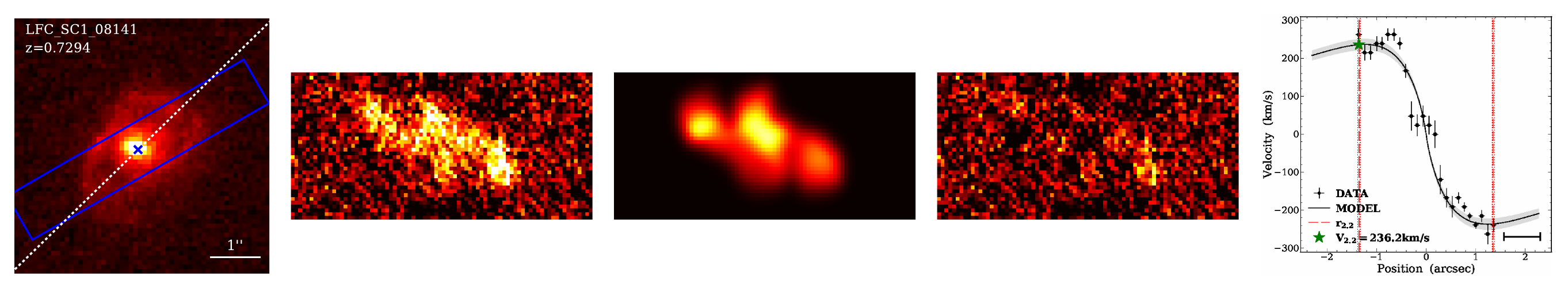}
\includegraphics[width=\textwidth]{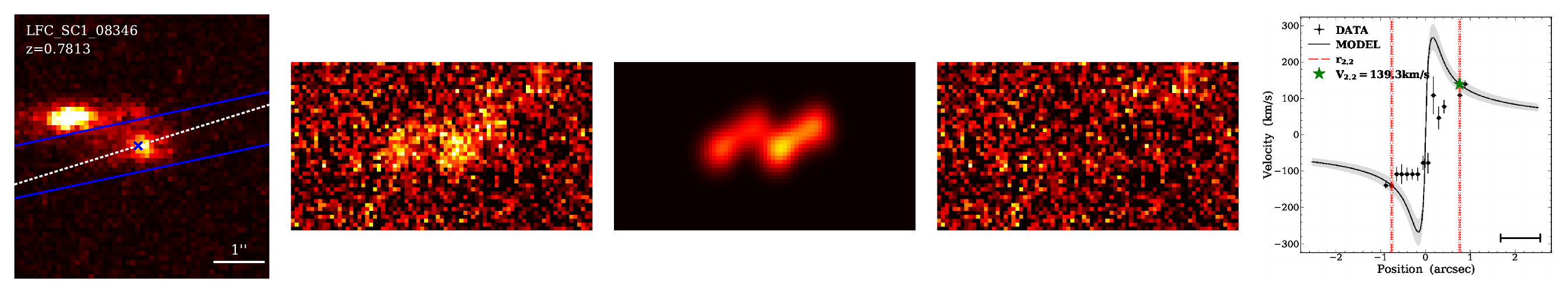}
\includegraphics[width=\textwidth]{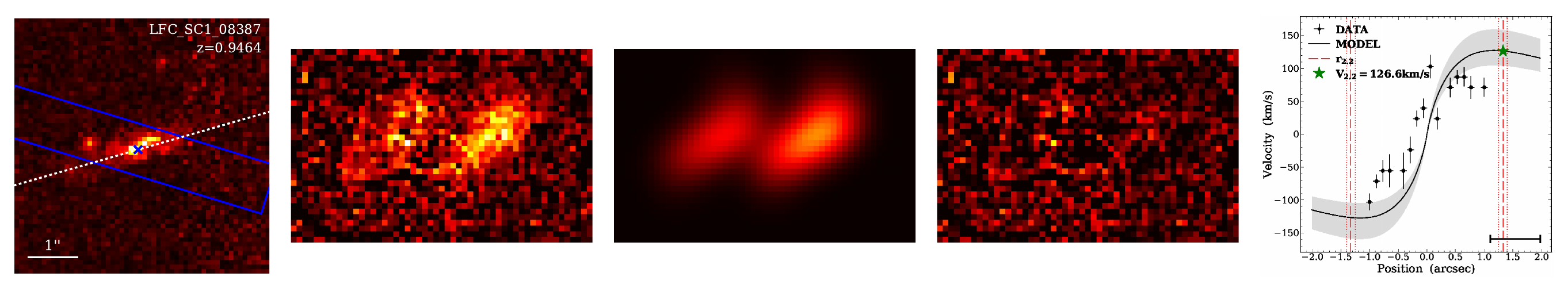}
\includegraphics[width=\textwidth]{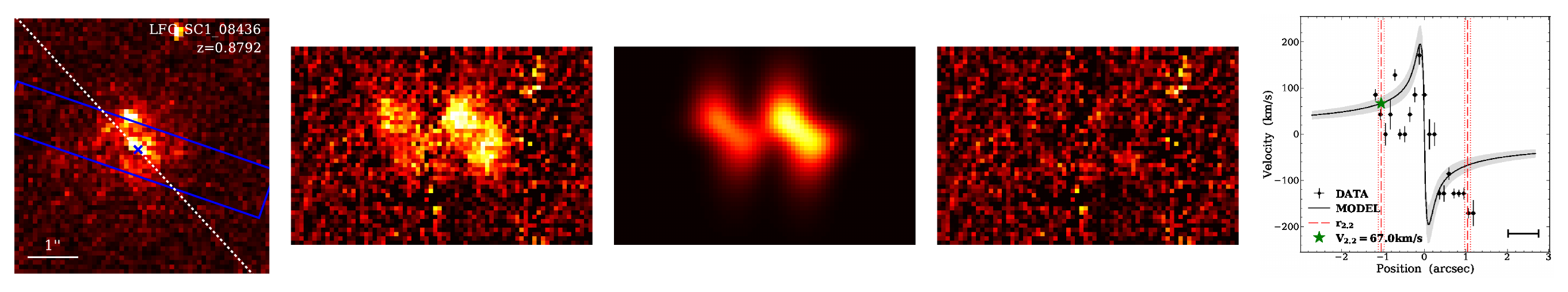}
\includegraphics[width=\textwidth]{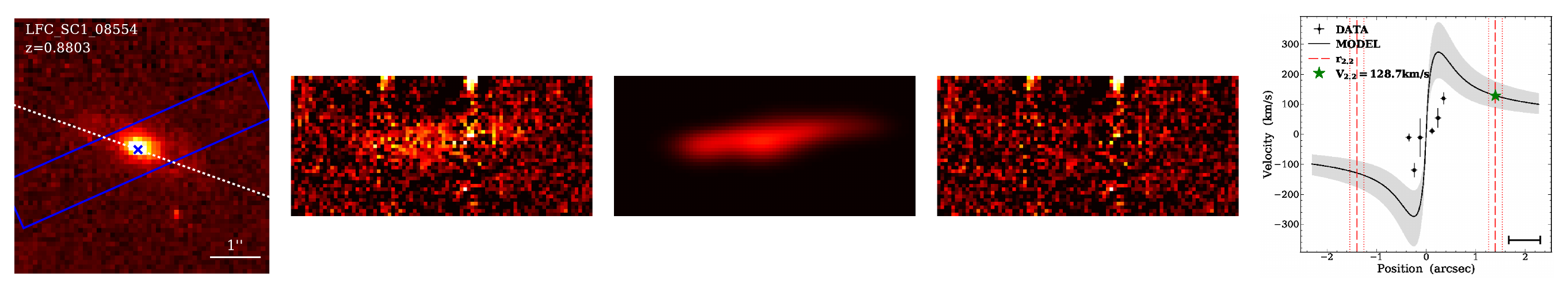}
\includegraphics[width=\textwidth]{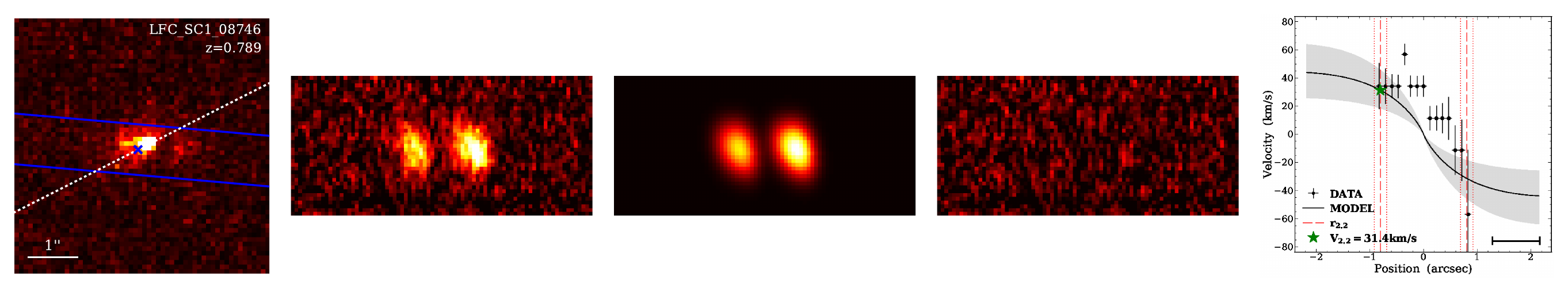}
\includegraphics[width=\textwidth]{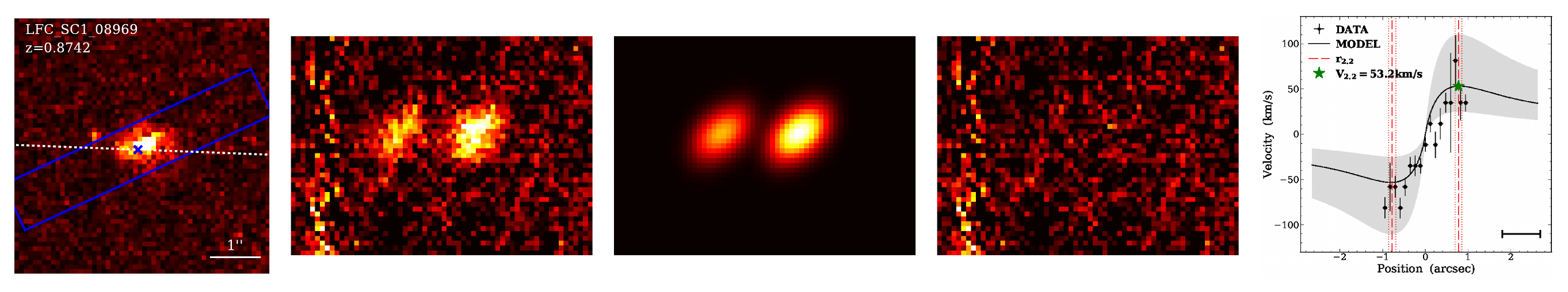}
\contcaption{}
\label{fig:appendix_kinemodel}
\end{figure*}

\begin{figure*}
\includegraphics[width=\textwidth]{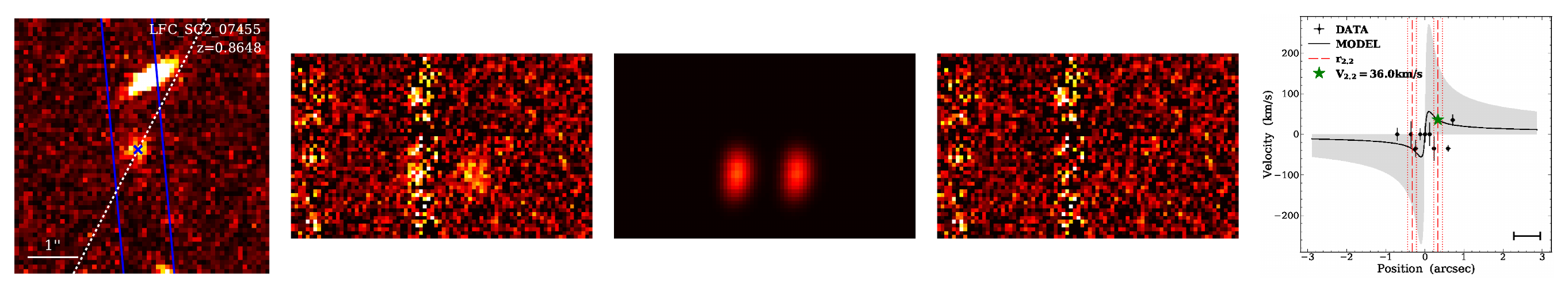}
\includegraphics[width=\textwidth]{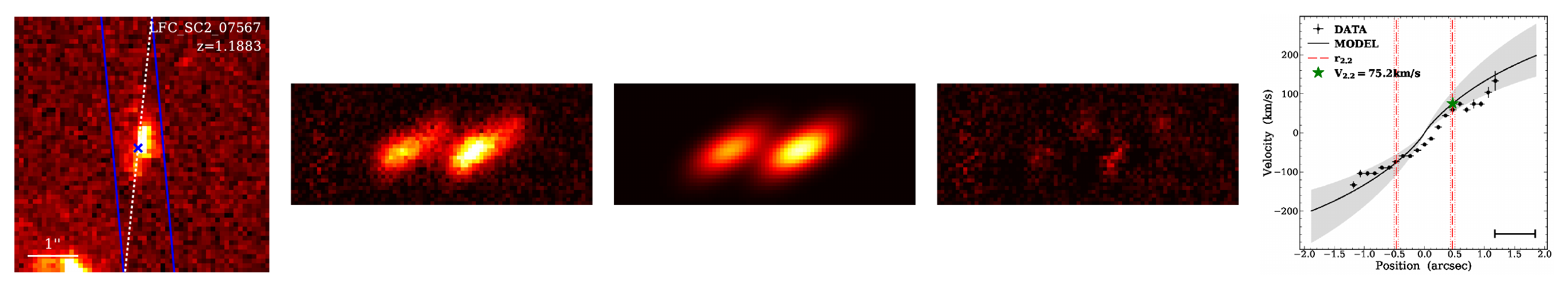}
\includegraphics[width=\textwidth]{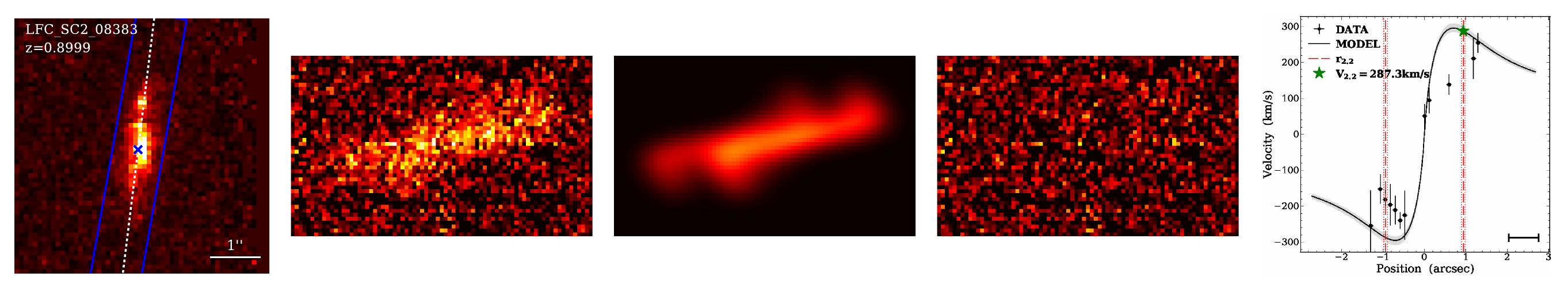}
\includegraphics[width=\textwidth]{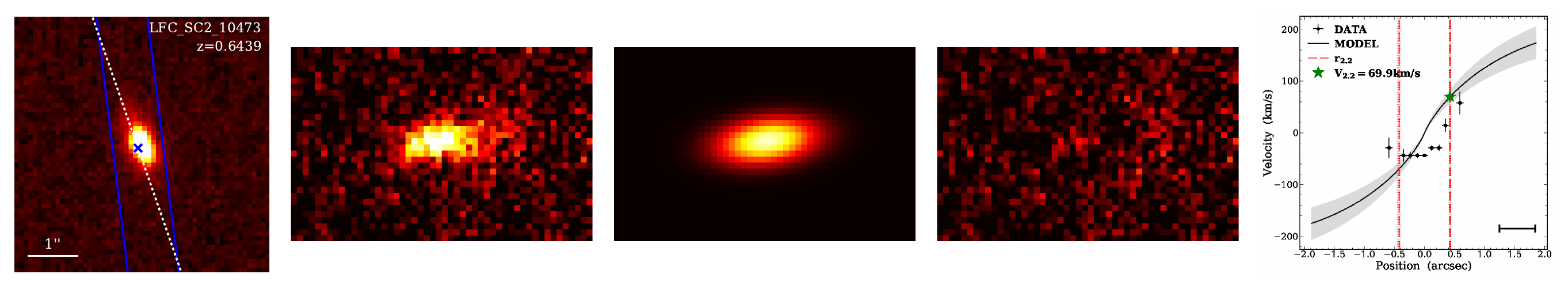}
    
\contcaption{}
\end{figure*}

\section{Galaxy Parameters Table} \label{app:tab}


\begin{landscape}
\begin{table}
\centering
\caption{Galaxy parameters.}
\label{tab_app:gal_par}
\begin{tabular}{cccccccccccccccc}
\hline\hline 
ID  & RA & Dec & \textit{z} & PA & Incl & $r_{2.2}$ & $log(M_\ast/M_\odot$) & $M_B$ & $log(1+\delta_{gal})$ & $log \eta$ & $V_{2.2}$  & $\sigma_0$\\[5pt]
 & deg & deg & & deg & deg & kpc &   & mag &  &  & km/s & km/s \\  [5pt]
 (1) & (2)  & (3)  & (4) & (5) &(6)  & (7)  & (8) & (9) & (10) & (11) & (12) & (13) \\
\hline

COS\_SC1\_00377 & 241.097630 & 43.095130 & 0.8926 &   -4.1$\pm$13.0 &  39.8$\pm$1.5 &   5.3$\pm$0.1 & $10.15_{-0.05}^{+0.05}$ & -20.84$\pm$0.03 &  0.76 & -0.35 &  264.8$_{-54.0}^{+59.2}$ & $  39.3_{-22.9}^{+36.5}$ \\[4pt]
COS\_SC1\_00394 & 241.094630 & 43.097474 & 0.6093 &  +15.0$\pm$13.1 &  60.4$\pm$0.8 &   6.7$\pm$0.2 & $ 9.72_{-0.05}^{+0.10}$ & -20.38$\pm$0.01 &  0.05 &  0.68 &  100.6$_{-5.7}^{+16.6}$ & $  26.5_{-3.2}^{+3.2}$ \\[4pt]
COS\_SC1\_00885 & 241.093460 & 43.129845 & 0.9093 &  +32.4$\pm$13.0 &  76.4$\pm$1.3 &   5.5$\pm$0.5 & $ 9.07_{-0.18}^{+0.36}$ & -18.94$\pm$0.32 &  0.05 &  0.47 &  136.6$_{-17.6}^{+17.9}$ & $  25.9_{-2.9}^{+8.3}$ \\[4pt]
COS\_SC1\_01218 & 241.138960 & 43.217272 & 0.8378 &  +54.8$\pm$13.1 &  51.4$\pm$2.4 &   3.7$\pm$0.5 & $ 9.05_{-0.18}^{+0.33}$ & -19.33$\pm$0.04 &  0.03 &  0.57 &   34.8$_{-26.8}^{+27.1}$ & $  52.4_{-4.5}^{+4.2}$ \\[4pt]
COS\_SC1\_01262 & 241.087130 & 43.208381 & 0.7736 &  -20.6$\pm$13.0 &  50.4$\pm$3.8 &   3.0$\pm$0.3 & $ 8.98_{-0.38}^{+0.20}$ & -18.87$\pm$0.08 &  0.41 &  1.15 &   54.3$_{-9.5}^{+12.9}$ & $  28.9_{-2.9}^{+3.1}$ \\[4pt]
COS\_SC1\_01642 & 241.137740 & 43.222466 & 1.2796 &  +49.5$\pm$13.0 &  77.8$\pm$1.6 &   5.6$\pm$0.9 & $ 9.49_{-0.26}^{+0.09}$ & -20.53$\pm$0.21 &  0.01 &  1.65 &  187.3$_{-47.9}^{+44.7}$ & $  15.8_{-2.3}^{+3.5}$ \\[4pt]
COS\_SC1\_01677 & 241.116670 & 43.220816 & 0.8957 &  -50.3$\pm$13.0 &  67.4$\pm$2.3 &   3.5$\pm$0.5 & $ 8.31_{-0.07}^{+0.21}$ & -18.19$\pm$0.17 &  0.15 &  0.42 &   29.5$_{-22.3}^{+8.0}$ & $  31.9_{-6.5}^{+7.0}$ \\[4pt]
COS\_SC1\_02010 & 241.134040 & 43.200556 & 1.2957 &  -10.9$\pm$13.0 &  42.8$\pm$4.0 &   7.1$\pm$1.1 & $10.27_{-0.18}^{+0.22}$ & -21.30$\pm$0.08 &  0.07 &  1.86 &  243.3$_{-47.2}^{+96.6}$ & $  58.7_{-11.9}^{+13.5}$ \\[4pt]
COS\_SC1\_02200 & 241.110100 & 43.187006 & 1.2931 &  +44.4$\pm$13.3 &  28.9$\pm$4.6 &   3.8$\pm$1.4 & $10.44_{-0.00}^{+0.06}$ & -21.33$\pm$0.09 &  0.14 &  1.94 &   22.3$_{-16.3}^{+6.1}$ & $  35.6_{-5.7}^{+6.9}$ \\[4pt]
COS\_SC2\_00032 & 241.087430 & 43.207257 & 0.9481 &  +46.1$\pm$30.2 &  27.8$\pm$2.0 &   5.8$\pm$0.1 & $10.36_{-0.05}^{+0.02}$ & -22.53$\pm$0.02 &  0.28 &  1.16 &  175.7$_{-31.9}^{+33.2}$ & $  43.1_{-3.2}^{+3.4}$ \\[4pt]
COS\_SC2\_00180 & 241.138120 & 43.225032 & 0.8597 &  -10.6$\pm$13.0 &  66.6$\pm$0.7 &   4.3$\pm$0.1 & $ 9.64_{-0.06}^{+0.12}$ & -20.85$\pm$0.01 &  0.21 & -0.23 &    5.7$_{-5.7}^{+1.1}$ & $  61.1_{-5.5}^{+4.9}$ \\[4pt]
COS\_SC2\_00526 & 241.116500 & 43.253207 & 0.7327 &  +89.8$\pm$13.3 &  60.8$\pm$0.7 &   2.7$\pm$0.1 & $ 9.64_{-0.16}^{+0.20}$ & -20.82$\pm$0.07 &  0.65 &  0.95 &   16.9$_{-10.7}^{+10.9}$ & $  52.8_{-4.8}^{+3.6}$ \\[4pt]
COS\_SC2\_00664 & 241.018350 & 43.262282 & 0.6014 &  -46.0$\pm$13.7 &  45.6$\pm$3.0 &   4.3$\pm$0.7 & $ 9.42_{-0.19}^{+0.20}$ & -18.93$\pm$0.02 & -0.08 & -0.35 &   72.0$_{-15.4}^{+8.9}$ & $  19.0_{-2.4}^{+3.1}$ \\[4pt]
COS\_SC2\_00885 & 241.057330 & 43.279894 & 0.7897 &  -32.7$\pm$13.0 &  71.4$\pm$0.9 &   5.6$\pm$0.2 & $ 9.57_{-0.15}^{+0.36}$ & -19.74$\pm$0.04 &  0.20 &  1.26 &   73.1$_{-14.7}^{+14.7}$ & $  32.3_{-6.1}^{+6.6}$ \\[4pt]
COS\_SC2\_01080 & 241.152230 & 43.352992 & 1.2743 &  +45.6$\pm$13.0 &  57.7$\pm$3.6 &   5.1$\pm$0.3 & $ 9.16_{-0.45}^{+0.92}$ & -20.17$\pm$0.22 &  0.19 &  1.22 &   81.8$_{-33.9}^{+44.4}$ & $  39.1_{-6.6}^{+5.9}$ \\[4pt]
COS\_SC2\_01132 & 241.132690 & 43.329046 & 0.9157 &  +22.7$\pm$13.2 &  43.7$\pm$4.6 &   1.7$\pm$0.0 & $ 8.89_{-0.23}^{+0.16}$ & -19.30$\pm$0.05 &  0.58 &  0.05 &   53.1$_{-18.3}^{+12.9}$ & $  25.5_{-2.7}^{+7.2}$ \\[4pt]
COS\_SC2\_01140 & 241.080770 & 43.326869 & 0.9248 &  +82.0$\pm$31.9 &  20.3$\pm$8.0 &   4.8$\pm$0.4 & $ 9.18_{-0.24}^{+0.70}$ & -19.85$\pm$0.19 &  0.90 & -0.17 &   57.2$_{-28.2}^{+29.7}$ & $  25.2_{-2.7}^{+4.2}$ \\[4pt]
COS\_SC2\_01224 & 241.153400 & 43.299811 & 0.6887 &   +4.8$\pm$13.2 &  52.0$\pm$2.6 &   3.1$\pm$0.8 & $ 8.54_{-0.06}^{+0.05}$ & -18.53$\pm$0.03 &  0.24 &  1.67 &   68.1$_{-12.6}^{+46.7}$ & $  16.2_{-2.5}^{+15.8}$ \\[4pt]
COS\_SC2\_01385 & 241.153790 & 43.355894 & 0.9240 &  -39.5$\pm$13.0 &  52.3$\pm$1.5 &   4.7$\pm$0.1 & $ 9.79_{-0.05}^{+0.00}$ & -20.04$\pm$0.06 &  1.43 & -1.11 &  154.8$_{-23.3}^{+55.4}$ & $  37.4_{-22.7}^{+8.0}$ \\[4pt]
COS\_SC2\_01519 & 241.140380 & 43.334894 & 1.1257 &   -1.5$\pm$14.1 &  34.7$\pm$4.9 &   4.8$\pm$0.2 & $ 9.63_{-0.14}^{+0.09}$ & -20.81$\pm$0.09 & -0.15 &  1.28 &  101.0$_{-24.1}^{+21.0}$ & $  40.9_{-5.9}^{+6.1}$ \\[4pt]
COS\_SC2\_01732 & 241.133900 & 43.328935 & 0.7899 &  +47.7$\pm$13.1 &  45.2$\pm$4.0 &   2.8$\pm$0.1 & $ 9.24_{-0.19}^{+0.35}$ & -18.78$\pm$0.10 & -0.04 &  1.34 &   26.2$_{-2.3}^{+2.3}$ & $  28.4_{-2.9}^{+2.9}$ \\[4pt]
COS\_SC2\_01738 & 241.128070 & 43.327621 & 0.8081 &  +38.1$\pm$13.1 &  50.0$\pm$1.0 &   3.2$\pm$0.1 & $ 9.64_{-0.00}^{+0.06}$ & -20.70$\pm$0.01 &  0.08 &  1.31 &   11.3$_{-4.2}^{+4.7}$ & $  43.9_{-3.8}^{+3.3}$ \\[4pt]
COS\_SC2\_01762 & 241.146320 & 43.327779 & 1.0357 &  +19.7$\pm$13.1 &  50.1$\pm$4.0 &   3.3$\pm$0.2 & $ 9.31_{-0.06}^{+0.13}$ & -20.02$\pm$0.02 & -0.11 &  1.26 &  141.6$_{-25.4}^{+31.0}$ & $  57.2_{-13.1}^{+12.8}$ \\[4pt]
COS\_SC2\_01956 & 241.124740 & 43.311720 & 0.9245 &  +23.5$\pm$13.0 &  37.1$\pm$6.4 &   1.4$\pm$0.5 & $ 8.75_{-0.27}^{+0.17}$ & -18.65$\pm$0.17 &  0.87 & -0.30 &   40.0$_{-21.3}^{+42.1}$ & $  25.2_{-2.7}^{+2.7}$ \\[4pt]
COS\_SC2\_02150 & 241.024320 & 43.306863 & 0.9189 &  -16.5$\pm$13.4 &  51.8$\pm$2.0 &   3.7$\pm$0.3 & $10.23_{-0.05}^{+0.04}$ & -19.77$\pm$0.08 &  1.56 &  0.95 &  159.9$_{-40.5}^{+35.8}$ & $  73.7_{-10.6}^{+14.6}$ \\[4pt]
FG2\_028s\_ACS & 241.122330 & 43.350782 & 1.2889 &  +68.1$\pm$13.0 &  61.2$\pm$3.0 &   3.6$\pm$0.4 & $ 9.27_{-0.49}^{+0.74}$ & -19.91$\pm$0.44 &  0.18 &  1.46 &   11.4$_{-6.7}^{+6.2}$ & $  49.2_{-4.2}^{+4.6}$ \\[4pt]
\hline
\end{tabular}\\
\begin{flushleft}
(1) ORELSE-SC1604 identification, (2) and (3) right ascension and declination J2000 coordinates, (4) spectroscopic redshift, (5) morphological galaxy position angle, defined as the angle measured counterclockwise (East of North) between the North direction in the sky and the galaxy major axis (Sec.~\ref{subsec:kinesample}), (6) inclination, defined as the angle between the line of sight and the normal to the plane of the galaxy (Sec.~\ref{subsec:kinesample}), (7) characteristic radius 2.2 times the galaxy disc scalelength (Sec.~\ref{subsec:smTFR}), (8) rotation velocity measured from the kinematic models at $r_{2.2}$ (Sec.~\ref{subsec:kinemeasure},\ref{subsec:smTFR}), (9) velocity dispersion from the kinematic models (Sec.~\ref{subsec:kinemeasure}).
\end{flushleft}
\end{table}
\end{landscape}

\begin{landscape}
\begin{table}
\centering
\contcaption{}
\label{tab_app:gal_par}
\begin{tabular}{cccccccccccccccc}
\hline\hline 
ID  & RA & Dec & \textit{z} & PA & Incl & $r_{2.2}$ & $log(M_\ast/M_\odot$) & $M_B$ & $log(1+\delta_{gal})$ & $log \eta$ & $V_{2.2}$  & $\sigma_0$\\[5pt]
 & deg & deg & & deg & deg & kpc &   & mag &  &  & km/s & km/s \\  [5pt]
 (1) & (2)  & (3)  & (4) & (5) &(6)  & (7)  & (8) & (9) & (10) & (11) & (12) & (13) \\
\hline
LFC\_SC1\_01267 & 241.077140 & 43.225489 & 0.6341 &   +3.3$\pm$13.0 &  67.4$\pm$0.9 &   5.2$\pm$0.1 & $ 9.56_{-0.10}^{+0.10}$ & -19.23$\pm$0.01 &  0.15 &  0.74 &  106.2$_{-4.7}^{+4.7}$ & $  17.9_{-2.3}^{+2.3}$ \\[4pt]
LFC\_SC1\_01306 & 241.088290 & 43.227328 & 0.9004 &  +86.2$\pm$14.7 &  39.6$\pm$2.1 &   5.2$\pm$0.2 & $ 9.63_{-0.06}^{+0.19}$ & -20.44$\pm$0.19 &  0.45 &  0.46 &   30.8$_{-6.6}^{+6.0}$ & $  25.9_{-2.7}^{+2.7}$ \\[4pt]
LFC\_SC1\_01366 & 241.055610 & 43.228766 & 0.8994 &  -53.5$\pm$13.0 &  71.6$\pm$0.7 &  10.6$\pm$0.4 & $10.31_{-0.06}^{+0.00}$ & -20.75$\pm$0.04 &  0.75 &  0.77 &   55.7$_{-12.0}^{+11.7}$ & $  61.4_{-16.8}^{+12.5}$ \\[4pt]
LFC\_SC1\_01472 & 241.105240 & 43.233046 & 0.8603 &   -9.1$\pm$15.5 &  42.6$\pm$1.4 &   3.6$\pm$0.1 & $ 9.94_{-0.21}^{+0.35}$ & -20.52$\pm$0.02 &  1.12 & -0.87 &  179.8$_{-16.4}^{+17.9}$ & $  35.7_{-5.0}^{+4.8}$ \\[4pt]
LFC\_SC1\_01504 & 241.034360 & 43.235418 & 0.9352 &  +34.7$\pm$13.0 &  44.6$\pm$3.0 &   3.7$\pm$0.2 & $ 8.98_{-0.06}^{+0.00}$ & -19.46$\pm$0.08 &  0.89 & -0.44 &   77.3$_{-16.1}^{+30.9}$ & $  40.3_{-4.0}^{+4.9}$ \\[4pt]
LFC\_SC1\_01549 & 241.089190 & 43.237260 & 0.8726 &   +5.7$\pm$13.1 &  45.4$\pm$2.3 &   2.2$\pm$0.1 & $ 9.20_{-0.14}^{+0.05}$ & -19.62$\pm$0.04 &  0.76 & -0.33 &  169.4$_{-24.4}^{+27.0}$ & $  28.2_{-6.2}^{+10.9}$ \\[4pt]
LFC\_SC1\_01710 & 241.090620 & 43.244222 & 0.8595 &  +45.5$\pm$13.6 &  44.5$\pm$3.0 &   2.2$\pm$0.2 & $ 8.78_{-0.00}^{+0.00}$ & -19.33$\pm$0.04 &  0.67 & -0.51 &   30.4$_{-30.4}^{+22.4}$ & $  38.5_{-16.3}^{+7.2}$ \\[4pt]
LFC\_SC1\_01870 & 241.082250 & 43.249738 & 0.9201 &  -85.2$\pm$13.1 &  36.1$\pm$2.9 &   5.7$\pm$0.2 & $ 9.56_{-0.06}^{+0.00}$ & -20.10$\pm$0.04 &  0.34 &  0.93 &   49.5$_{-16.7}^{+22.7}$ & $  25.5_{-2.8}^{+6.2}$ \\[4pt]
LFC\_SC1\_01894 & 241.021780 & 43.251321 & 0.9374 &  +40.1$\pm$13.0 &  55.8$\pm$2.2 &   4.8$\pm$0.6 & $ 9.09_{-0.07}^{+0.06}$ & -19.35$\pm$0.09 &  1.04 & -0.10 &   29.1$_{-12.2}^{+11.9}$ & $  44.2_{-3.7}^{+4.0}$ \\[4pt]
LFC\_SC1\_01931 & 241.102220 & 43.251363 & 0.9107 &  +44.1$\pm$13.0 &  66.3$\pm$0.6 &   8.7$\pm$0.2 & $10.39_{-0.11}^{+0.08}$ & -21.19$\pm$0.02 &  0.40 &  1.27 &  155.8$_{-37.9}^{+61.1}$ & $  87.2_{-14.1}^{+11.4}$ \\[4pt]
LFC\_SC1\_02006 & 241.053950 & 43.251663 & 0.8638 &  -16.6$\pm$13.1 &  31.8$\pm$3.0 &   2.5$\pm$0.1 & $10.04_{-0.05}^{+0.29}$ & -20.29$\pm$0.02 &  0.50 & -0.74 &  167.0$_{-37.6}^{+30.4}$ & $  27.0_{-2.9}^{+6.2}$ \\[4pt]
LFC\_SC1\_02510 & 241.047820 & 43.271859 & 1.2934 &   +4.3$\pm$13.0 &  63.2$\pm$1.7 &   9.9$\pm$0.8 & $ 9.69_{-0.14}^{+0.00}$ & -20.93$\pm$0.10 &  0.20 &  1.64 &  171.7$_{-7.3}^{+4.5}$ & $  15.5_{-2.2}^{+2.6}$ \\[4pt]
LFC\_SC1\_02783 & 241.117870 & 43.276426 & 0.6437 &   +2.9$\pm$13.6 &  56.4$\pm$0.7 &   5.6$\pm$0.2 & $10.25_{-0.05}^{+0.23}$ & -20.68$\pm$0.01 &  0.46 &  1.24 &  244.1$_{-17.6}^{+17.4}$ & $  23.1_{-13.0}^{+16.8}$ \\[4pt]
LFC\_SC1\_03063 & 241.108030 & 43.284153 & 0.9601 &  +70.8$\pm$13.0 &  51.3$\pm$1.7 &   3.9$\pm$0.2 & $ 9.78_{-0.13}^{+0.01}$ & -20.55$\pm$0.03 &  0.13 &  1.23 &  276.3$_{-20.9}^{+22.0}$ & $  34.9_{-6.9}^{+7.3}$ \\[4pt]
LFC\_SC1\_03257 & 241.088390 & 43.287589 & 0.9189 &   -6.8$\pm$14.3 &  35.8$\pm$1.4 &   7.2$\pm$0.1 & $10.34_{-0.05}^{+0.06}$ & -21.39$\pm$0.02 &  0.80 &  1.01 &  525.9$_{-134.7}^{+332.0}$ & $ 113.8_{-28.1}^{+13.1}$ \\[4pt]
LFC\_SC1\_03266 & 241.016860 & 43.286258 & 1.1820 &  -25.6$\pm$13.0 &  52.6$\pm$1.7 &   5.5$\pm$0.3 & $10.93_{-0.06}^{+0.00}$ & -24.07$\pm$0.00 &  0.33 & -0.68 &  126.3$_{-25.2}^{+30.8}$ & $  56.4_{-14.1}^{+18.4}$ \\[4pt]
LFC\_SC1\_03415 & 241.072210 & 43.290971 & 0.6213 &  -12.2$\pm$14.6 &  38.4$\pm$1.3 &   5.2$\pm$0.2 & $ 9.68_{-0.00}^{+0.05}$ & -20.16$\pm$0.01 &  0.15 &  0.91 &   36.7$_{-12.4}^{+26.0}$ & $  25.9_{-3.0}^{+3.0}$ \\[4pt]
LFC\_SC1\_03496 & 241.119730 & 43.293805 & 0.9152 &  +24.2$\pm$13.0 &  74.2$\pm$0.7 &   7.4$\pm$0.3 & $ 9.67_{-0.20}^{+0.29}$ & -20.10$\pm$0.04 &  0.71 &  0.47 &   93.2$_{-15.9}^{+16.7}$ & $  33.7_{-17.8}^{+7.6}$ \\[4pt]
LFC\_SC1\_03626 & 241.120470 & 43.297133 & 0.7156 &  -34.6$\pm$13.2 &  41.0$\pm$2.3 &   1.3$\pm$0.1 & $ 8.89_{-0.00}^{+0.00}$ & -19.21$\pm$0.03 & -0.15 &  1.59 &   24.9$_{-13.4}^{+17.2}$ & $  50.9_{-3.9}^{+4.7}$ \\[4pt]
LFC\_SC1\_03933 & 241.094440 & 43.303735 & 0.8138 &   -4.1$\pm$13.1 &  50.0$\pm$2.6 &   2.8$\pm$1.1 & $ 8.66_{-0.05}^{+0.32}$ & -19.01$\pm$0.03 &  0.12 &  1.16 &    3.3$_{-3.3}^{+34.5}$ & $  42.8_{-8.5}^{+4.9}$ \\[4pt]
LFC\_SC1\_04470 & 241.073000 & 43.314769 & 0.8079 &  -22.5$\pm$13.0 &  63.0$\pm$0.6 &   7.0$\pm$0.1 & $10.37_{-0.06}^{+0.00}$ & -21.09$\pm$0.01 &  0.36 &  1.30 &   78.8$_{-12.8}^{+12.4}$ & $  36.3_{-5.4}^{+5.2}$ \\[4pt]
LFC\_SC1\_05193 & 241.099760 & 43.333074 & 1.0569 &  -40.0$\pm$13.1 &  48.2$\pm$2.8 &  18.2$\pm$3.8 & $ 9.51_{-0.14}^{+0.06}$ & -20.78$\pm$0.04 &  0.02 &  1.24 &  220.2$_{-48.3}^{+78.1}$ & $  21.4_{-3.0}^{+5.3}$ \\[4pt]
LFC\_SC1\_05250 & 241.227680 & 43.333795 & 0.8665 &  +32.5$\pm$18.9 &  29.0$\pm$1.7 &   5.4$\pm$0.1 & $10.33_{-0.06}^{+0.00}$ & -20.96$\pm$0.02 &  0.08 & -0.03 &    9.7$_{-9.7}^{+26.0}$ & $  45.9_{-3.5}^{+3.6}$ \\[4pt]
LFC\_SC1\_05281 & 241.093530 & 43.334808 & 1.1877 &  -29.1$\pm$13.1 &  34.4$\pm$2.0 &   7.4$\pm$0.2 & $10.41_{-0.13}^{+0.00}$ & -22.26$\pm$0.03 &  0.69 & -0.12 &   98.4$_{-22.5}^{+31.7}$ & $  29.2_{-9.3}^{+7.0}$ \\[4pt]
LFC\_SC1\_05297 & 241.106420 & 43.334128 & 0.8088 &  -73.0$\pm$15.3 &  44.7$\pm$0.4 &   1.0$\pm$0.0 & $10.63_{-0.00}^{+0.05}$ & -22.77$\pm$0.00 &  0.62 &  1.37 &  141.7$_{-50.1}^{+53.3}$ & $ 143.6_{-8.6}^{+7.6}$ \\[4pt]
LFC\_SC1\_05319 & 241.101690 & 43.335719 & 0.9223 &  -50.6$\pm$13.0 &  70.5$\pm$0.8 &  11.2$\pm$0.4 & $10.60_{-0.10}^{+0.05}$ & -21.31$\pm$0.03 &  1.15 & -1.08 &  194.3$_{-26.7}^{+32.0}$ & $ 121.1_{-9.1}^{+11.8}$ \\[4pt]
LFC\_SC1\_05419 & 241.111080 & 43.339032 & 0.7288 &  -87.7$\pm$13.0 &  45.4$\pm$1.8 &   3.4$\pm$0.2 & $ 9.33_{-0.00}^{+0.00}$ & -19.67$\pm$0.01 &  0.29 &  1.53 &  135.1$_{-6.3}^{+7.0}$ & $  32.8_{-9.7}^{+6.8}$ \\[4pt]
LFC\_SC1\_05441 & 241.079990 & 43.336862 & 1.1792 &  -22.2$\pm$21.3 &  25.8$\pm$4.8 &   0.8$\pm$1.1 & $ 9.69_{-0.06}^{+0.07}$ & -21.00$\pm$0.01 &  0.73 & -0.58 &  110.6$_{-107.2}^{+102.0}$ & $  58.1_{-16.5}^{+14.6}$ \\[4pt]
LFC\_SC1\_05501 & 241.181540 & 43.339918 & 0.6718 &  +87.2$\pm$13.2 &  49.6$\pm$0.9 &   4.5$\pm$0.2 & $10.05_{-0.13}^{+0.20}$ & -20.21$\pm$0.01 & -0.16 &  1.71 &  205.1$_{-8.2}^{+12.0}$ & $  18.3_{-2.5}^{+3.2}$ \\[4pt]
LFC\_SC1\_05513 & 240.996480 & 43.339669 & 0.7003 &  +59.1$\pm$15.3 &  54.0$\pm$0.6 &   2.5$\pm$0.5 & $10.08_{-0.00}^{+0.09}$ & -20.36$\pm$0.01 &  0.16 &  1.75 &  115.2$_{-23.1}^{+22.5}$ & $  68.8_{-4.5}^{+3.9}$ \\[4pt]

\hline
\end{tabular}
\end{table}
\end{landscape}

\begin{landscape}
\begin{table}
\centering
\contcaption{}
\label{tab_app:gal_par}
\begin{tabular}{cccccccccccccccc}
\hline\hline 
ID  & RA & Dec & \textit{z} & PA & Incl & $r_{2.2}$ & $log(M_\ast/M_\odot$) & $M_B$ & $log(1+\delta_{gal})$ & $log \eta$ & $V_{2.2}$  & $\sigma_0$\\[5pt]
 & deg & deg & & deg & deg & kpc &   & mag &  &  & km/s & km/s \\  [5pt]
 (1) & (2)  & (3)  & (4) & (5) &(6)  & (7)  & (8) & (9) & (10) & (11) & (12) & (13) \\
\hline

LFC\_SC1\_05698 & 241.221370 & 43.344166 & 0.6241 &  +15.2$\pm$13.0 &  68.0$\pm$1.3 &   3.6$\pm$0.1 & $ 8.87_{-0.00}^{+0.06}$ & -18.83$\pm$0.02 &  0.06 &  1.30 &   45.2$_{-15.0}^{+26.4}$ & $  20.7_{-6.0}^{+11.8}$ \\[4pt]
LFC\_SC1\_05854 & 241.172270 & 43.345116 & 0.7687 &  +89.0$\pm$13.0 &  62.4$\pm$0.8 &   6.6$\pm$0.1 & $10.45_{-0.06}^{+0.05}$ & -23.29$\pm$0.01 &  0.48 &  1.39 &  161.1$_{-10.7}^{+20.4}$ & $  34.1_{-5.1}^{+4.9}$ \\[4pt]
LFC\_SC1\_05873 & 241.209710 & 43.347089 & 0.9363 &  -68.8$\pm$13.1 &  64.9$\pm$0.5 &  11.8$\pm$0.3 & $10.88_{-0.14}^{+0.17}$ & -21.90$\pm$0.01 &  0.35 & -0.14 &  200.0$_{-23.8}^{+24.0}$ & $  67.2_{-3.4}^{+3.6}$ \\[4pt]
LFC\_SC1\_05897 & 240.991920 & 43.348995 & 0.9438 &  +76.5$\pm$14.2 &  42.9$\pm$1.4 &   3.7$\pm$0.1 & $ 9.78_{-0.06}^{+0.00}$ & -20.79$\pm$0.02 &  0.18 &  1.15 &   62.1$_{-16.7}^{+10.9}$ & $  25.4_{-3.6}^{+5.9}$ \\[4pt]
LFC\_SC1\_06080 & 240.957530 & 43.353295 & 0.9626 &  -80.1$\pm$14.6 &  43.8$\pm$2.2 &   5.7$\pm$0.6 & $ 9.53_{-0.06}^{+0.06}$ & -20.16$\pm$0.06 &  0.11 &  1.61 &   60.3$_{-17.7}^{+13.5}$ & $  24.9_{-3.6}^{+5.9}$ \\[4pt]
LFC\_SC1\_06220 & 241.139920 & 43.353319 & 0.9192 &  -12.3$\pm$14.2 &  45.2$\pm$0.8 &   4.7$\pm$0.1 & $ 9.89_{-0.05}^{+0.07}$ & -21.82$\pm$0.00 &  1.74 & -1.69 &   79.1$_{-56.0}^{+77.3}$ & $  39.2_{-3.7}^{+3.7}$ \\[4pt]
LFC\_SC1\_06261 & 241.145960 & 43.356009 & 0.6019 &   -7.6$\pm$13.2 &  39.8$\pm$3.4 &   6.4$\pm$0.2 & $ 8.90_{-0.11}^{+0.00}$ & -18.87$\pm$0.02 &  0.28 &  0.24 &   51.7$_{-3.9}^{+5.8}$ & $  19.0_{-2.4}^{+2.4}$ \\[4pt]
LFC\_SC1\_06319 & 241.125510 & 43.356700 & 0.8214 &  -76.6$\pm$18.3 &  28.4$\pm$1.3 &   1.3$\pm$0.1 & $10.60_{-0.05}^{+0.00}$ & -21.61$\pm$0.01 &  0.13 &  2.26 &  166.4$_{-106.9}^{+105.7}$ & $ 141.7_{-19.5}^{+18.7}$ \\[4pt]
LFC\_SC1\_06351 & 241.224870 & 43.357840 & 0.8660 &  +21.8$\pm$13.0 &  75.6$\pm$0.8 &   8.9$\pm$0.5 & $ 9.99_{-0.06}^{+0.22}$ & -19.84$\pm$0.05 &  0.34 & -0.12 &   64.2$_{-25.1}^{+24.7}$ & $  44.9_{-4.6}^{+4.5}$ \\[4pt]
LFC\_SC1\_06398 & 240.966790 & 43.359102 & 0.9153 &  +82.7$\pm$13.0 &  55.9$\pm$0.8 &   4.6$\pm$0.1 & $ 9.68_{-0.06}^{+0.31}$ & -20.72$\pm$0.02 &  0.26 &  0.85 &  204.3$_{-28.3}^{+28.3}$ & $  25.0_{-2.7}^{+2.9}$ \\[4pt]
LFC\_SC1\_06445 & 241.000310 & 43.359738 & 0.8204 &  -39.8$\pm$13.2 &  42.7$\pm$1.0 &   3.6$\pm$0.1 & $10.00_{-0.05}^{+0.30}$ & -20.93$\pm$0.01 & -0.03 &  1.93 &  161.7$_{-27.1}^{+27.0}$ & $  52.8_{-3.3}^{+3.1}$ \\[4pt]
LFC\_SC1\_06447 & 241.109970 & 43.360221 & 0.7150 &  +30.2$\pm$14.4 &  33.7$\pm$4.1 &   3.9$\pm$0.3 & $ 9.04_{-0.15}^{+0.31}$ & -19.01$\pm$0.03 & -0.02 &  1.51 &   47.4$_{-10.2}^{+11.2}$ & $  15.2_{-2.2}^{+2.2}$ \\[4pt]
LFC\_SC1\_06566 & 241.182370 & 43.361757 & 0.7742 &  +27.0$\pm$13.1 &  77.4$\pm$0.7 &   5.9$\pm$0.0 & $ 9.22_{-0.11}^{+0.40}$ & -19.86$\pm$0.01 &  0.20 &  1.40 &   36.6$_{-13.9}^{+23.4}$ & $  61.1_{-3.4}^{+3.5}$ \\[4pt]
LFC\_SC1\_06630 & 241.129140 & 43.364469 & 1.2114 &  +36.7$\pm$13.0 &  66.5$\pm$1.5 &   7.8$\pm$0.2 & $10.24_{-0.00}^{+0.16}$ & -20.91$\pm$0.08 &  0.75 &  1.08 &   76.8$_{-16.2}^{+25.7}$ & $  17.7_{-3.0}^{+31.4}$ \\[4pt]
LFC\_SC1\_06692 & 241.031000 & 43.363715 & 1.2456 &  +76.6$\pm$13.1 &  72.2$\pm$1.1 &   6.5$\pm$0.4 & $ 9.76_{-0.14}^{+0.16}$ & -21.55$\pm$0.48 &  0.35 &  1.40 &  100.3$_{-24.2}^{+27.2}$ & $  31.5_{-24.9}^{+23.3}$ \\[4pt]
LFC\_SC1\_06953 & 241.191330 & 43.370455 & 0.9327 &  +89.5$\pm$13.1 &  63.3$\pm$1.2 &   7.3$\pm$0.4 & $ 9.31_{-0.07}^{+0.06}$ & -20.00$\pm$0.04 &  1.32 & -1.54 &  144.3$_{-26.2}^{+25.8}$ & $  41.3_{-25.2}^{+11.8}$ \\[4pt]
LFC\_SC1\_07038 & 241.158390 & 43.372251 & 1.2688 &   -6.8$\pm$13.2 &  50.6$\pm$1.4 &  11.6$\pm$0.2 & $10.59_{-0.19}^{+0.02}$ & -22.03$\pm$0.06 &  0.66 &  1.64 &  232.8$_{-24.3}^{+25.0}$ & $  25.0_{-4.2}^{+4.2}$ \\[4pt]
LFC\_SC1\_07098 & 241.169340 & 43.373581 & 1.1764 &   +8.0$\pm$13.0 &  77.6$\pm$1.0 &  11.0$\pm$0.4 & $ 9.95_{-0.37}^{+0.05}$ & -21.02$\pm$0.08 &  0.83 &  0.49 &  157.1$_{-9.1}^{+9.5}$ & $  29.2_{-9.6}^{+6.5}$ \\[4pt]
LFC\_SC1\_07191 & 240.782510 & 43.372464 & 0.7886 &  +75.4$\pm$13.7 &  25.0$\pm$1.2 &   4.2$\pm$0.2 & $10.93_{-0.05}^{+0.00}$ & -22.10$\pm$0.01 &  0.44 &  1.83 &  314.2$_{-148.2}^{+265.5}$ & $ 156.5_{-18.1}^{+15.5}$ \\[4pt]
LFC\_SC1\_07212 & 240.954850 & 43.376347 & 0.9502 &  -10.8$\pm$13.0 &  45.1$\pm$1.2 &   6.9$\pm$0.2 & $10.15_{-0.00}^{+0.05}$ & -21.16$\pm$0.02 &  0.18 &  1.37 &  155.4$_{-13.5}^{+16.4}$ & $  24.6_{-2.6}^{+2.6}$ \\[4pt]
LFC\_SC1\_07506 & 240.933350 & 43.382706 & 0.9495 &  -60.9$\pm$13.0 &  64.9$\pm$2.5 &   6.7$\pm$1.2 & $ 8.88_{-0.00}^{+0.00}$ & -19.58$\pm$0.09 &  0.40 &  1.41 &   62.0$_{-9.8}^{+9.3}$ & $  25.3_{-3.9}^{+4.7}$ \\[4pt]
LFC\_SC1\_07581 & 241.080550 & 43.383325 & 0.7782 &  -64.0$\pm$13.0 &  61.2$\pm$1.5 &   3.4$\pm$0.3 & $ 9.29_{-0.06}^{+0.00}$ & -19.37$\pm$0.01 &  0.20 &  1.70 &   79.6$_{-13.2}^{+62.5}$ & $  28.9_{-3.3}^{+14.6}$ \\[4pt]
LFC\_SC1\_07604 & 240.929690 & 43.385125 & 0.7874 &  +77.9$\pm$13.0 &  76.2$\pm$0.8 &  11.6$\pm$0.7 & $ 9.93_{-0.06}^{+0.06}$ & -20.06$\pm$0.02 &  0.34 &  1.33 &   84.6$_{-8.6}^{+8.6}$ & $  28.6_{-3.3}^{+8.5}$ \\[4pt]
LFC\_SC1\_07772 & 241.023590 & 43.388684 & 0.8763 &  +28.1$\pm$13.0 &  37.8$\pm$3.9 &   6.8$\pm$0.6 & $ 9.11_{-0.00}^{+0.06}$ & -19.38$\pm$0.06 &  0.18 &  1.27 &   38.0$_{-12.8}^{+9.5}$ & $  26.5_{-2.7}^{+2.7}$ \\[4pt]
LFC\_SC1\_07823 & 240.906640 & 43.388358 & 0.8998 &  -25.8$\pm$13.0 &  22.1$\pm$2.4 &   5.6$\pm$0.1 & $10.81_{-0.06}^{+0.05}$ & -21.67$\pm$0.02 &  0.71 & -0.34 &  160.9$_{-47.9}^{+66.7}$ & $  36.3_{-8.5}^{+8.3}$ \\[4pt]
LFC\_SC1\_07838 & 241.016040 & 43.390109 & 0.6125 &  -76.5$\pm$13.0 &  60.5$\pm$2.2 &   6.3$\pm$0.8 & $ 8.71_{-0.05}^{+0.19}$ & -18.22$\pm$0.02 &  0.14 &  1.00 &   61.2$_{-5.2}^{+7.2}$ & $  18.6_{-2.4}^{+2.5}$ \\[4pt]
LFC\_SC1\_08073 & 240.944320 & 43.395143 & 0.9666 &  -48.1$\pm$13.0 &  64.7$\pm$1.2 &   3.2$\pm$0.2 & $ 9.47_{-0.19}^{+0.30}$ & -19.85$\pm$0.05 &  0.08 &  1.82 &   87.6$_{-16.9}^{+20.2}$ & $  24.1_{-2.6}^{+26.0}$ \\[4pt]
LFC\_SC1\_08141 & 240.897370 & 43.393261 & 0.7294 &  -45.8$\pm$13.0 &  42.1$\pm$0.7 &   9.8$\pm$0.2 & $10.68_{-0.00}^{+0.05}$ & -21.73$\pm$0.01 &  0.69 &  1.64 &  236.2$_{-14.1}^{+14.1}$ & $  30.2_{-3.0}^{+3.0}$ \\[4pt]
LFC\_SC1\_08346 & 240.918560 & 43.400770 & 0.7813 &  -72.6$\pm$13.8 &  27.5$\pm$3.7 &   5.6$\pm$0.2 & $ 9.90_{-0.12}^{+0.24}$ & -20.82$\pm$0.01 &  0.80 &  0.92 &  139.3$_{-18.0}^{+19.1}$ & $  28.7_{-3.2}^{+12.2}$ \\[4pt]
LFC\_SC1\_08387 & 240.929440 & 43.402396 & 0.9464 &  -73.9$\pm$13.0 &  74.8$\pm$1.2 &  10.5$\pm$0.6 & $ 9.42_{-0.06}^{+0.00}$ & -20.15$\pm$0.07 &  0.49 &  1.38 &  126.6$_{-22.0}^{+31.6}$ & $  39.5_{-7.3}^{+7.5}$ \\[4pt]

\hline
\end{tabular}
\end{table}
\end{landscape}

\begin{landscape}
\begin{table}
\centering
\contcaption{}
\label{tab_app:gal_par}
\begin{tabular}{cccccccccccccccc}
\hline\hline 
ID  & RA & Dec & \textit{z} & PA & Incl & R$_{2.2}$ & $log(M_\ast/M_\odot$) & $M_B$ & $log(1+\delta_{gal})$ & $log \eta$ & $V_{2.2}$  & $\sigma_0$\\[5pt]
 & deg & deg & & deg & deg & kpc &   & mag &  &  & km/s & km/s \\  [5pt]
 (1) & (2)  & (3)  & (4) & (5) &(6)  & (7)  & (8) & (9) & (10) & (11) & (12) & (13) \\
\hline

LFC\_SC1\_08436 & 240.978700 & 43.402296 & 0.8792 &  +42.6$\pm$13.2 &  42.1$\pm$2.6 &   8.1$\pm$0.6 & $ 9.74_{-0.07}^{+0.00}$ & -20.50$\pm$0.02 &  0.24 &  0.94 &   67.0$_{-14.2}^{+13.8}$ & $  40.3_{-4.4}^{+3.7}$ \\[4pt]
LFC\_SC1\_08554 & 240.937540 & 43.405199 & 0.8803 &  +70.2$\pm$13.0 &  65.3$\pm$0.7 &  10.8$\pm$1.1 & $10.67_{-0.40}^{+0.00}$ & -20.75$\pm$0.04 &  0.44 &  0.23 &  128.7$_{-41.0}^{+46.9}$ & $  63.8_{-24.1}^{+25.6}$ \\[4pt]
LFC\_SC1\_08746 & 241.119470 & 43.409787 & 0.7890 &  -63.1$\pm$13.4 &  47.7$\pm$2.4 &   6.0$\pm$0.8 & $ 8.84_{-0.06}^{+0.15}$ & -19.55$\pm$0.02 &  0.40 &  1.70 &   31.4$_{-13.1}^{+14.4}$ & $  34.1_{-4.9}^{+4.9}$ \\[4pt]
LFC\_SC1\_08969 & 240.790910 & 43.415146 & 0.8742 &  +87.8$\pm$14.3 &  43.7$\pm$2.5 &   6.0$\pm$0.6 & $ 9.16_{-0.08}^{+0.18}$ & -19.63$\pm$0.05 &  0.08 &  1.26 &   53.2$_{-28.1}^{+49.4}$ & $  32.4_{-8.5}^{+5.6}$ \\[4pt]
LFC\_SC2\_07455 & 241.131820 & 43.109358 & 0.8648 &  -27.5$\pm$13.3 &  28.9$\pm$11.1 &   2.5$\pm$0.9 & $ 9.08_{-0.18}^{+0.21}$ & -18.31$\pm$0.02 & 0.19 &  0.88 &   36.0$_{-36.0}^{+66.2}$ & $  32.4_{-14.1}^{+7.9}$ \\[4pt]
LFC\_SC2\_07567 & 241.134450 & 43.112453 & 1.1883 &   -6.0$\pm$13.3 &  56.2$\pm$1.4 &   3.9$\pm$0.3 & $ 9.75_{-0.21}^{+0.16}$ & -21.06$\pm$0.03 &  0.08 &  1.02 &   75.2$_{-20.5}^{+30.7}$ & $  41.4_{-2.7}^{+2.5}$ \\[4pt]
LFC\_SC2\_08383 & 241.064220 & 43.137042 & 0.8999 &   -7.0$\pm$13.0 &  72.7$\pm$0.7 &   7.3$\pm$0.3 & $10.79_{-0.11}^{+0.05}$ & -20.64$\pm$0.04 &  0.09 & -0.29 &  287.3$_{-10.2}^{+12.4}$ & $  47.0_{-10.4}^{+12.5}$ \\[4pt]
LFC\_SC2\_10473 & 241.158780 & 43.256062 & 0.6439 &  +19.2$\pm$13.1 &  58.1$\pm$0.7 &   2.9$\pm$0.1 & $ 9.84_{-0.10}^{+0.05}$ & -19.88$\pm$0.01 &  0.07 &  1.28 &   69.9$_{-12.3}^{+12.6}$ & $  53.8_{-4.3}^{+3.3}$ \\[4pt]

\hline
\end{tabular}
\end{table}
\end{landscape}


\bsp	
\label{lastpage}
\end{document}